\documentclass{aa}[twocolumn]  

\usepackage{graphicx}
\usepackage{txfonts}
\usepackage{amsmath}
\usepackage{xcolor}
\usepackage{color}
\usepackage{placeins}
\usepackage{capt-of}
\definecolor{dblue}{rgb}{0.0, 0.0, 0.55}
\usepackage[colorlinks=true,urlcolor=dblue,citecolor=dblue,
            linkcolor=dblue]{hyperref}
\usepackage[version=4]{mhchem}
\usepackage[normalem]{ulem}

\def\Tg{T_{\rm g}}
\def\Td{T_{\rm d}}
\def\nH{n_{\langle\rm H\rangle}}

\begin{document}

\title{Molecules in the shadows of the protoplanetary disk HD 143006}
\subtitle{Evidence for C/O\,$\gtrsim$\,1 from simultaneously modeling fourteen ALMA lines}

\author{S.\,E.\ van Terwisga\inst{1}
        \and 
        P.\ Woitke\inst{1}
        \and 
        A.\,M.\ Weiss\inst{1,2}
        \and 
        A. D. Bosman\inst{3}
        \and
        L. Trapman\inst{4}
}

\institute{Space Research Institute, Austrian Academy of
           Sciences, Schmiedlstr. 6, 8042 Graz, Austria,
           {\sf sierkeyse.vanterwisga@oeaw.ac.at}
           \and 
           Leibniz Institute for Astrophysics Potsdam (AIP), 
           An der Sternwarte 16, 14482 Potsdam, Germany
           \and
           Space Research Organisation Netherlands, Niels Bohrweg 4, 2333 CA Leiden, The Netherlands
           \and
           Department of Astronomy, University of Wisconsin-Madison, 475 N,
           Charter St., Madison, WI 53706, USA
}

\date{Received June 16, 2026 / accepted September 21, 2026; accepted version prior to language editing.}

\abstract
  {The inner disk of the T\,Tauri star HD\,143006 is inclined and casts a broad shadow on a large portion of the outer disk.  We use this extraordinary geometry to study the impact of stellar ultraviolet (UV) irradiation on disk chemistry, temperature, and line formation.  We present new ALMA observations of this disk, covering CO, $^{13}$CO, C$^{18}$O, CN, HCN, HCO$^+$, CS and C$_3$H$_2$.  Combined with previously published C$_2$H, N$_2$H$^+$ and H$_2$CO observations, these data constitute an extensive dataset of fourteen ALMA lines from nine different molecules. We subdivide these data into two disk halves: the illuminated part and the shadowed part.  We develop ProDiMo models for these two disk halves by optionally including an inclined inner disk that blocks the stellar illumination of the upper layers of the outer disk while it does not hinder the stellar illumination of the outer disk along the midplane. The observed line flux contrasts between the illuminated and the shadowed disk sides are not very pronounced.  We explain this phenomenon by introducing ``equilibrated models'' which consider a slow heating/cooling relaxation of gas parcels orbiting in and out of the shadow.  Our results fit the radial line intensity profiles of all ALMA lines, both in flux and radial shape, with an overall reduced $\chi$ of about 2.3, both from the illuminated and the shadowed disk sides, along with two ALMA continuum images, where we see evidence for radial drift, and the spectral energy distribution (SED) beyond 20\,$\mu$m, all with one model.  In most cases, the line intensities only increase slightly on the illuminated side because the gas warms up slowly, although fast UV photochemistry significantly reduces the observable molecular concentrations. The dust temperatures increase even more rapidly, but this has little effect on the line formation.  However, for the lines of N$_2$H$^+$ and CS, which probe deeper layers where the gas and dust temperatures are coupled, slow UV-desorption of ices, such as NH$_3$ and H$_2$S, plays a crucial role.  This can result in a brightening of these lines on the shadowed side.  Our analysis suggests with high confidence that the carbon-to-oxygen ratio in the outer disk of HD\,143006 is  slightly greater than one ($\sim$\,1.05). The disk mass is fitted to be as low as $0.0012\,M_{\odot}$, however, almost equally well-fitting models with a fixed disk mass of $0.021 M_\odot$ can be obtained after re-adjusting the disk shape, dust, and gas heating/cooling parameters.} 


\maketitle

\section{Introduction}

Understanding the fundamental properties of protoplanetary disks, such as mass, temperature, density structure, ionization, chemical composition, and turbulence, is essential to understand planet formation. However, constraining these quantities is difficult since many of them lack a direct observational tracer, and the typical disk is compact, distant, and faint \citep{Miotello2021, Miotello2023, Trapman2025c, Trapman2025b}. Therefore, it is essential to understand how to link observable quantities, such as molecular line luminosities and emitting areas, to physical disk parameters. This, in turn, requires a deep understanding of how radiation transport, heating and cooling, and chemistry interact in nature, and capturing these processes in sophisticated thermo-chemical models. The outcomes of such models can then be compared to - ideally - sensitive, high-resolution data of molecular emission from telescopes such as the Atacama Large (sub-)Millimeter Array (ALMA).

Fitting line observations from a larger number of molecules (in particular, from species beyond CO and its main isotopologues) is immediately challenging to thermo-chemical disk models. This is because they are slow to evaluate, and because once multiple other molecules are included, the observations start to rapidly challenge the chemical rate networks used in the models. Some recent examples include, e.g.,~\citet{Kama2016,Fedele2017,Woitke2019,Podio2020} and \citet{Furuya22}.

A further challenge in disk modeling is to advance from fitting integrated line fluxes to fitting the radial intensity profiles (often simplified to column densities), which can show gaps and rings, possibly related to the radial positions of ice-lines.  \cite{Rab2020} have fitted ALMA continuum and CO line data of HD\,163296 with the thermo-chemical ProDiMo modeling code.  \cite{Zhang2021} have used DALI (Dust And LIne,~\citealp{Bruderer2009}) models to explain ALMA continuum and CO isotopologue data. \cite{Cleeves2018} have used CO, \ce{C2H} and HCN observations to constrain the gas-phase C/O and N/O abundance ratios, and ~\citet{Leemker2024} fitted spatially-resolved line data to constrain C/O abundances in the gas of HD 100546 using CO isotopologues, [C I], HCN, CN, \ce{C2H}, NO, and \ce{HCO+} with DALI. \citet{Rivieremarichalar2026} similarly studied the S/O abundances in AB Aurigae with resolved observations of CO isotopologues, \ce{H2CO}, \ce{HCO+}, HCN, \ce{H2S}, SO and CS. While these models are chemically rich, they are usually targeted toward a specific scientific question and seek answers based on e.g. ALMA continuum and 3–4 line observations, often modeling the lines species by species. In contrast, this paper aims to constrain and explain the entire disk system of HD\,143006 with a single model, based on a large data set of ALMA continuum images, 14 lines and auxiliary data such as SED fluxes and molecular line emission heights. 

Protoplanetary systems with a misaligned inner disk, which casts a broad shadow over one side of the outer disk, are of particular interest in this context. As material in the outer disk orbits in and out of the shadow, it is exposed to a varying stellar radiation field in terms of the ultraviolet (UV), optical, near infrared (IR) and X-ray irradiation. Many of the underlying gas and dust properties should not change during these orbits, such as element abundances, dust properties and gas-to-dust ratio. Moreover, the heating and cooling behavior of the gas and the dust is in principle predictable. Thus, these disks allow us to eliminate some of the key unknowns in protoplanetary disk chemistry and evolution. In disks where the shadow spans a wide range of azimuth, we expect that the orbital timescales are comparable to the heating and cooling times of the disk gas, and that any chemical effects can be spatially resolved by observations.

So far, only a small number of such broad-shadowed disks have been discovered, see \citet{Benisty2023} for a review, and fewer have been targeted by high-resolution observations of molecular lines, but these observations have already produced a number of fascinating insights. In the HD 100546 system, a broad shadow was proposed to drive the observational asymmetry between CS and CO by \citet{Keyte2023}. The authors of that paper attribute this asymmetry to an azimuthally varying C/O ratio in the gas phase\footnote{Our result do not support this hypothesis. See also footnote~\ref{foot:C/O} for a discussion of the definition and meaning of C/O.}, as a result of more efficient freeze-out in the shadow. Other molecules in this disk also show notable azimuthal asymmetries, such as HCO$^+$, HCN, CN and C$_2$H. These asymmetries occur at similar radii and azimuths for all molecules, and have likewise been linked to the possible presence of a broad shadowed region~\citep{Booth2024}. The geometry of this shadow, however, remains somewhat unclear. Scattered-light observations reveal a complex outer-disk morphology~\citep{Garufi2016}, and suggest the shadow towards the south-southwest of the disk is possibly interrupted by a bright lane.

In other systems, clear chemical asymmetries may not be caused (primarily) by shadows. While~\citet{Temmink2023} suggested shadows might be responsible for CS $J$=7-6 and and HCN $J$=4-3 asymmetries in HD 142527, further observations led ~\citet{Temmink2026} to conclude that infall, rather than shadows, are the dominant cause for these features.

Some molecular emission lines are therefore sensitive to the presence of disk shadows. Others, however, show no significant impact at all, such as $^{12}$CO in HD\,100546. In that latter disk, an asymmetric underlying continuum emission may affect (some of) the observed azimuthal variations in molecular abundance. Another puzzling observation is that the line intensity contrasts between the illuminated and the non-illuminated locations are often rather small: in HD\,100546, the contrast between the brightest and faintest CN emission at radii of $\sim\!200$\,au is only of the order of $\sim\!10\%$.

This result is in some contrast to the models by~\citet{Young2021}, who combined hydrodynamical models with chemical modeling and ray-tracing. These authors predict that the most obvious differences in molecular line intensity between the shadowed and illuminated side are expected in $^{12}$CO and HCO$^+$, as well as SO. However, their model assumes a small ($12^\circ$) misalignment in the inner and outer disks and, as the authors point out, their results are not scale-free and the relevant chemical processes depend on the density and (average) temperature profiles of the disk as well as the specific parameters of the warp.

In general, even for the well-studied HD\,100546 system, it has proven difficult to explain the different molecular line behaviors in a single model. For instance, it is unclear if the proposed azimuthal C/O variation is consistent with the other molecular data. Partly, this is due to the lack of (high-resolution) data. In particular, information on the vertical emitting height of some of the molecular gas lines could provide a very strong constraint, but was not (yet) available to these authors, and some data are only available at quite low resolution.

In this study, we present new ALMA observations for a very well-characterized system with an azimuthally broad shadow in the outer disk: HD\,143006. We develop a physical-chemical disk model for this object using ProDiMo~\citep{Woitke2009,Woitke2016}. Our modeling aims at reproducing the line intensities as function of radius of nine different molecules observed with ALMA, inside and outside of the shadow cast by an inclined inner disk.  Our model also needs to be consistent with previous observations of the heights of the $^{12}$CO emitting surface over the midplane, and with the spectral energy distribution (SED). In this way, we can self-consistently model some key quantities that are responsible for the line formation: the lack of UV photons, the lower temperatures, and the associated chemical and cooling relaxation timescales.

HD\,143006 is a system uniquely suited for this type of analysis. Its broad shadow has been characterized by SPHERE and SCExAO scattered light images taken at multiple epochs between 2016 and 2022 \citep{Benisty2018,Ren2023,Mullin2026}. Analysis of these images confirms that the outer disk shadow does not change significantly in depth or azimuthal width on short ($\sim\!5$\,yr) timescales. In parallel, the system has been observed at high resolution by the DSHARP Large Program (2016.1.00484.L; ~\citealt{Andrews2018,Perez2018}) and the exoALMA Large Program (2021.1.01123.L; ~\citealt[e.g.][]{Teague2025}). Additionally, \citet{Codron2025} have studied the complex inner disk structure with infrared interferometry. Consequently, we have an incredibly rich observational dataset on which to base our modeling of the molecular lines. 

While our analysis does not sensitively depend on the physical cause for the misaligned inner and outer disks in the HD 143006 system, it is nonetheless important to consider. ~\citet{Ballabio2021} favor a scenario where an embedded planet orbits a misaligned binary star, and predict the shadow may move on longer timescales than can currently be tested. However, alternative explanations, such a flyby-induced misalignment~\citep{Nealon2020} or the infall and capture of interstellar material with a different net angular momentum than the original disk~\citep[e.g.][]{Dullemond2019, Kuffmeier2021, Gupta2023, Krieger2024} can also lead to misalignments similar to those seen in HD 143006. No strong predictions have been made in the literature of how, if at all, these different mechanisms for inducing misalignment may influence the chemical abundances of the inner and outer disk gas.

\begin{table*}
\caption{Beam parameters and measured molecular line fluxes from the illuminated left disk side and the shaded right disk side of HD\,143006.
\label{tab:fluxes_mol}}
\vspace*{-2mm}
\begin{tabular}{ccccccccccc}
\hline
molecule & transition & $\nu$ & $b_{\rm maj}$   & $b_{\rm min}$ & PA & $\epsilon$\tablefootmark{a} & 2$\times$Flux left & 2$\times$Flux right & left / right\\
         &            & [GHz] & $[\rm{arcsec}]$ & $[\rm{arcsec}]$ & $[\rm{deg}]$ & & [mJy\,km/s] & [mJy\,km/s] & \\
\hline
&&&&&&\\*[-2.0ex]
$^{12}$CO & $J$=3-2 & 345.796 & 0.16 & 0.12 & -61.7 & 0.44 & $6149 \pm 20$ & $5434 \pm 20$ & $1.13 \pm 0.01$ \\
$^{13}$CO & $J$=3-2 & 330.588 & 0.21 & 0.18 & 71.7 & 0.49 & $1684 \pm 16$ & $1546 \pm 16$ & $1.09 \pm 0.02$ \\
C$^{18}$O & $J$=3-2 & 329.331 & 0.30 & 0.25 & 87.2 & 0.35 & $396 \pm 5$ & $392 \pm 5$ & $1.01 \pm 0.02$ \\
CN & $N$=3-2 $J$=5/2-3/2 & 340.035\tablefootmark{b} & 0.20 & 0.17 & 64.7  & 0.48 & $1715 \pm 17$ & $1459 \pm 17$ & $1.18 \pm 0.02$ \\
CN & $N$=3-2 $J$=7/2-5/2 & 340.248\tablefootmark{c} & 0.20 & 0.17 & 64.1 & 0.47 & $2169 \pm 13$ & $1906 \pm 13$ & $1.14 \pm 0.01$ \\
HCN & $J$=4-3 & 354.506\tablefootmark{d} & 0.16 & 0.12 & -61.3 & 0.44 & $1943 \pm 36$ & $1683 \pm 36$ & $1.15 \pm 0.03$ \\
CS & $J$=7-6 & 342.883 & 0.28 & 0.23 & 80.3 & 0.36 & $354 \pm 5$ & $376 \pm 5$ & $0.94 \pm 0.02$ \\
HCO$^+$ & $J$=4-3 & 356.734 & 0.15 & 0.11 & -62.7 & 0.43 & $2425 \pm 36$ & $1903 \pm 35$ & $1.27 \pm 0.03$ \\
N$_2$H$^+$ &$J$=3-2 & 279.517 & 0.69 & 0.44 & -89.0 & 0.51 & 325 $\pm$ 4 & 331 $\pm$ 4 & 0.98 $\pm$ 0.02\\
c-C$_3$H$_2$ & $4_{3,2}$-$3_{0,3}$ & 354.240 & 0.24 & 0.20 & -81.0 & 0.28 & $21 \pm 4$ & $11 \pm 4$ & $1.93 \pm 0.81$ \\
\hline
&&&&&&&\\[-2.1ex]
C$^{18}$O & $J$=2-1 & 219.560 &
  \multicolumn{3}{l}{\citet{Pegues20}} &  
  \multicolumn{2}{c}{$135\pm 17$} & --\\
HCN & $J$=3-2       & 265.886\tablefootmark{e} &
  \multicolumn{3}{l}{\citet{Bergner19}} &  
  \multicolumn{2}{c}{$2060\pm 210$} & --\\       
C$_2$H & $N$=3-2 $J$=7/2-5/2 $F$=4-3 & 262.004 &
  \multicolumn{3}{l}{\citet{Bergner19}} &  
  \multicolumn{2}{c}{$435\pm 56$} & --\\
p-H$_2$CO & $3_{0,3}$-$2_{0,2}$ & 218.222 &
  \multicolumn{3}{l}{\citet{Pegues20}} &  
  \multicolumn{2}{c}{$16\pm 2$} & --\\
\hline
\end{tabular}
\tablefoot{The entries below are total line fluxes taken from the literature. 
\tablefoottext{a}{Beam volume ratio up to the first null for JvM correction following~\citet{Czekala2021}.}
\tablefoottext{b}{Fluxes in this paper correspond to the sum of the hyperfine lines at 340.0196, 340.0315, and 340.035\,GHz}
\tablefoottext{c}{Fluxes in this paper correspond to the sum of the blended hyperfine lines at 340.24777 and 340.248544\,GHz}
\tablefoottext{d}{Fluxes in this paper correspond to the sum of the hyperfine lines at 354.5053665, 354.5054773, 354.5055229, and 354.5058459\,GHz.}
\tablefoottext{e}{This molecule has six fully blended hyperfine transitions at 265.886\,GHz.}
}
\vspace*{2mm}
\end{table*}

\begin{table*}
\caption{Left/right integrated fluxes for the continuum images. \label{tab:fluxes_cont}}
\vspace*{-2mm}
\begin{tabular}{ccccc ccc}
\hline
Continuum & $\lambda\,[\mu\rm m]$ & $b_{\rm maj}$ & $b_{\rm min}$ & PA & 2$\times$Flux left [Jy] & 2$\times$Flux right [Jy] & left / right\\
\hline
&&&&&&\\*[-2.0ex]
DSHARP & 1255 & 0.05'' & 0.04'' & 84.0$^\circ$ & 0.069 $\pm$ 0.003 & 0.059 $\pm$ 0.003 & 1.180  $\pm$ 0.070 \\
exoALMA & 908 & 0.14'' & 0.10'' & -75.9$^\circ$ & 0.183 $\pm$ 0.005 & 0.131 $\pm$ 0.005 & 1.405  $\pm$ 0.064 \\
\hline
\end{tabular}
\end{table*}

\smallskip
In Sect.\,\ref{sec:observations} we describe the new ALMA observations of HD\,143006 and our data reduction approach. We also describe our auxiliary data collection in this section, which includes previous ALMA observations, photometric, UV and X-ray data, and how we determined the stellar properties.  Section~\ref{sec:obsdata} describes how we subdivided the data into an illuminated and a shadowed disk side, and extracted azimuthally averaged radial intensity profiles from these data both continuum and line.  In Sect.\,\ref{sec:ProDiMo} we describe our modeling strategy with ProDiMo, including the introduction of a quasi-inclined inner disk, how we take into account the effects of long cooling timescales, and the fitting algorithm. In Sect.\,\ref{sec:MainModel} we present the main, best-fitting model and in Sect.\,\ref{sec:discussion} we discuss our results for the disk mass and the C/O ratio in relation to previous work. Sect.\,\ref{sec:conclusion} contains our conclusions.

\section{Observations and data reduction}
\label{sec:observations}

\begin{figure}
    \vspace*{-1mm}\hspace*{-1mm}
    \includegraphics[width=\linewidth, trim=5 0 5 0, clip]{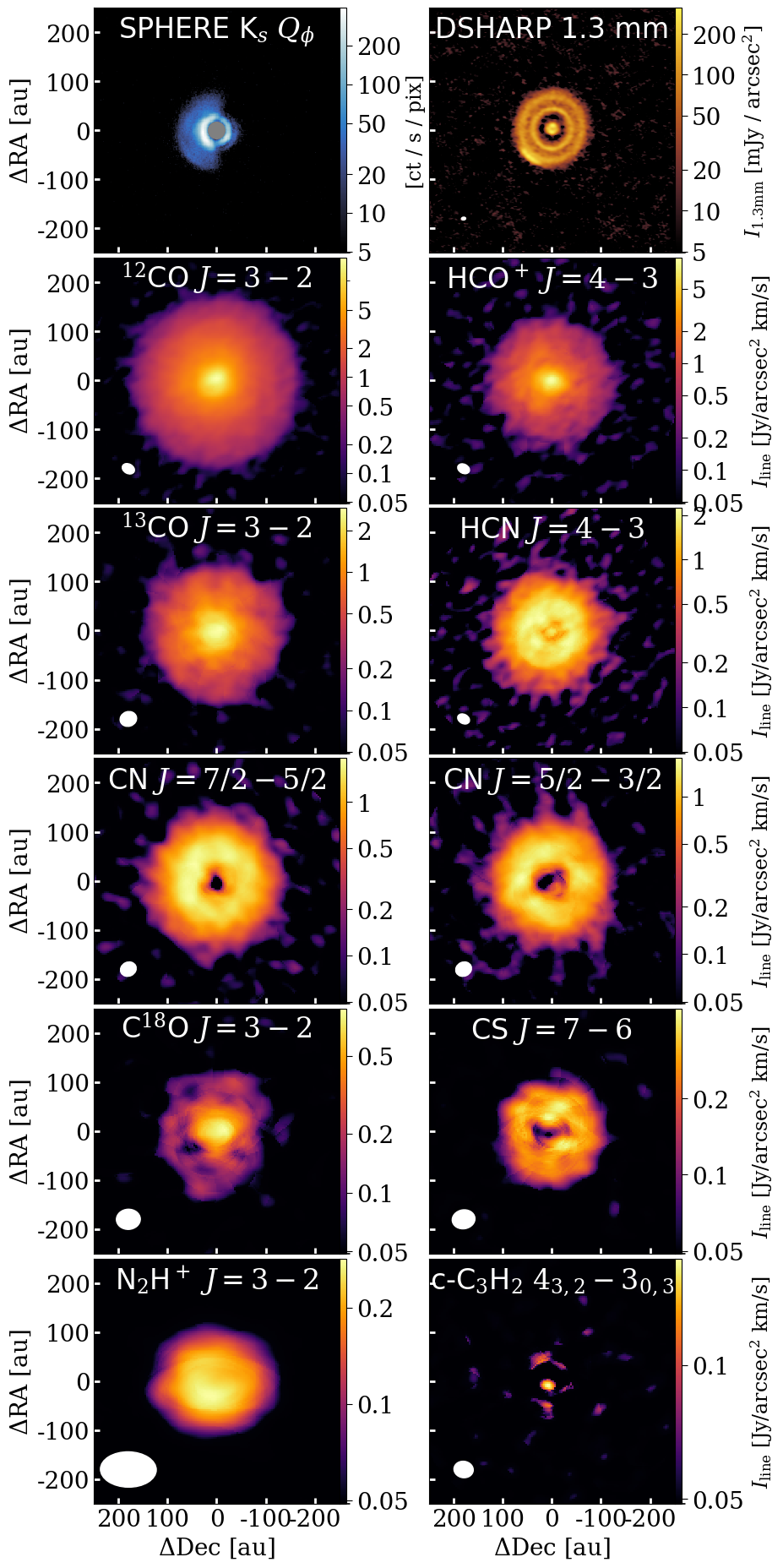}
    \vspace*{-6mm}
    \caption{SPHERE K$_s$ \citep{Ren2023}, ALMA 1.3\,mm \citep{Perez2018} and moment-zero maps for the molecular lines in ALMA proposals 2021.1.01683.S and 2021.1.00334.S that were used for this paper, sorted by peak intensity. The western (right) part of the disk is in the scattered-light shadow.
    \label{fig:all_mol}}
    \vspace*{-2mm}
\end{figure}

\begin{figure}
    \vspace*{-2mm}\hspace*{-2mm}
    \includegraphics[width=93mm,height=164mm]{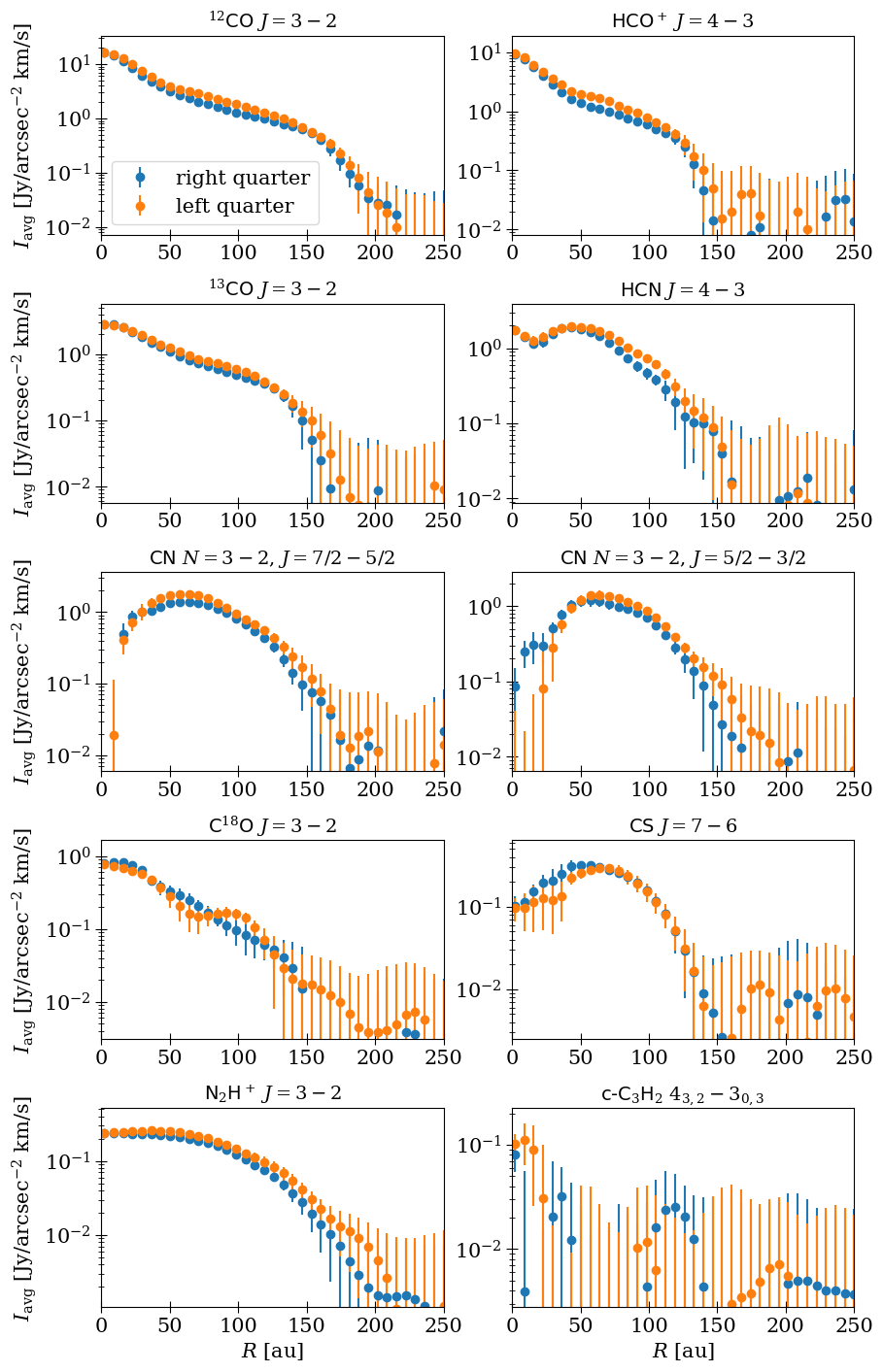}\\*[-6mm]
    \caption{Radial intensity profiles for the lines shown in Fig~\ref{fig:all_mol} for the illuminated right (blue) and shadowed left (orange) side. The used masks are shown in Fig.\,\ref{fig:example_radialprofile_mask}.}
    \label{fig:radial_line_intensity_profiles}
    \vspace*{-3mm}
\end{figure}

\subsection{ALMA observations}
\label{sec:alma_observations}

The primary test of how irradiation conditions affect disk chemistry comes from a large dataset consisting of both previously-published and newly-presented ALMA observations of molecular lines towards HD 143006. Additionally, we used archival continuum data from the DSHARP~\citep{Andrews2018, Perez2018} and exoALMA~\citep{Teague2025,Curone2025}. In this section, we discuss the previously unpublished ALMA data and the properties of the archival observations.

\subsubsection{Calibration and reduction of the new ALMA data}
\label{sec:obsdata}

To be sensitive to the presence of molecular line emission gradients from the shadowed to the illuminated side of the disk, HD 143006 was observed at high resolution in Band 7 by ALMA project 2021.1.01683.S. To achieve a nominal resolution of $\sim 0.2''$ without losing sensitivity to total flux, observations were carried out in both short- and long-baseline configurations in December 2019 and April 2021. To observe a large number of lines, including the photochemically interesting species CN, HCN, and CO, two different spectral setups were necessary, both with six spectral windows. In Setup A, spectral windows are centered at 340.020\,GHz for CN $N$=3-2,\,$J$=5/2-3/2 at 340.269\,GHz, for CN $N$=3-2,\,$J$=7/2-5/2 at 342.887\,GHz, for CS $J$=7-6 at 329.334\,GHz, and for C$^{18}$O $J$=3-2 at 330.592\,GHz, for $^{13}$CO $J$=3-2, with a continuum spectral window covering these CO isotopologue transitions at low spectral resolution at 330.004\,GHz. Setup B spectral windows centered at 356.738\,GHz target HCO$^+$ $J$=4-3 at 354.244\,GHz, c-C$_3$H$_2$ $4_{3,2} - 3_{0,3}$ at 354.509\,GHz, HCN $J$=4-3 at 345.800\,GHz, $^{12}$CO $J$=3-2 at 345.800\,GHz, and H$^{13}$CN at 345.344
,GHz. The continuum spectral window (at 343.020\,GHz) contains CS $J$=7-6 with low velocity resolution. For details on the observations (baselines, observation dates, and calibrators) per spectral setup, please refer to Table~\ref{tab:calibration}.

These observations were pipeline-calibrated with \texttt{casa} and subsequently self-calibrated for phase and amplitude in order to maximize the final image signal-to-noise ratio with \texttt{casa} version 6.7.2-42~\citep{casa}. Highly accurate calibrated data are essential to this paper. First, to combine with previously-published line fluxes, we need to achieve flux calibration accuracy to the best possible level ($\sim 10\%$ at Band 7, \citealt{almatechhb}). Moreover, it was essential for this paper that the individual execution blocks (EBs) of the observations are aligned correctly, in order to prevent spurious asymmetries in the reconstructed image. For the self-calibration of these data, we have closely followed the methods presented by the exoALMA collaboration in \citep{Loomis25}. In particular, we start from a spectrally-averaged dataset with the lines flagged out. We perform a similar initial self-calibration step before aligning the data to a common phase center. After this, we concatenated the short-baseline data and performed phase calibration (down to 20\,s). We subsequently aligned and concatenated the long-baseline data to the short-baseline data, and self-calibrated the joint data down to 30\,s (Setup A) and 120\,s (Setup B) respectively, and performed a round of amplitude self-calibration. This is a slight difference to the exoALMA scheme, where the short-baseline data undergo flux scaling before concatenating, but as those authors also point out, this is in principle also possible. Finally, the self-calibration solutions were applied to the full data and the continuum subtracted using the ~\textit{contsub} task in \texttt{casa} (to first order). For this, we flagged out all channels within $\pm 10$\,km\,s$^{-1}$ of the lines except for the spectral window targeting CN $N$=3-2, $J$=5/2-3/2, where only the positions of the three brightest hyperfine lines ($F$=3/2-1/2, $F$=7/2-5/2, $F$=3/2-3/2) were flagged. We verified that this did not negatively affect the quality of the continuum subtraction for this species.

Because the $^{12}$CO, $^{13}$CO and CS lines observed in this paper are the same as those observed with exoALMA at comparable spatial (but not spectral) resolution, we could verify the absolute flux scaling of our observations after self-calibration by processing the continuum-subtracted data cubes with the same pipeline. For all of these overlapping lines, the final integrated fluxes agree to within $10\%$, as expected for ALMA \citep{almatechhb}.

Unlike the exoALMA program, whose science goals primarily emphasize resolution over recovering the total flux, this paper's analysis is crucially dependent on the accurate comparison of model fluxes with observational data over an area of the disk. For many of the lines presented here, moreover, the emission in individual channels is often faint compared to $^{12}$CO. This means that it is necessary to treat the data carefully. First, we use the combination of two Keplerian masks following the parameters derived by~\citet{Perez2018}, one corresponding to the position angle ($PA = 164.3^\circ$) and inclination ($i = 24.1^\circ$) of the inner disk and one for the outer disk with position angle $PA = 176^\circ$ and inclination $i = 17.02^\circ$. The outer radii of these masks are out to 50 AU (for the inner disk) and out to 250 AU respectively, to ensure covering all beam-convolved emission from both regions. The final clean mask is an inclusive or of the inner- and outer disk masks at each pixel.

All cubes were cleaned down to a depth of $3\sigma$. For all molecules with faint lines (C$^{18}$O, CS, and c-C$_3$H$_2$) we imaged the lines with a Briggs robust parameter of 0.5 in \texttt{tclean}; for the faint lines, a robust parameter of 1.0 was used instead for higher S/N in the resulting images. The resulting beam properties are listed in Table~\ref{tab:fluxes_mol}. Finally, in order to achieve consistent units in the image plane, we corrected for the so-called Jorsater-Van Moorsel (or JvM) effect by rescaling the residuals by the ratio of the clean beam to the dirty beam volume~\citep{JvM1995,Czekala2021}. This is particularly important for retrieving the total flux of the faint lines. Finally, we used \texttt{bettermoments}~\citep{bettermoments} to generate moment-0 maps of the data, using the Keplerian masks previously defined for the cleaning procedure. These maps are shown in Figure~\ref{fig:all_mol} for all the lines. On the same spatial scale, we also show the DSHARP 1.25\,mm continuum~\citep{Perez2018} and SPHERE K$_s$ Q$_\phi$ data from~\citep{Ren2023}.

\begin{figure}
    \vspace*{-1mm}\hspace*{-2mm}
    \includegraphics[width=\linewidth, trim={0cm 9mm 9mm 0cm}]{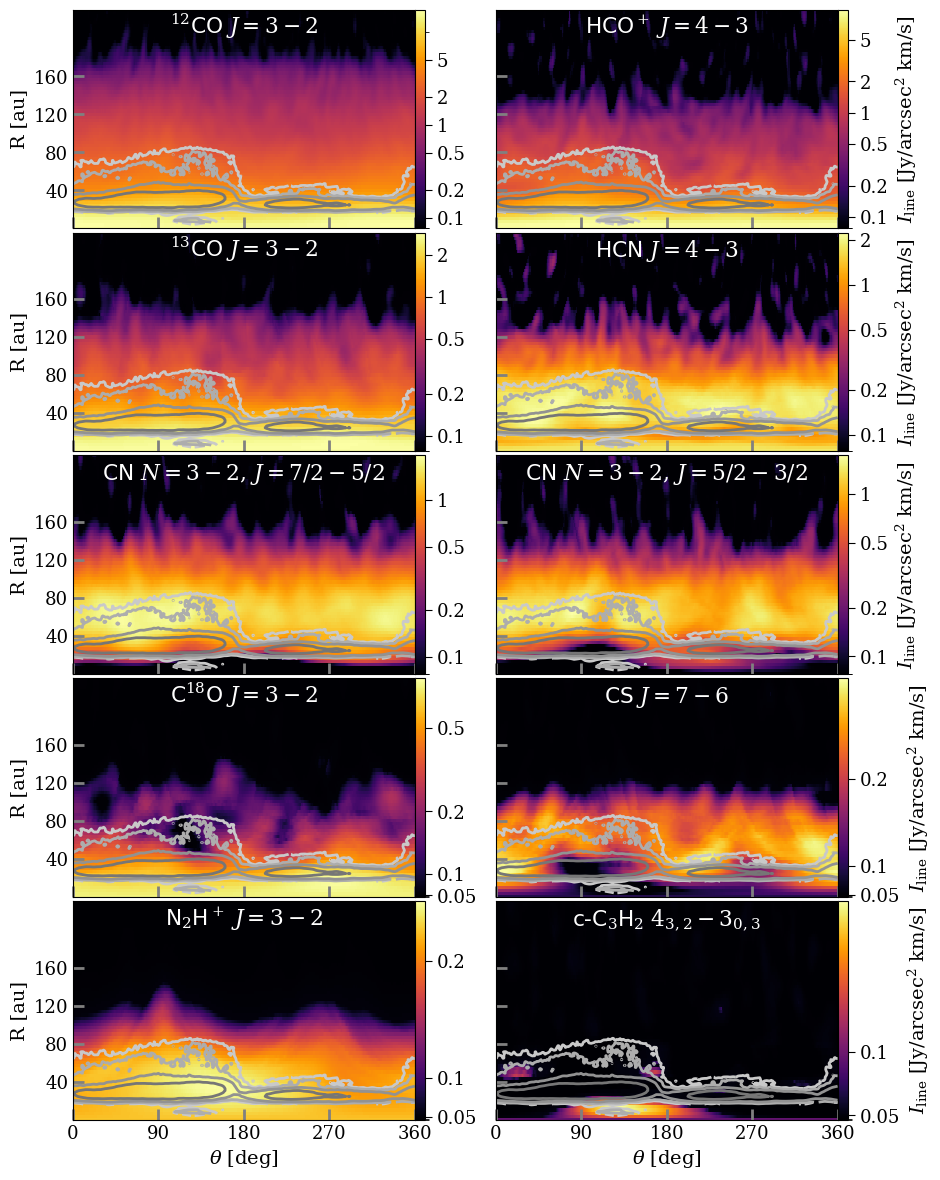}
    \vspace*{1mm}
    \caption{Polar deprojections of the moment-0 maps in Figure~\ref{fig:all_mol} in the disk frame. Overplotted contours correspond to the $[20, 32, 64, 150, 383]$\,counts pix$^{-1}$ s$^{-1}$ in the $K_s$-band $Q_\phi$ data for the 2021-07-24 epoch presented in~\citep{Ren2023}. \label{fig:polar_scattered_light}}
    \vspace*{-2mm}
\end{figure}

\begin{figure}
    \centering
    \includegraphics[width=75mm]{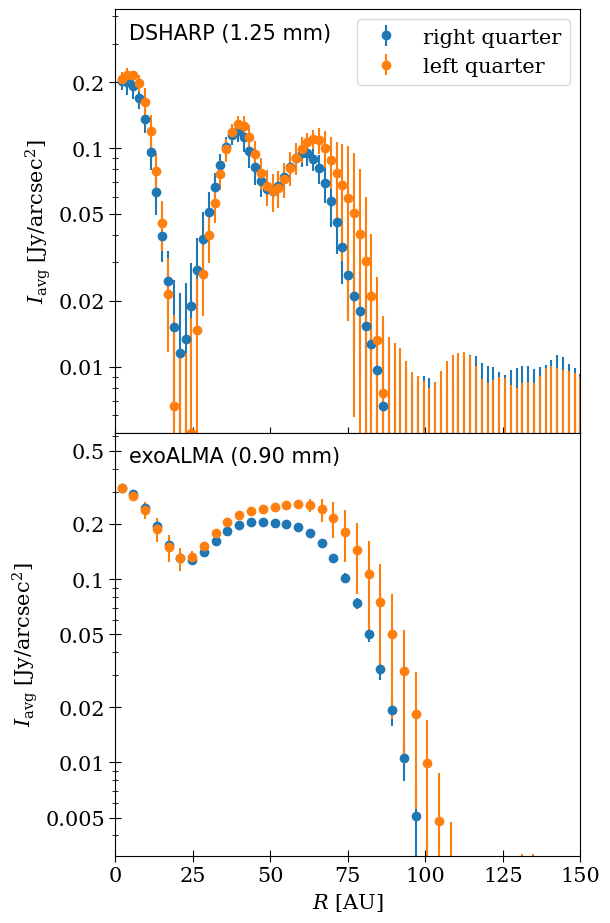}\\[-2mm]
    \caption{Continuum radial intensity profiles for DSHARP (top) and exoALMA (bottom), showing the left (orange) and right (blue) quarters of the disk).}
    \label{fig:cont_radial_profiles}
\end{figure}

\subsubsection{Archival ALMA data}
In addition to the newly presented molecular line data, we also leveraged archival and previously published images, both for the continuum and for additional spectral lines, to make sure we covered the important tracers of disk chemistry. We used the data cube of N$_2$H$+$ $J$=3-2 that was produced and published by~\citet{Trapman2025a} following a scheme very similar to the one used above for the newly presented ALMA data, including the JvM-effect correction, and treated it exactly as our observations. In addition, we take the fluxes - but not the image cubes - for C$^{18}$O $J$=2-1 and p-H$_2$CO $3_{0,3}$-$2_{0,2}$ from~\citet{Pegues20} and for HCN $J$=3-2 and C$_2$H $N$=3-2, $J$=7/2-5/2, $F$=4-3 from \citet{Bergner19} as a constraint to our models.

For the millimeter continuum data, we preferred the highest-resolution images available. In Band 6, extremely high-resolution images from DSHARP at 1.3\,mm have been published~\citep{Perez2018} and were used directly. We obtain some spectral leverage by also considering the exoALMA continuum data in Band 7 at 0.9\,mm, presented in~\citet{Curone2025}. The continuum fluxes and flux contrasts, as well as the beam parameters, are presented in Table~\ref{tab:fluxes_cont}. In the near-infrared, SPHERE $K_s$-band data exist~\citep[e.g.][]{Benisty2018,Ren2023}, but we did not use them for the fitting procedure here to limit the number of assumptions needed on the dust particle properties. Instead, we used these observations to qualitatively compare the scattered-light and molecular data in and out of the shadow.

\subsubsection{Data treatment for ProDiMo modeling}
As Figure~\ref{fig:all_mol} shows, the resolved molecular span a large range in intensity (displayed throughout in velocity-integrated, beam-independent units of Jy\,arcsec$^{-2}$\,km\,s$^{-1}$, for consistency). As expected, $^{12}$CO and HCO$^+$ are the brightest lines. Some of the molecules show centrally-peaked emission (CO and its isotopologues, as well as HCO$^+$) while CN and CS display central cavities, and HCN shows a dip at $\sim 25$\,au. N$_2$H$^+$ is difficult to interpret due to the low resolution of the data needed to retrieve this faint line at high S/N. Although the scattered-light shadow is sharp, the images immediately show that any contrast between the left (illuminated) and right (shadowed) sides is faint in molecular lines. To quantify this contrast, Table~\ref{tab:fluxes_mol} shows the difference between the left (illuminated) and right (shadowed) sides of the image split along the disk's axis (at a PA of $176.2^{\circ}$), following the~\citet{Perez2018} parameters for the outer disk geometry and the scattered-light shadow geometry~\citep{Benisty2018}. 
To visualize the molecular abundances in and out of the shadow, we compared the moment-0 maps with the scattered light data directly. For this purpose, we deprojected the moment-0 maps according to the inclination of the outer disk and de-rotated them. We treated the contours of the scattered light data in the same way and overplot them in Figure~\ref{fig:polar_scattered_light}.

Our ProDiMo models are two-dimensional. To provide quantities that can directly be compared from the model to the data, and prevent effects such as beam smearing, heating and cooling processes, and chemical timescales comparable to the local Keplerian timescale, we extract radial profiles in an arc of $90^{\circ}$ on the left and right of the images (see Figure~\ref{fig:example_radialprofile_mask}). This maximizes the S/N and minimizes any of these effects. The resulting radial intensity profiles for the left (illuminated) and right (shadowed) parts of the disk are shown in Fig.~\ref{fig:radial_line_intensity_profiles} on a logarithmic axis. For the continuum data, the same profiles are shown in~\ref{fig:cont_radial_profiles}. We did not deproject the disk for this purpose, but verified that - given the nearly face-on orientation of the outer disk - this choice does not affect the retrieved line profiles. Like in ProDiMo, the data are presented after beam convolution and in units of Jy\,arcsec$^{-2}$\,km\,s$^{-1}$.

\subsection{Photometric data collection}
\label{sec:Photometry}

Table~\ref{tab:photo} lists our photometric data collection that we obtained by means of the \href{http://vizier.cds.unistra.fr/vizier/sed/}{VizieR Photometry viewer}. Unfortunately, we could not find any Herschel PACS or SPIRE data for this source. Three millimeter points are added from different ALMA observations.

\subsection{UV data}
\label{sec:UV}
\begin{figure}
  \includegraphics[width=90mm,height=72mm,trim=3 6 3 3,clip]{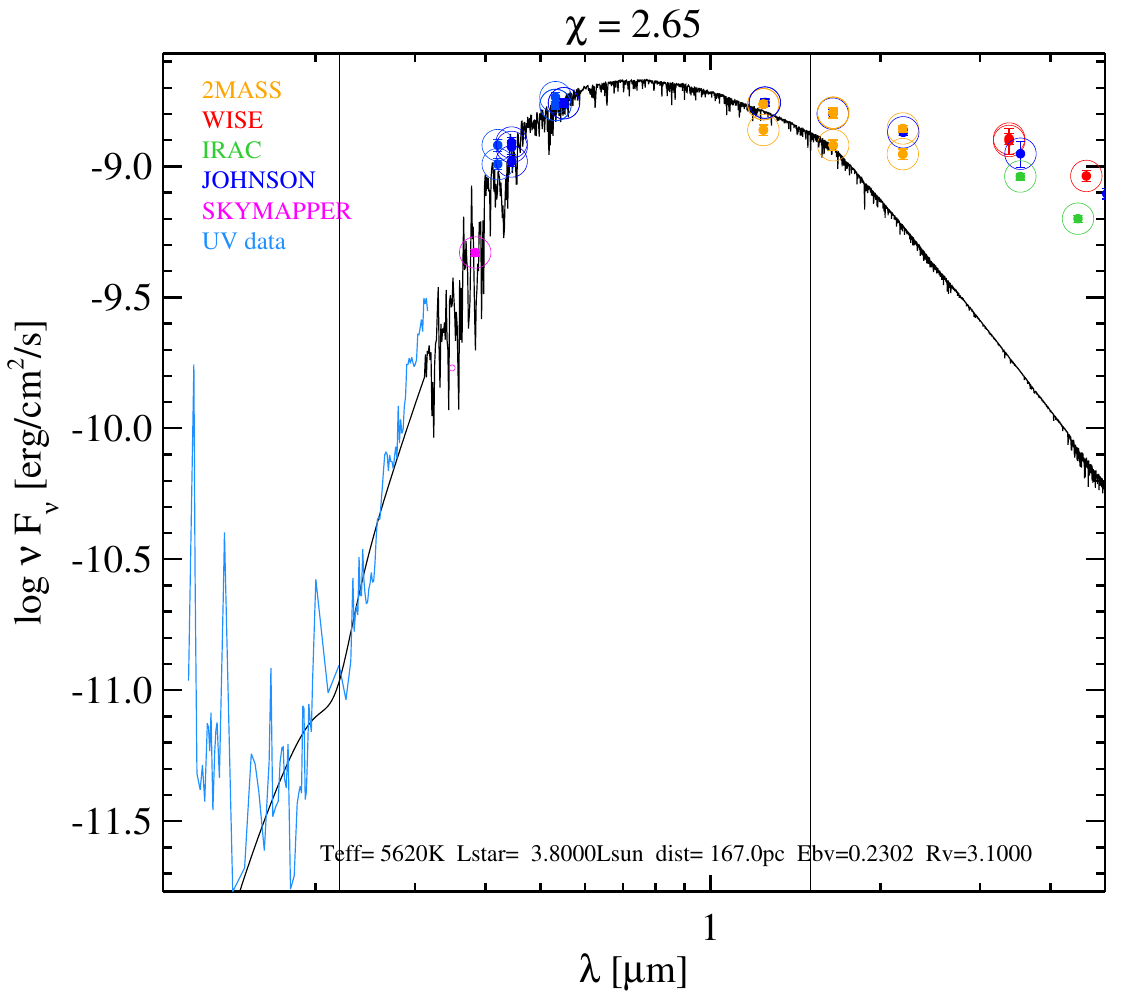}
  \vspace*{-4mm}\hspace*{-1mm}
  \caption{Fitting the reddening parameter $E_{B-V}$ to our photometric and UV data collection between 220\,nm and 1500\,nm, using fixed stellar parameters, see text. The figure title reports the $\chi^2$-value achieved by the best fit model in the title.  The model spectrum is reddened and then convolved with the respective filter transmission functions before comparing to the photometric data listed in Table~\ref{tab:photo}. The smooth black line shows our powerlaw-fit to the UV data between 220\,nm and the intersection point with the photospheric spectrum at 285\,nm. This graph appears curved in this plot because of the interstellar reddening.}
  \label{fig:SED_star}
\end{figure}

Two IUE observations {\tt lwp25951.mxlo} and {\tt lwp25951.mxlo} for HD\,143006 from the IUE program QC071 of \cite{Valenti2003} have been processed as described in \cite{Dionatos2019}.  The resulting UV spectrum shows a smooth transition to our photospheric model spectrum around 285\,nm, and a qualitative change to emission lines short-ward of about 220\,nm, most prominently the Ly\,$\alpha$ line at 121.6\,nm, see Fig.\,\ref{fig:SED_star}. To the best of our knowledge, there is no UV-data for this source for wavelengths shorter than 119\,nm.  This part of the UV-spectrum, down to 91.2\,nm, is estimated from a power-law fit shown as black line in Fig.\,\ref{fig:SED_star}.

\subsection{Stellar properties}
\label{sec:star}
The values for the stellar effective temperature $T_{\rm eff}\!=\!5620\,$K, the stellar luminosity $L_\star\!\!=\!3.8\rm\,L_\odot$, the stellar mass $M_\star\!=\!1.56\rm\,M_\odot$ and the distance 167\,pc are adopted from \cite{Trapman2025a}.  We then fitted our photometric and UV spectral data between 0.22\,$\mu$m and 1.5\,$\mu$m with a standard interstellar reddening law \citep{Fitzpatrick1999}, see Fig.\,\ref{fig:SED_star}, resulting in $A_V\!=\!0.72$ (or $E_{B-V}\!=\!0.23$ with $R_V\!=\!3.1$).  These data are consistent with a stellar radius $R_\star\!=\!2.06\rm\,R_\odot$ and $\log g\!=\!4.0$. According to the pre-main sequence stellar models of \citet{Siess2000}, the adopted stellar properties suggest a spectral type G7 and an age of about 10\,Myrs. 
The fluxes beyond 1.5\,$\mu$m are already contaminated by the disk and should not be used to determine stellar properties.

\subsection{X-ray data}
\label{sec:Xray}
HD\,143006 was observed by ROSAT to have an X-ray flux of 0.0136 counts/sec \citep{Freund2022}.  Using the count-to-flux conversion factor of \cite{Schmitt2004}, this flux corresponds to an X-ray luminosity of about $1.1\times10^{30}\rm\,erg/s$ at 167\,pc in the range 0.1-10\,keV. We use a correction factor of 1.85, inferred from a two-temperature coronal plasma model and $E_{B-V}\!=\!0.23$~\citep{Vuong2003} like those presented in XSPEC~\citep{Guedel2007}. This gives an unabsorbed X-ray luminosity of about $2\times10^{30}\rm\,erg/s$.  This value is in line with the observed relations of X-ray luminosity with stellar mass and stellar luminosity \citep{Telleschi2007}, which result in about $(2.5-4.2)\times10^{30}\rm\,erg/s$ for the stellar parameters assumed for HD\,143006.

\begin{table}
\caption{Photometric data collection for HD\,143006.}
\vspace*{-2mm}
\label{tab:photo}
\resizebox{90mm}{!}{\begin{tabular}{c|cc|c|p{34mm}}
  \hline
  &&&&\\[-2.2ex]
  $\lambda\,[\mu\rm m]$ & $F_\nu\,[\rm Jy]$ & $\sigma\,[\rm Jy]$ & filter & reference\\
  \hline
  \hline
  &&&&\\[-2.2ex]
 0.350 & 0.0198   & 0.0003  & SKYMAPPER.U & J/ApJ/925/164/catalog   \\
 0.387 & 0.060    & 0.0007  & SKYMAPPER.V & J/ApJ/925/164/catalog   \\
 0.420 & 0.169    & 0.009   & HIP.BT      & II/346/jsdc\_v2        \\
 0.420 & 0.143    & 0.007   & HIP.BT      & V/136/tycall          \\
 0.444 & 0.155    & 0.007   & JOHNSON.B   & I/305/out       \\
 0.444 & 0.183    & 0.008   & JOHNSON.B   & IV/34/epic      \\
 0.532 & 0.326    & 0.013   & HIP.VT      & II/346/jsdc\_v2         \\
 0.532 & 0.305    & 0.01    & HIP.VT      & V/136/tycall           \\
 0.532 & 0.304    & 0.011   & HIP.VT      & J/PASP/120/1128/catalog\\
 0.554 & 0.32     & 0.012   & JOHNSON.V   & J/A+A/663/A4/meanpast   \\
 1.239 & 0.715    & 0.016   & 2MASS.J     & I/317/sample    \\
 1.239 & 0.716    & 0.016   & 2MASS.J     & I/353/gsc242    \\
 1.239 & 0.57     & 0.026   & 2MASS.J     & J/other/NatAs/6.89/danc \\
 1.250 & 0.731    & 0.016   & JOHNSON.J   & II/346/jsdc\_v2  \\
 1.250 & 0.729    & 0.013   & JOHNSON.J   & J/PASP/120/1128/catalog \\
 1.630 & 0.873    & 0.032   & JOHNSON.H   & J/PASP/120/1128/catalog\\
 1.649 & 0.882    & 0.038   & 2MASS.H     & I/339/hsoy      \\
 1.649 & 0.663    & 0.031   & 2MASS.H     & J/other/NatAs/6.89/danc \\
 2.164 & 1.02     & 0.03    & 2MASS.KS    & I/317/sample    \\
 2.164 & 0.814    & 0.037   & 2MASS.KS    & J/other/NatAs/6.89/danc \\
 2.19  & 0.988    & 0.027   & JOHNSON.K   & J/PASP/120/1128/catalog \\
 3.35  & 1.41     & 0.16    & WISE.W1     & IV/38/tic       \\
 3.35  & 1.44     & 0.07    & WISE.W1     & J/ApJ/758/31/table1\\
 3.40  & 1.32     & 0.15    & JOHNSON.L   & II/346/jsdc\_v2  \\
 3.6   & 1.08     & 0.02    & IRAC.36     & J/ApJ/758/31/table1\\
 4.5   & 0.943    & 0.017   & IRAC.45     & J/ApJ/758/31/table1\\
 4.60  & 1.42     & 0.07    & WISE.W2     & IV/38/tic       \\
 5.03  & 1.32     & 0.06    & JOHNSON.M   & II/346/jsdc\_v2  \\
 8.0   & 0.771    & 0.021   & IRAC.80     & J/ApJ/758/31/table1\\
 11.6  & 0.677    & 0.009   & WISE.W3     & IV/38/tic\\
 11.6  & 0.689    & 0.013   & WISE.W3     & J/ApJ/758/31/table1\\
 18.4  & 1.83     & 0.18    & AKARI.L18W  & J/MNRAS/471/770/table1\\
 22.1  & 2.53     & 0.06    & WISE.W4     & IV/38/tic\\
 22.1  & 2.48     & 0.05    & WISE.W4     & J/ApJ/758/31/table1\\
 24.0  & 3.39     & 0.12    & MIPS.24     & J/ApJ/758/31/table1\\
 25.0  & 3.16     & 0.3     & IRAS.F25    & I/270/cpirss01\\
 60.0  & 6.57     & 0.6     & IRAS.F60    & I/270/cpirss01\\
 85.0  & 3.74     & 0.28    & AKARI.WIDES & II/298/fis\\
 100   & 4.82     & 0.4     & IRAS.F100   & I/270/cpirss01\\
 140   & 3.61     & 0.45    & AKARI.WIDEL & II/298/fis\\
 160   & 3.74     & 0.4     & AKARI.N160  & II/298/fis\\ 
 870   & 0.175    & 0.0003  & ALMA        & \cite{Testi2022}\\   
 1052  & 0.102    & 0.0004  & ALMA        & \cite{Trapman2025a}\\
 1300  & 0.059    & 0.001   & ALMA        & \cite{Trapman2025a}\\
  \hline
\end{tabular}}\\
\tablefoot{Photometric data collected from the 
\href{http://vizier.cds.unistra.fr/vizier/sed/}{VIZIER website}. 
The right column lists the references to the catalogs as generated by VIZIER.} 
\vspace*{-1mm}
\end{table}

\subsection{Spitzer IRS spectrum}
\label{sec:Spitzer}

A Spitzer/IRS low-resolution spectrum of HD\,143006, reduced by \cite{Pontoppidan2010}, PI Jeroen Bouwman, was downloaded from the \href{https://spexodisks.com/}{SpExoDisks website}. We used the data up to 31\,$\mu$m, beyond which the spectrum increases rapidly, which is inconsistent with the models and the photometric data.

\section{ProDiMo modeling of HD 143006}
\label{sec:ProDiMo}

{\sc ProDiMo} is a radiation thermo-chemical disk model introduced by \cite{Woitke2009} to simulate the chemistry, the heating and cooling balance of dust and gas, and the line and continuum radiative transfer in protoplanetary disks.  The code considers an axisymmetric (2D) disk structure with various options for multiple disk zones and different physical treatments of the important physical and chemical processes, such as dust size distribution and opacities, dust settling, non-LTE treatment of atoms and molecules, and chemistry \citep{Woitke2016}. 
The chemical network used in this paper is based on the large DIANA standard chemical network \citep{Kamp2017}.  However, in this paper we have disregarded all Si- and Fe-bearing species and used very low element abundances for Mg and Na, see Table~\ref{tab:parameter}. We have done this in order to reduce the abundances of potential electron donors to a minimum, assuming that Si, Fe, Mg and Na are mostly incorporated in the refractory grains.  Furthermore, we disregarded frozen PAH molecules (called PAH\# in ProDiMo) in this work. PAH molecules can efficiently absorb free electrons, but when they freeze out on dust surfaces, they loose that ability in the model, which is likely an artifact. All these measures aim at establishing low electron concentrations in the lower disk regions, where carbon and sulfur have mostly recombined, to avoid the fast recombination of \ce{HCO+} and \ce{N2H+} with free electrons.

With respect to the large DIANA standard chemical network \citep{Kamp2017}, we have only 11 elements (H, He, C, N, O, Ne, Na, Mg, S, Ar, and PAHs), 197 species (38 less -- no Si- and Fe-species, and no PAH\#), and 2764 reactions. 2159 reactions are from the UMIST~2022 database \citep{Millar2024}, and 605 reactions from a ProDiMo-specific collection of reactions that include a simple freeze-out and ice desorption chemistry \citep{Woitke2009}, excited molecular hydrogen chemistry \citep{Kamp2017}, low-temperature tunneling rates for a few key reactions \citep{Meisner2019}, polycyclic aromatic hydrocarbons in five different charging states \citep{Thi2019}, 288 photo-reactions with cross-sections from the Leiden photo-dissociation database \citep{vanDishoeck2006, vanHemert2008, Heays2017}, 94 X-ray reactions \citep{Aresu2011}, and a few updates concerning collider and three-body reactions by \cite{Kanwar2023,Kanwar2025}.

ProDiMo considers a large number of heating and cooling processes in the radiation field created by the star, the disk and the background environment.  In the current work, we are using 120 heating and 112 cooling processes, with 59 atomic and molecular species contributing of the line heating and cooling. Recent updates described by \cite{Woitke2024} have introduced a new escape probability method, a new treatment of dust settling according to \cite{Riols2018}, and a new recipe for smoothly increasing column densities at the inner rim and secondary disk walls.  All these improvements are used in this paper as well.  We also use a new 2D treatment of molecular shielding of UV photo-processes, where the UV radiation is traced on a common 1\,\AA\ grid into the disk, both radially and vertically, using the local molecular UV opacities, such that every mutual combination of shielding molecule and shielded photo-process is included. All {\sc ProDiMo} models were run with git-version {\tt 65d9530c}\,\ from May 5$^{\rm th}$, 2026. 

\subsection{Disk setup and modeling strategy}
\label{sec:model_approach}
From the observations, HD\,143006 appears to have an inclined or warped inner disk \citep{Benisty2018,Codron2025} and an outer disk that shows at least two concentric rings in continuum mm-images \citep{Perez2018}.  The inclined inner disk casts a shadow onto the upper layers of the outer disk \citep{Benisty2018} from where the scattered light and the molecular emissions primarily come from, however the inner wall of the outer disk still seems well-illuminated by the star along the midplane, evident from the scattered light images which show a continuous first ring around 30\,au \citep{Ren2023}.  We model the outer disk by a continuous disk which starts at about 30\,au and extends to about 200\,au as evident from the $^{12}$CO data.  We interpret the second ring seen at about 70\,au in mm-continuum images as an accumulation of large dust particles that have moved inward by radial drift to about $70\,$au in the past, see Sect.\,\ref{sec:RadialDrift}.  

Since {\sc ProDiMo} cannot handle truly 3D disk structures, we have designed an axisymmetric disk structure with an inner disk that is ``inclined'', i.e. centered along constant relative height $\tan(\alpha_w)\!=\!h/r$. The figure shows the geometry of the inner disk that looks like an inclined inner disk in this cut. However, since the model is axisymmetric, it is rather a cone with a wide opening angle, and that cone is mirrored both to negative $r$ and negative $z$.  Its purpose in the model is to cast a shadow solely on the upper layers of the outer disk, similar to what an inclined inner disk would do. In order to simulate the illuminated and the shadowed disk sides, we run two disk models: one model for the illuminated disk side where we remove the inner disk, and one model for the opposite disk side where the shadow-casting inner disk is present. This does mean that our models implicitly assume the contribution from the back of the disk is not included. The validity of this assumption is discussed in Section~\ref{sec:backillum}.

\begin{figure}  
  \vspace*{0mm}
  \includegraphics[page=1,width=90mm,height=72mm,trim=35 30 65 335,clip]
  {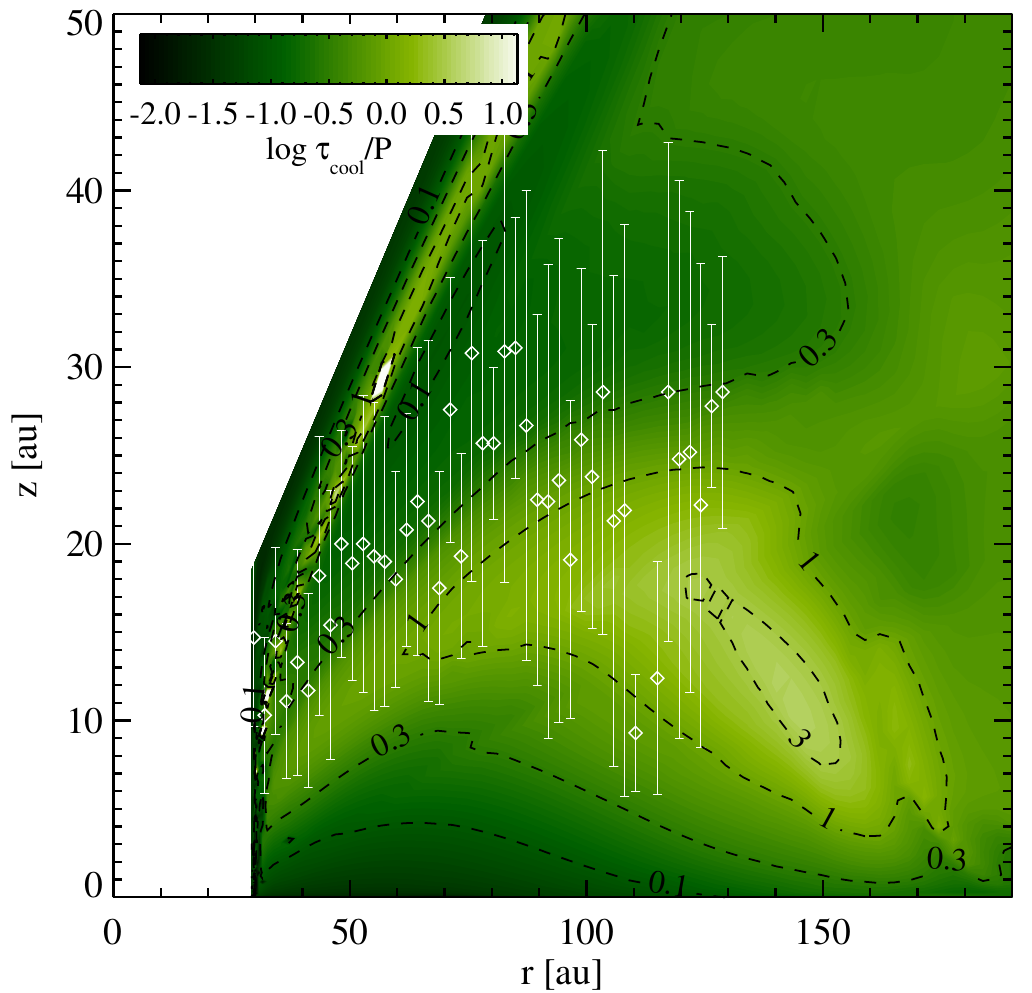}
  \vspace*{-5mm}
  \caption{The heating/cooling relaxation timescale of the gas divided by the orbital period in the one-zone model. The errorbars show the $^{13}$CO emission surface data from \cite{Galloway2025}. 
  }
  \label{fig:taucool}
  \vspace*{2mm}
  \includegraphics[width=86mm]{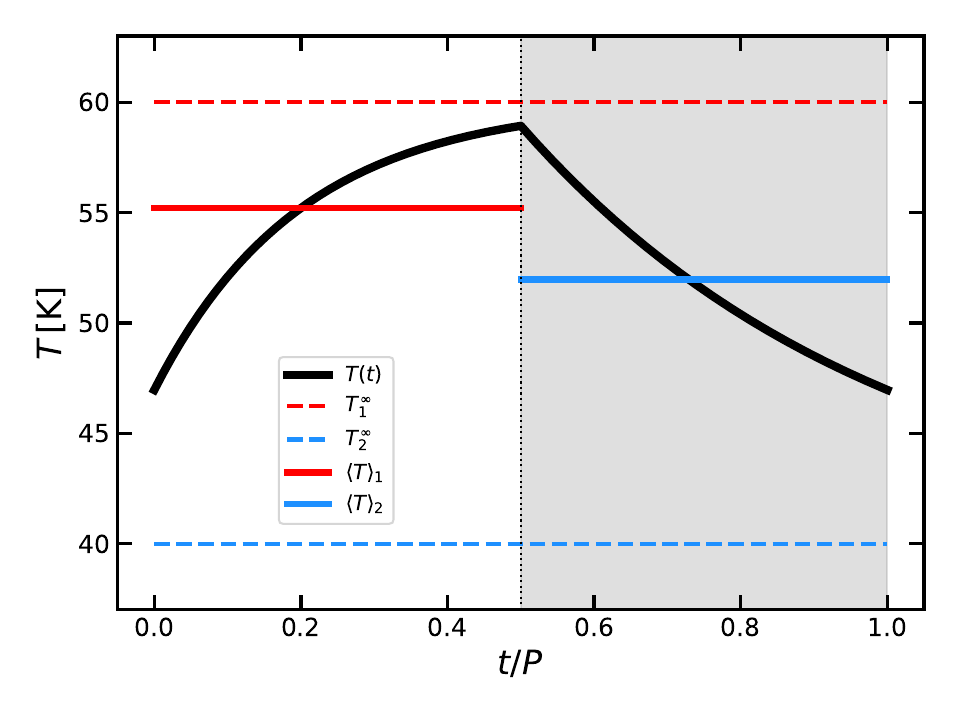}
  \vspace*{-5mm}
  \caption{Variation of gas temperature with orbital phase. The shaded box indicates where the gas parcel is in the shadow. The dashed lines indicate the gas temperatures in heating \& cooling balance in the irradiated part $T_1^\infty$ (red dashed) and in the shadowed part $T_2^\infty$ (blue dashed). The actual temperature $T(t)$ (black) is assumed to relax toward these limiting values with given cooling relaxation timescales $\tau_1^{\rm cool}$ and $\tau_2^{\rm cool}$, respectively. The full red and blue lines indicate the average temperatures, $\langle T\rangle_1$ and $\langle T\rangle_2$ used in the equilibrated one-zone and two-zone models, respectively.}
  \label{fig:T-relaxation}
  \vspace*{-2mm}
\end{figure}

After the first experience with such pairs of models, it soon became clear that using these models directly results in too large temperature and line flux contrasts between the illuminated and the non-illuminated disk sides. The observations (Table\,\ref{tab:fluxes_mol}) show left/right molecular line flux contrasts of about $0.95-1.15$, whereas the models that assume the gas to be in heating/cooling balance show line flux contrast factors up to 1.9, depending on the line, see Table~\ref{tab:result_linefluxes}.  In the model, when the UV-illumination by the star reaches the outer disk, it triggers important heating processes, for example photo-dissociation heating, heating by exothermal reactions, and photoeffect on PAH molecules, that cause a strong warming of the line-emitting molecular gas in the illuminated disk surface.

To refine our modeling approach, we realized that the gas heating/cooling relaxation timescale, see definition in Sect.\,8.3 of \cite{Woitke2009}, is typically longer than the Kepler period in most parts of the disk responsible for the line emissions, see Fig.\,\ref{fig:taucool}.  Therefore, when a gas parcel periodically rotates in and out of the shadow, the temperature relaxation is expected to be incomplete, leading to less pronounced temperature contrasts between the illuminated and shadowed disk sides.  Figure~\ref{fig:taucool} shows more details.  In the $^{13}$CO line emitting regions at $h/r\!\approx\!0.2-0.5$, we find $\tau_{\rm cool}/P\!\approx\!0.1-3$. In higher disk regions the gas becomes warmer with more active radiative processes and shorter relaxation timescales. In the lower, more optically thick disk regions, the gas temperature starts to couple to the dust temperature, which also results in shorter cooling relaxation timescales.  It is important to note that most parts of the line emitting molecular gas is energetically decoupled from the dust, and that the cooling relaxation timescale in those regions is driven by line emission -- not by continuum emission, as e.g.\ assumed by \cite{Codron2025}.

Knowing the temperatures in heating/cooling balance, $T_1^\infty$ and $T_2^\infty$, where index~1 belongs to the illuminated disk side and index~2 belongs to the shadowed disk side, and the respective cooling relaxation timescales on each disk side, $\tau_1^{\rm cool}$ and $\tau_2^{\rm cool}$, we can formulate the following time-dependent relaxation ansatz for the gas temperature $T$ as function of time $t$
\begin{equation}
  T(t) = \left\{\!\!\begin{array}{ll}
  T_1^\infty + \Big(T(0)-T_1^\infty\Big)
  \,\exp\left(-\frac{t\,-\,0}{\tau_1^{\rm cool}}\right) 
  \!\!\!&,\ 0\!\leq\!t\!<\!P/2\\
  T_2^\infty + \Big(T(P/2)-T_2^\infty)\Big)
  \,\exp\left(-\frac{t\,-\,P/2}{\tau_2^{\rm cool}}\right) 
  \!\!\!&,\ P/2\!\leq\!t\!<P
  \end{array}\right. 
  \label{eq:Trelax}              
\end{equation}
Equation~(\ref{eq:Trelax}) can be solved for the unknowns $T(0)$ and $T(P/2)$ when the thermal relaxation process is assumed to be periodic, $T(0)\!=\!T(P)$. The resulting $T(t)$ is visualized in Fig.\,\ref{fig:T-relaxation} for an example with $T_1^\infty\!=\!60$\,K, $T_2^\infty\!=\!40$\,K, $\tau_1^{\rm cool}/P\!=\!0.2$ and $\tau_2^{\rm cool}/P\!=\!0.5$, i.e. for a case where the relaxation timescales are actually quite short. Even in this case, the contrast between the averaged temperatures
\begin{equation}
\langle T_1\rangle = \frac{2}{P} \int_0^{P/2} T(t)\,dt
\quad,\quad
\langle T_2\rangle = \frac{2}{P} \int_{P/2}^{P} T(t)\,dt 
\label{eq:Taverage}
\end{equation}
is much smaller ($\rm 55.2\,K/52.0\,K\!=\!1.06$) than the contrast between the relaxed temperatures ($\rm 60\,K/40.0\,K\!=\!1.5$). Hence, the gas heat capacity plays an important role for understanding the magnitude of the observed line flux contrasts.

Thus, we need to run four disk models to test one set of model parameters against the observations. The first pair of disk models are called the {\em ``original one-zone and two-zone''} models in the following. These exclude or include the inner disk, respectively, and treat the gas in heating/cooling balance.  From these models we read off $T_1^\infty(r,z)$, $T_2^\infty(r,z)$, $\tau_1^{\rm cool}(r,z)$ and $\tau_2^{\rm cool}(r,z)$. Then we apply Eq.\,(\ref{eq:Taverage}) to obtain $\langle T_1\rangle(r,z)$ and $\langle T_2\rangle(r,z)$, and these averaged gas temperatures are then fed back to the original models to compute a second pair of models called the {\em ``equilibrated one-zone and two-zone''} models. In those models, we do not change the continuum radiation field nor the dust temperatures, but only re-calculate the chemistry and line formation for a given gas temperature structure, see Sects.\,\ref{sec:TdustRelax} and \ref{sec:ChemRelax} for further discussion. 

The left/right observations of HD\,143006 are finally compared to the one/two zone equilibrated models by computing a reduced $\chi^2$ that evaluates the mean deviations between all model predictions and observations, which includes the following data:
\begin{enumerate}
\item the photometric fluxes (see Table\,\ref{tab:fluxes_cont}),
\item the Spitzer/IRS spectrum (see Sect.\,\ref{sec:Spitzer}),
\item the two ALMA continuum radial intensity profiles,
\item the 14 ALMA lines, out of which 10 have radial line intensity profiles and left/right contrasts (the other four lines only have integrated line fluxes),
\item the $^{12}$CO and $^{13}$CO surface emission heights from \cite{Galloway2025}.
\end{enumerate}
The total $\chi^2$ of a model is computed from these components as
\begin{equation}
  \chi^2 = w_{\rm ph}\,\chi^2_{\rm ph}
         + w_{\rm Sp}\,\chi^2_{\rm Sp}
         + w_{\rm im}\,\chi^2_{\rm im}
         + w_{\rm lin}\,\chi^2_{\rm lin}
         + w_{\rm hei}\,\chi^2_{\rm hei} \ ,
\end{equation}
with weights 
$w_{\rm ph}\!=\!0.1$,      
$w_{\rm Sp}\!=\!0.05$,     
$w_{\rm im}\!=\!0.35$,     
$w_{\rm lin}\!=\!0.3$, and
$w_{\rm hei}\!=\!0.2$.    
These weights reflect our assessment of the urgency  to fit the various kinds of observational data. They are also used to balance out the relative contributions to the total $\chi^2$. Concerning the photometry and the Spitzer/IRS spectrum, we ignore the (1.6\,-\,13)\,$\mu$m data for the one-zone models, because the outer disk does not contribute significantly to this spectral range, see Fig.\,\ref{fig:ContFit}. Concerning the two-zone models, we do not consider the inner disk in form of two cones to be realistic, therefore we put $w_{\rm ph}\!=\!w_{\rm Sp}\!=\!0$ for the two-zone models. Concerning the continuum images and line observations, we use the averaged radial intensity data separately obtained from the left and right disk sides, as explained in Sect.\,\ref{sec:obsdata}, and compare them to the results of the equilibrated one-zone and two-zone models, respectively.  The model continuum images and moment-0 maps need to be convolved with the respective ALMA beams, before we average the intensities in concentric rings to get $\langle I\rangle(r)$ that are then compared to the data. Concerning the CO emission heights, we ignore any left/right differences and assume that these data are not affected much by the illumination. At every given radius, we use the mean value of the two heights $z(15\%)$ and $z(85\%)$ shown in Figs.\,\ref{fig:LineFormation1} and \ref{fig:LineFormation2}. These are the heights over the midplane where the cumulative line intensity in a vertical column at given radius reaches 15\% and 85\%, using our escape probability theory.

Finally, the two $\chi$ values obtained for the equilibrated one-zone and two-zone models are averaged. We note that all stellar, dust and outer disk model parameters are always the same for all four models, they only differ in terms of the inner disk, which is missing in the one-zone models.

\subsection{A simple recipe for dust radial drift}
\label{sec:RadialDrift}
The (sub-)mm continuum intensities (both left and right side) show a very steep decline beyond about 75\,au, which are consistent with a sharp outer edge after beam convolution. This is in contrast to the slow decrease of all observed line intensities, which extend to about 150-200\,au, depending on the line.  All attempts to fit these intensity profiles for both continuum and lines, only with optical depth effects, failed.  Therefore, we interpret this sudden end of the observable dust disk at (sub-)mm wavelengths as a consequence of the action of radial drift in the past, where larger grains have moved inward toward the first radial pressure maximum as seen from the outside.   In order to mimic the effects of radial drift on the local dust size distribution function we introduce four parameters $\{R_{\rm drift},g_{\rm drift},a_{\rm drift},w_{\rm drift}\}$. First, we define a linear function of radius that slowly switches between 0 and 1 
\begin{equation}
  f(r) = \max\bigg\{0,\min\Big\{1,\frac{1}{2}
       +\frac{r-R_{\rm drift}}{g_{\rm drift}}\Big\}\bigg\} \ ,
  \label{eq:fr}
\end{equation}
where $R_{\rm drift}$ is the drift radius and $g_{\rm drift}$ the drift width. Second, we define a critical size above which the dust particles are affected by radial drift, which varies between $a_{\rm max}$ and the $3^{\rm rd}$ parameter $a_{\rm drift}$ 
\begin{equation}
  a_{\rm crit}^{\rm drift}(r) = \Big(a_{\rm max}^p 
  + f(r)\,\big(a_{\rm drift}^p - a_{\rm max}^p\big)\Big)^{1/p} \ ,
  \label{eq:amaxdrift}        
\end{equation}
such that $a_{\rm crit}^{\rm drift}(r)=a_{\rm max}$ for $r\!\ll\!R_{\rm drift}$, and $a_{\rm crit}^{\rm drift}(r)=a_{\rm drift}$ for $r\!\gg\!R_{\rm drift}$.
Third, we remove all dust particles with $a>a_{\rm crit}^{\rm drift}(r)$ from the model and put them into a size-dependent reservoir.  Fourth, for every dust size bin, we redistribute the particles in the reservoir with a Gaussian probability distribution in radius 
\begin{equation}
   p_a(r) \propto \exp\left(-\frac{\big(r-R_{\rm drift}^{\rm max}(a)\big)^2}{w_{\rm drift}^2}\right)
\end{equation}
where $R_{\rm drift}^{\rm max}(a)$ is the largest radius where grains of size $a$ are unaffected by drift (derived from Eq.\,\ref{eq:amaxdrift} with $a_{\rm crit}^{\rm drift}\!=\!a$) and $w_{\rm drift}$ is the redistribution width. The implemented numerical scheme is mass-conservative. This way, the large grains beyond $R_{\rm drift}$ are "brushed inwards" and accumulate around $R_{\rm drift}^{\rm max}$, whereas the small grains remain unaffected and stay radially well-mixed with the gas. We do this separately for all vertical layers in the model after settling. Useful results are obtained with $p\!=\!-\,0.2$. 
Table~\ref{tab:parameter} shows the best-fit values of the four drift parameters. The impact of these parameters on the disk structure and dust/gas ratio are further discussed in Sect.~\ref{sec:MainModel} and the Figures therein.

\begin{table}
\begin{center}
\caption{Parameters of the best-fitting ProDiMo model of HD\,143006.}
\vspace*{-1mm}
\label{tab:parameter}
\resizebox{88mm}{!}{\begin{tabular}{l|c|c}
\hline 
\multicolumn{3}{c}{}\\[-2.2ex]
\multicolumn{2}{l}{\hspace*{5mm} stellar and system parameter\tablefootmark{(1)}} &\\
\hline
&&\\[-2.1ex]
stellar mass                      & $M_{\star}$   & $1.56\,M_\odot$\\
stellar luminosity                & $L_{\star}$   & $3.8\,L_\odot$\\ 
effective temperature             & $T_{\star}$   & $5620\,$K\\
UV excess                         & $f_{\rm UV}$  & (0.013)$\,^{(5)}$\\
X-ray luminosity                  & $L_X$         & $2\times 10^{30}\rm erg/s$\\
X-ray emission temperature        & $T_X$         & $20\times10^6$\,K\\
accretion rate                    & $\dot{M}_{\rm acc}$ & $2\times10^{-8}M_\odot$/yr\\ 
optical extinction                & $A_{\rm V}$   & 0.72\\
distance                          & $d$           & 167\,pc\\
disc inclination                  & $i$           & 17\degr\\
disc position angle               & PAobj         & 176.2\degr\\[0.3ex]
\hline
\multicolumn{3}{c}{}\\[-2.2ex]
\multicolumn{2}{l}{\hspace*{5mm} environmental parameter} &\\
\hline
&&\\[-2.1ex]
strength of interstellar UV       & $\chi^{\rm ISM}$ & 1\\
cosmic ray H$_2$ ionization rate  & $\zeta_{\rm CR}$ & 
                                  $\rm 1.7\times10^{-17}\,s^{-1}$\\[0.3ex]
\hline 
\multicolumn{3}{c}{}\\[-2.2ex]
\multicolumn{2}{l}{\hspace*{5mm} dust parameter} &\\
\hline
&&\\[-2.1ex]
minimum dust particle radius\tablefootmark{$\star$} & $a_{\rm min}$   & $0.0623\,\mu$m\\
maximum dust particle radius\tablefootmark{$\star$} & $a_{\rm max}$   & $2.20\,$mm\\
dust size dist.\ powerlaw index        & $a_{\rm pow}$   & -3.5\\
max.\ hollow volume ratio             & $V_{\rm hollow}^{\rm max}$  & 80\%\\
dust settling parameter\tablefootmark{(6)} \tablefootmark{$\star$} & $\alpha_{\rm set}$ & $2.26\times10^{-4}$\\
&&\\[-2.5ex]
dust composition                      & $\rm Mg_{0.7}Fe_{0.3}SiO_3$ & 65.0\%\\
(volume fractions)\tablefootmark{$\star$}          & amorph.\,carbon  & 20.0\%\\
                                      & porosity         & 25\%\\
\hline 
\multicolumn{3}{c}{}\\[-2.2ex]
\multicolumn{2}{l}{\hspace*{5mm} inner disk parameter} &\\
\hline
&&\\[-2.1ex]
quasi warp angle\tablefootmark{$\star$}           & $\alpha_{\rm w}$  & $15.1^\circ$\\
disc gas mass\tablefootmark{$\star$}              & $M_{\rm gas}$     & $7.01\times10^{-5}\,M_\odot$\\
disc dust mass\tablefootmark{$\star$}             & $M_{\rm dust}$    & $6.13\times10^{-7}\,M_\odot$\\
inner disk radius\tablefootmark{$\star$}          & $R_{\rm in}$      & 0.0962\,au\\
outer radius\tablefootmark{$\star$}               & $R_{\rm out}$     & 19.0\,au\\
column density power index\tablefootmark{$\star$} & $\epsilon$        & 0.767\\
extension of inner rim\tablefootmark{(2)}     & raduc             & 1.2\\
maximum $\Sigma$ reduction\tablefootmark{2} & reduc             & $10^{-4}$\\
reference gas scale height\tablefootmark{$\star$} & $H(1\,{\rm au})$  & 0.0993\,au\\
flaring power index\tablefootmark{$\star$}        & $\beta$           & 1.21\\[0.3ex] 
\hline 
\multicolumn{3}{c}{}\\[-2.2ex]
\multicolumn{2}{l}{\hspace*{5mm} outer disk parameter} &\\
\hline
&&\\[-2.1ex]
disc gas mass\tablefootmark{$\star$}              & $M_{\rm gas}$     & $1.22\times10^{-3}\,M_\odot$\\
disc dust mass\tablefootmark{$\star$}             & $M_{\rm dust}$    & $4.05\times10^{-5}\,M_\odot$\\
inner disk radius\tablefootmark{$\star$}          & $R_{\rm in}$      & 29.2\,au\\
tapering-off radius\tablefootmark{$\star$}        & $R_{\rm tap}$     & 11.1\,au\\
column density power index\tablefootmark{$\star$} & $\epsilon$        & -4.29\\
tapering off power index \tablefootmark{$\star$}  & $\gamma$          & 1.02\\
extension of inner wall\tablefootmark{(2)} \tablefootmark{$\star$} & raduc          & 1.06\\
maximum $\Sigma$ reduction \tablefootmark{(2)} & reduc             & $10^{-3}$\\
reference gas scale height\tablefootmark{$\star$} & $H(30\,{\rm au})$ & 3.81\,au\\
flaring power index\tablefootmark{$\star$}        & $\beta$           & 0.971\\ 
drift radius\tablefootmark{$\star$}               & $R_{\rm drift}$   & 69.6\,au\\
minimum drift particle size\tablefootmark{$\star$} &$a_{\rm drift}$   & 84.5\,$\mu$m\\
drift width\tablefootmark{$\star$}                & $g_{\rm drift}$   & 3.75\,au\\
drift redistribution width           & $w_{\rm drift}$   & 8\,au\\[0.3ex]  
\hline 
\multicolumn{3}{c}{}\\[-2.2ex]
\multicolumn{3}{l}{\hspace*{5mm} gas parameter, and element abundances} \\
\hline
&&\\[-2.1ex]
C/O ratio\tablefootmark{(3)} \tablefootmark{$\star$}            & $\epsilon_{\rm C}/\epsilon_{\rm O}$ & 1.05\\
C/N ratio\tablefootmark{(3)} \tablefootmark{$\star$}            & $\epsilon_{\rm C}/\epsilon_{\rm N}$ & 0.251\\
C/S ratio\tablefootmark{(3)} \tablefootmark{$\star$}            & $\epsilon_{\rm C}/\epsilon_{\rm S}$ & 26000\\
Na, Mg abundances\tablefootmark{(4)}          & $\epsilon_{\rm Na}=\epsilon_{Mg}$ & $1\times10^{-11}$\\
PAH abundance rel.\ to ISM\tablefootmark{$\star$} & $f_{\rm PAH}$        & 0.00524\\
chemical heating efficiency\tablefootmark{$\star$} & $\gamma^{\rm chem}$ & 0.120\\
heat per photodissociation\tablefootmark{(2)} \tablefootmark{$\star$} & $\Delta E_{\rm pd}$ & 0.580\,eV\\
turbulent Doppler width\tablefootmark{$\star$}    & $v_{\rm turb}$       & 0.0362\,km/s\\[0.3ex]
\hline   
\end{tabular}}
\end{center}
\vspace*{-3mm}
\tablefoottext{1}{See \citet{Woitke2016} for the definitions of parameters.}\\
\tablefoottext{2}{See \citet{Woitke2024} for the definitions of parameters.}\\
\tablefoottext{3}{The carbon abundance$\,^{(4)}$ is assumed to be fixed: $\epsilon_{\rm C}=1.38\times10^{-4}$.}\\
\tablefoottext{4}{with respect to hydrogen.}\\
\tablefoottext{5}{derived from the IUE data; that spectrum is then used as input.}\\ 
\tablefoottext{6}{only applies to the outer disk according to \cite{Riols2018}  -- the inner disk is unsettled.}\\ 
\tablefoottext{$\star$}{Optimised during the fitting, final value given with three digits.}
\end{table}

\subsection{Fitting process}
In order to adjust our 31 free dust, disk shape and chemical parameters to the observational data of HD\,143006, we minimize $\chi^2$ using the $(1,14)$ genetic algorithm of \citet{Rechenberg2000}, see further explanations in \citet{Woitke2019}. The fitted parameter values obtained this way are listed and marked with a star in Table~\ref{tab:parameter}.  This fitting strategy does not allow us to explore the full parameter space.  Using for example a Markov-Chain-Monte-Carlo (MCMC) fitting method would require us to run millions of disk models, but since one disk model takes about 5 CPU hours, and we need 4 of them to test one set of parameter values against the observations, this is simply not feasible.

\begin{figure}
  \vspace*{-1mm}\hspace*{-2mm}  \includegraphics[page=1,width=90mm,height=75mm,trim=50 28 60 340,clip] 
    {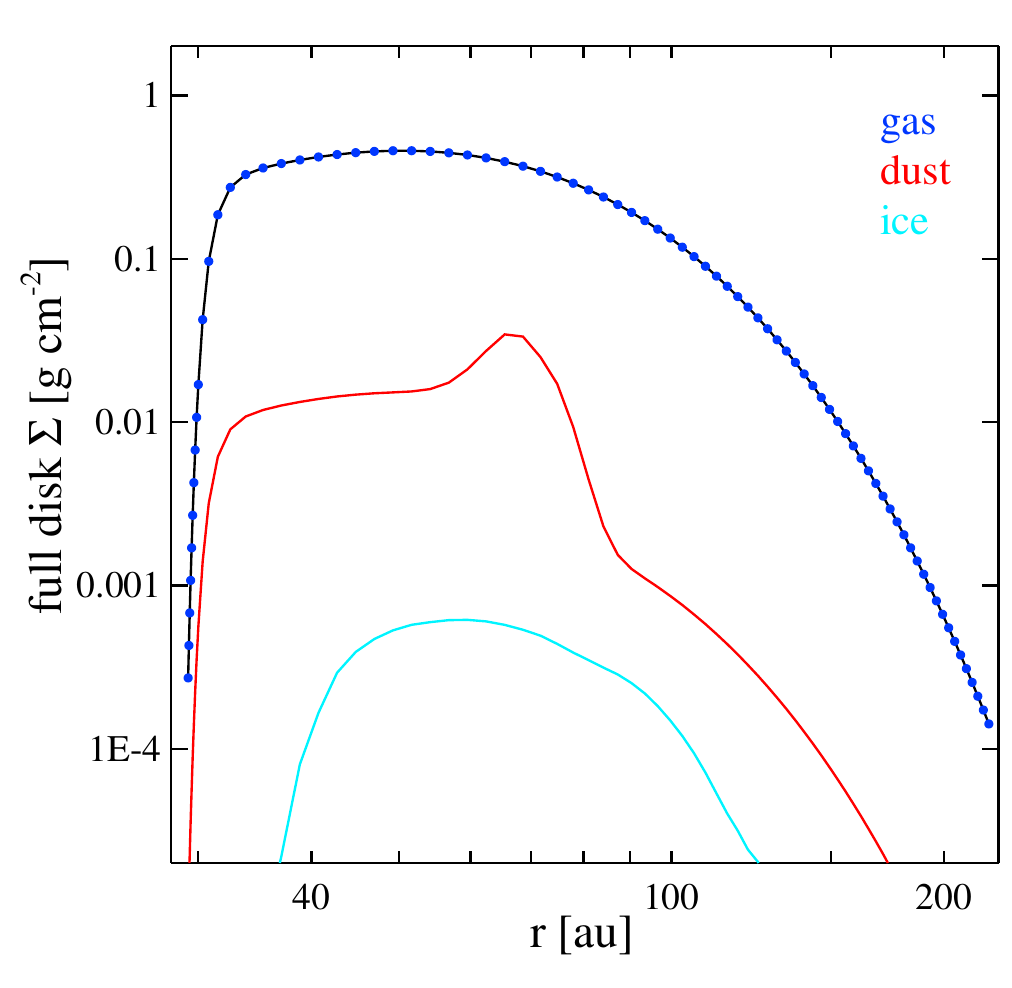}\\*[-7mm]
  \caption{Mass column density structure in the model for the outer disk. The gas and dust column densities  are derived directly from the model parameters listed in Table~\ref{tab:parameter}, whereas the ice column densities are a consistent result of the thermo-chemical simulations.  The maximum of the dust column density around 70\,au is due to the accumulation of large particles by radial drift.} 
  \label{fig:coldens}
  \vspace*{3mm}\hspace*{-4mm}
\includegraphics[page=1,width=91mm,height=72mm,trim=40 40 81 350,clip] 
  {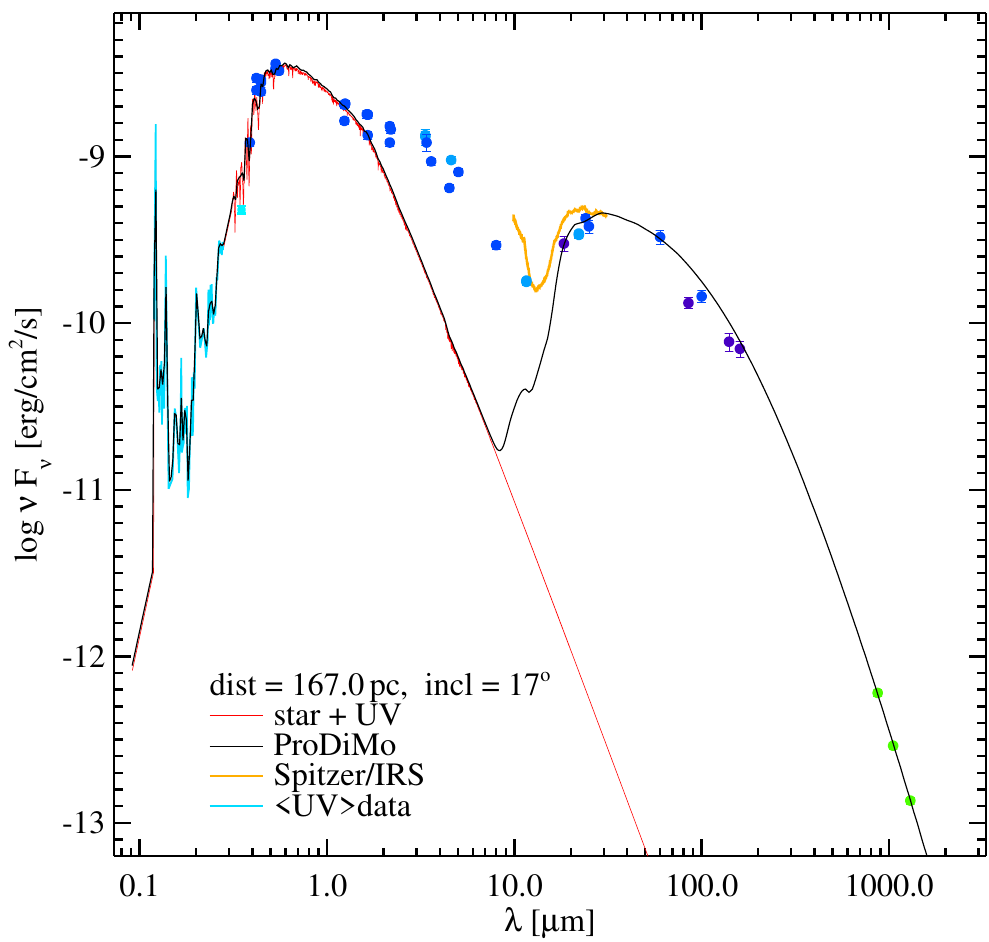}
  \vspace*{-1mm}
  \caption{Spectral energy distribution (SED) of the one-zone model for HD\,143006 in comparison to our photometric data collection and a Spitzer/IRS spectrum. The outer disk is responsible for the far-IR excess beyond about 20\,$\mu$m.}
  \label{fig:SED}
  \vspace*{-2mm}
\end{figure}

\section{Main modeling results}
\label{sec:MainModel}

Table~\ref{tab:parameter} lists the parameter values of our best-fitting model, which was found after 193 generations of the genetic fitting algorithm that required about $\rm 200 \times 14 \times (4\times 5\ CPUh) = 56000\ CPUh$. All parameters marked with $^\star$ were treated as free parameters, and the resulting values describe the fitted disk structure, the fitted dust size distribution, the settling and drift properties, and the fitted element abundance ratios such as the C/O and C/N. The ``quasi-warp'' parameter $\alpha_W$ is also a free parameter in this fit for models with an inner disk. The resulting physical structure of the HD\,143006 model disk is described and visualized in the following Sections.

\subsection{Column density structure and disk extension}
Figure~\ref{fig:coldens} shows the gas, dust and ice column density structure. The inner disk is found to be too warm to host any significant amounts of ices. The outer disk is cooler, but the UV irradiation along the midplane prevents the formation of any ices, via UV-photo desorption, until the UV is entirely absorbed at about 35\,au. Beyond that radius, many ices form in unison in the midplane, including \ce{H2O}, \ce{NH3} and CO ices. The radial drift of all particles larger than about $a_{\rm drift}\!\approx\!80\,\mu$m causes a local maximum of dust/gas around 70\,au -- its effect on the mm-opacity is very pronounced (not shown here), making the outer disk beyond 70\,au suddenly very transparent at mm-wavelengths.  

\begin{figure*}
  \vspace*{-3mm}
  \centering
  \begin{tabular}{cc}
  gas density structure & dust density structure\\
    \hspace*{-4mm}
    \includegraphics[page=1,width=84mm,height=66mm,trim=30 30 65 340,clip] 
    {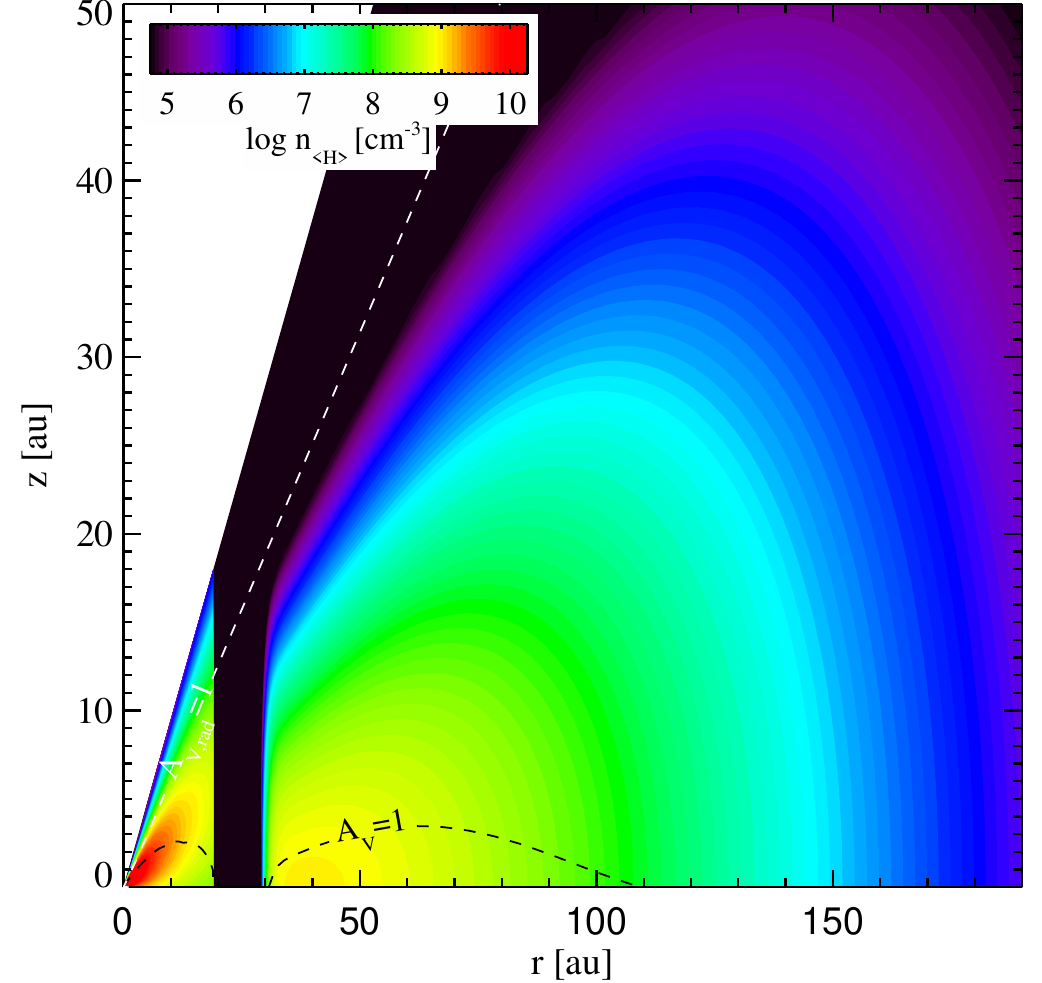} &
    \hspace*{-6mm}
    \includegraphics[page=2,width=84mm,height=66mm,trim=30 30 65 340,clip] 
    {Figs/out_twoZone_Fig10.pdf} \\[0mm]
  gas/dust mass ratio & mean particle size\\
    \hspace*{-4mm}
    \includegraphics[page=3,width=84mm,height=66mm,trim=30 30 65 340,clip] 
    {Figs/out_twoZone_Fig10.pdf} &
    \hspace*{-6mm}
    \includegraphics[page=4,width=84mm,height=66mm,trim=30 30 65 340,clip] 
    {Figs/out_twoZone_Fig10.pdf} \\[-1mm]
  \end{tabular}
  \caption{Fitted dust and gas disk structure of HD\,143006 as function of distance from the symmetry axis $r$ and height over the midplane $z$. Top left: total hydrogen nuclei particle density $\nH$. Top right: dust mass density $\rho_{\rm dust}$. Bottom left: the gas/dust mass ratio. Bottom right: the mean dust particle radius weighted by mass $\langle a^3\rangle^{1/3}$, emphasizing the location of the big grains. The contour lines show the vertical optical extinction $A_V\!=\!1$ (black dashed), and the radial optical extinction $A_{V,\rm rad}\!=\!1$ (white dashed). 
  We note the different scalings of the $z$-axes.} 
  \label{fig:DiskStructure}
\end{figure*}

\begin{figure*}
  \vspace*{-3mm}
  \centering
  \begin{tabular}{cc}
  $\chi_{\rm UV}$\ \ \ one-zone model & $\chi_{\rm UV}$\ \ \ two-zone model\\
    \hspace*{-4mm}
    \includegraphics[page=1,width=81mm,height=63mm,trim=30 30 65 340,clip] 
    {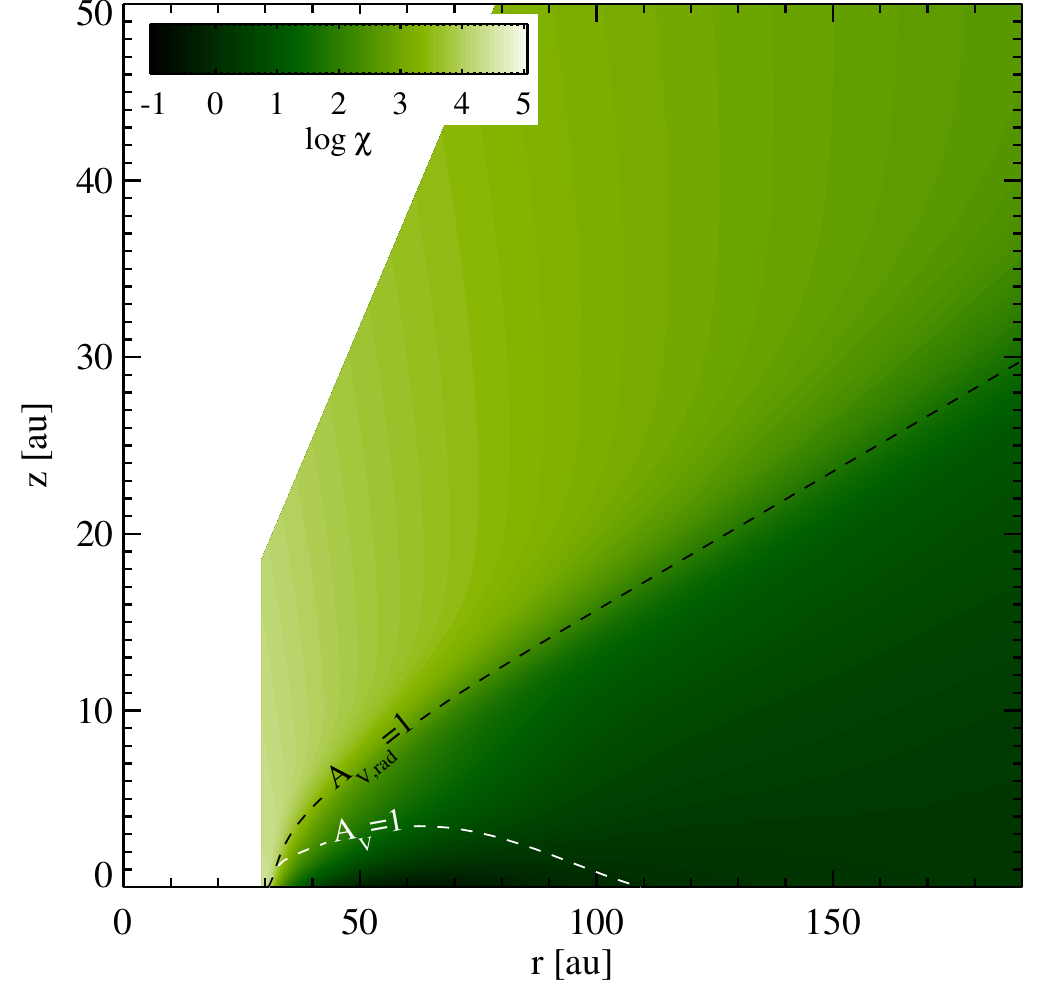} &
    \hspace*{-6mm}
    \includegraphics[page=1,width=81mm,height=63mm,trim=30 30 65 340,clip] 
    {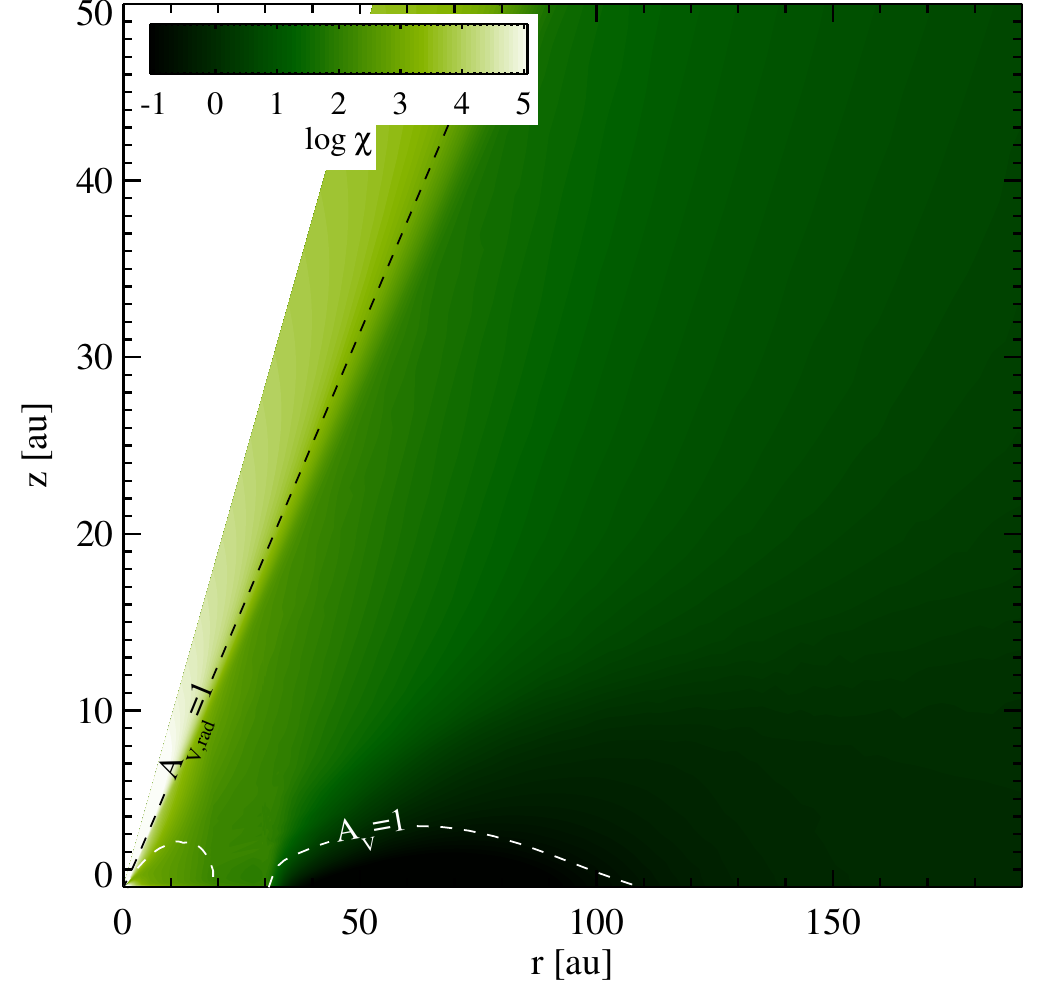} \\[0mm]
  $\zeta_{\rm X}$\ \ \ one-zone & $\zeta_{\rm X}$\ \ \ two-zone\\
    \hspace*{-4mm}
    \includegraphics[page=2,width=81mm,height=63mm,trim=30 30 65 340,clip] 
    {Figs/out_oneZone_Fig11.pdf} &
    \hspace*{-6mm}
    \includegraphics[page=2,width=81mm,height=63mm,trim=30 30 65 340,clip] 
    {Figs/out_twoZone_Fig11.pdf} \\[-1mm]
  \end{tabular}
  \caption{UV- and X-ray radiation field conditions. The upper row shows $\chi_{\rm UV}$, the UV flux normalized to a Draine field with $1.92\times10^8\rm\,photons/cm^2/s$ between 91.2 and 205\,nm, disregarding molecular shielding. The radial and vertical optical depths contours, $A_{\rm V,rad}\!=\!1$ and $A_{\rm V,ver}\!=\!1$ are shown with black and white dashed lines. The lower row shows $\zeta_{\rm X}$, the \ce{H2}-ionization rate due to stellar X-rays.  The white contour line at $2\times10^{-17}\rm\,s^{-1}$ marks where cosmic ray ionization becomes more important than X-ray ionization in the model.} 
  \label{fig:Irradiation}
  \vspace*{-1mm}
\end{figure*}

\begin{table*}
\caption{Comparison of observed and simulated line fluxes\tablefootmark{(2)}, and left/right line flux contrast ratios\tablefootmark{(4)}.} 
\label{tab:result_linefluxes}
\vspace*{-3mm}\hspace*{-1mm}
\resizebox{180mm}{!}{\begin{tabular}{cc|c|c|c|c}
\hline
&&&&&\\[-2ex]
\!\!\!Molecule\!\!\!\! & \!\!\!Transition\!\!\! & 
\!left/right observations\!   & \!one/two zone original\! & 
\!one/two zone equilibrated\! & \!one/two fast-cooling \! \\
&& $\rm 2\times F_{\rm line}$ [Jy km/s] 
 & models $\rm F_{\rm line}$ [Jy km/s] 
 & models $\rm F_{\rm line}$ [Jy km/s] 
 & \!models\tablefootmark{(1)} $\rm F_{\rm line}$ [Jy km/s]\!\! \\
\hline
&&&&&\\[-2ex]
$^{12}$CO & $J$=3-2 & $6.15\,/\,5.43 = 1.13 \pm 0.01$
                  & $7.77\,/\,4.13 = 1.87$ 
                  & $5.62\,/\,6.18 = 0.91$
                  & $6.95\,/\,6.21 = 1.12$ \\
$^{13}$CO & $J$=3-2 & $1.68\,/\,1.55 = 1.09 \pm 0.02$
                  & $3.83\,/\,1.80 = 2.12$
                  & $2.74\,/\,2.64 = 1.04$
                  & $3.24\,/\,2.72 = 1.19$ \\        
C$^{18}$O & $J$=3-2 & $0.40\,/\,0.39 = 1.01 \pm 0.02$
                  & $2.07\,/\,1.27 = 1.68$
                  & $1.77\,/\,1.61 = 1.10$
                  & $1.91\,/\,1.61 = 1.19$ \\
CN   &$J$=5/2-3/2 & $1.72\,/\,1.46 = 1.18 \pm 0.02$
                  & $4.25\,/\,1.49 = 2.85$
                  & $3.50\,/\,2.63 = 1.33$
                  & $3.94\,/\,2.92 = 1.35$ \\
CN   &$J$=7/2-5/2 & $2.17\,/\,1.91 = 1.14 \pm 0.01$
                  & $4.53\,/\,1.63 = 2.77$
                  & $3.69\,/\,2.86 = 1.29$
                  & $4.18\,/\,3.17 = 1.32$ \\
HCN     & $J$=4-3 & $1.94\,/\,1.68 = 1.15 \pm 0.03$
                  & $2.37\,/\,1.46 = 1.62$
                  & $2.13\,/\,2.45 = 0.86$
                  & $2.33\,/\,2.87 = 0.81$ \\
HCO$^+$  &$J$=4-3 & $2.42\,/\,1.90 = 1.27 \pm 0.03$
                  & $3.50\,/\,1.49 = 2.34$
                  & $2.28\,/\,1.97 = 1.16$
                  & $3.09\,/\,1.98 = 1.56$ \\
CS       &$J$=7-6 & $0.35\,/\,0.38 = 0.94 \pm 0.02$
                  & $0.60\,/\,0.48 = 1.26$
                  & $0.58\,/\,0.55 = 1.04$
                  & $0.59\,/\,0.58 = 1.02$ \\
N$_2$H$^+$ &$J$=3-2 & $0.32\,/\,0.33 = 0.98 \pm 0.02$
                  & $0.38\,/\,0.26 = 1.43$
                  & $0.36\,/\,0.27 = 1.34$
                  & $0.38\,/\,0.27 = 1.42$ \\
C$_3$H$_2$ &\!\!\!$4_{3,2}$-$3_{0,3}$\!\!
                  & $0.021\,/\,0.011 = 1.9 \pm 0.8$\ \ \ \ \ 
                  & $0.081\,/\,0.054 = 1.49$
                  & $0.085\,/\,0.10  = 0.84$
                  & $0.079\,/\,0.096 = 0.82$ \\[0.3ex]
\hline
&&&&&\\[-2.1ex]
C$^{18}$O & $J$=2-1 & $0.135 \pm 0.017$ 
          & $0.77\,/\,0.53 = 1.46$
          & $0.71\,/\,0.63 = 1.12$
          & $0.73\,/\,0.63 = 1.16$\\
HCN       & $J$=3-2 & $2.06 \pm 0.21$\tablefootmark{(3)}   
          & $1.34\,/\,1.05 = 1.28$
          & $1.23\,/\,1.48 = 0.83$
          & $1.31\,/\,1.63 = 0.80$\\
C$_2$H    & 7/2-5/2 & $0.435 \pm 0.056$ 
          & $0.088\,/\,0.088 = 1.00$
          & $0.094\,/\,0.10 = 0.94$
          & $0.089\,/\,0.10 = 0.88$\\
H$_2$CO & $3_{0,3}$-$2_{0,2}$ & $0.016 \pm 0.002$ 
          & $0.017\,/\,0.042 = 0.41$
          & $0.018\,/\,0.035 = 0.51$
          & $0.017\,/\,0.034 = 0.50$ \\[0.3ex]
\hline
\end{tabular}}\\[1mm]
\tablefoottext{1}{assuming $5\times$ shorter cooling relaxation timescales $\tau_{\rm cool}$ than found in the original one/two zone models.}\\
\tablefoottext{2}{The one-zone models have no inner disk and are hence missing the fluxes generated by the inner disk.  In order to correct for this effect, we multiply the one-zone line fluxes by $F_{\rm tot}/F_{>28\rm\,au}$ where $F_{\rm tot}$ and $F_{>28\rm\,au}$ are total and outer line fluxes read off from the respective two-zone model.}\\
\tablefoottext{3}{We note that according to the data, the HCN $J$=3-2 line is stronger than the HCN $J$=4-3 line, which is puzzling.}\\
\tablefoottext{4}{For the spectral lines below the dividing line, we only have integrated line flux data.}
\vspace*{-1mm}
\end{table*}

\subsection{Spectral Energy Distribution (SED)\label{sec:SEDdisc}}
Figure~\ref{fig:SED} shows the Spectral Energy Distribution (SED) of the one-zone model, which fits the photometric data and the Spitzer/IRS spectrum beyond about 13\,$\mu$m reasonably well ($\chi_{\rm ph}\!\sim\!1.4$ and $\chi_{\rm Sp}\!\sim\!4$, see Table~\ref{tab:Mdisk_var}). The far-IR excess depends critically on the radial position and the height of the inner wall of the outer disk, so this is an important confirmation of the geometry of a strongly settled outer disk. The inner disk is not present in this model, hence the near-IR excess between about 2\,$\mu$m and 13\,$\mu$m is missing.  Our approach to use an axisymmetric ``quasi-warped'' inner disk (see Sect.\,\ref{sec:model_approach}), which appears a second time below the midplane, is not realistic, therefore, we do not expect the inner disk to fit the near-IR excess.  The purpose of the inner disk in the model primarily is to cast a realistic shadow on the upper parts of the outer disk.

\subsection{2D disk structure}
Figure~\ref{fig:DiskStructure} shows the 2D gas and dust structure of the disk, the dust/gas ratio, and the mean dust size as affected by settling and radial drift.  The dust settling is strong in our best-fitting model ($\alpha_{\rm set}\!\approx\!2\times10^{-4}$), making the spatial dust distribution much flatter than the gas distribution. With the radial drift recipe, big grains are also accumulating radially out to 70\,au. The strong settling allows our model to fit both the measured heights of the CO line emissions around $h/r\!\approx\!0.2-0.5$ and the effective height of the inner wall of the outer disk, which is derived from $A_{\rm V,rad}\!=\!1$ to be about $h/r\!\approx\!0.16$, in good agreement with the value of $h/r\!\approx\!0.13$ measured by \cite{Codron2025}.  This height is also pivotal for the far-IR excess between about 30 to 200\,$\mu$m, see Sect.\,\ref{sec:SEDdisc}. A higher wall would intercept more star light and would result in a too strong far-IR excess. From these results we conclude that most of the molecular line emissions come from layers over the inner wall of the outer disk, and therefore, this wall is not capable of shielding the line emitting molecules from the stellar irradiation, whereas the inclined inner disk is. The upper, observable surface of the inclined inner disk surface is located along $h/r\!\approx\!0.57$, corresponding to a measurable misalignment angle of about $30^\circ$.

\subsection{UV and X-ray irradiation}
Figure~\ref{fig:Irradiation} shows the calculated UV radiation field and the H$_2$ X-ray ionization rate in the disk, separately for the illuminated one-zone and the shadowed two-zone disk models. The inner disk reduces the UV irradiation of the line-emitting regions of the outer disk by several orders of magnitude, although some UV photons still reach those layers via dust scattering.  The effect of the inner disk on the X-ray ionization is significant (we neglect X-ray scattering here).  In the irradiated case (without the inner disk) the ionization of the line-emitting gas in the outer disk, as probed by ALMA, is rather due to X-rays than due to cosmic rays.  In the shadowed case, this relation is reversed. As we discuss in Section~\ref{sec:ChemRelax}, this has important consequences for the relaxation times of the outer-disk chemistry.

\begin{figure}[!htbp]
  \vspace*{1mm} \hspace*{-3.5mm}
  \begin{tabular}{cc}
    \includegraphics[page=1,width=46.3mm,trim=0 20 27 2,clip] 
    {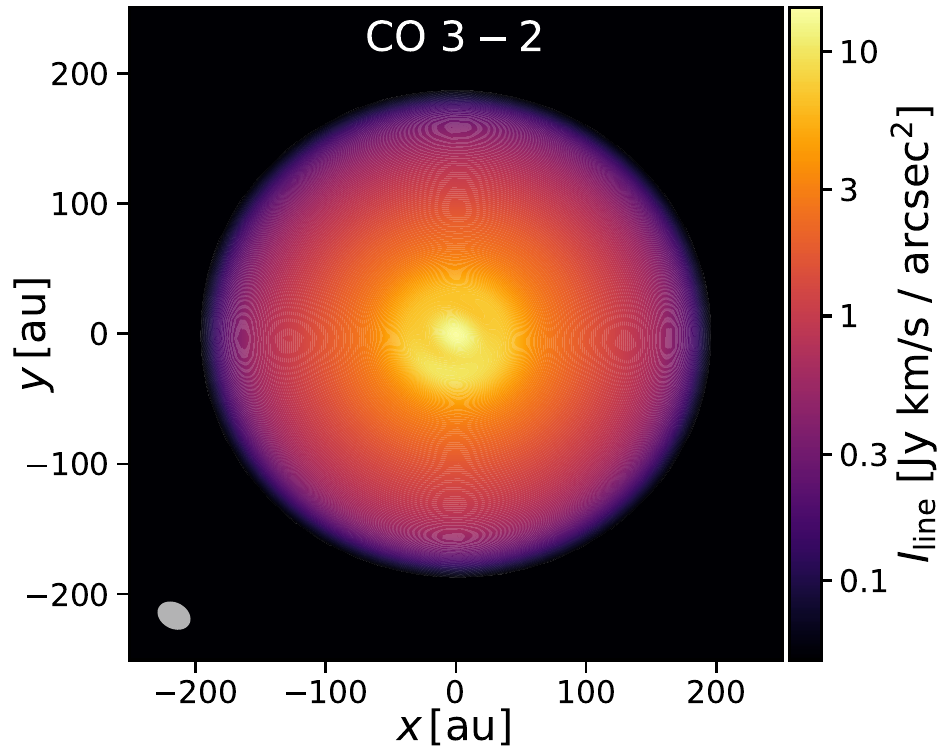} & \hspace*{-4mm}
    \includegraphics[page=7,width=42.6mm,trim=62 20 0 2,clip] 
    {Figs/moment0.pdf} \\[-0.7mm]
    \includegraphics[page=2,width=46.3mm,trim=0 20 27 2,clip] 
    {Figs/moment0.pdf} & \hspace*{-4mm}
    \includegraphics[page=6,width=42.6mm,trim=62 20 0 2,clip] 
    {Figs/moment0.pdf} \\[-0.7mm]
    \includegraphics[page=5,width=46.3mm,trim=0 20 27 2,clip] 
    {Figs/moment0.pdf} & \hspace*{-4mm}
    \includegraphics[page=4,width=42.6mm,trim=62 20 0 2,clip] 
    {Figs/moment0.pdf} \\[-0.7mm]
    \includegraphics[page=3,width=46.3mm,trim=0 20 27 2,clip] 
    {Figs/moment0.pdf} & \hspace*{-4mm}
    \includegraphics[page=8,width=42.6mm,trim=62 20 0 2,clip] 
    {Figs/moment0.pdf} \\[-0.7mm]
    \includegraphics[page=9,width=46.3mm,trim= 0 0 27 2,clip] 
    {Figs/moment0.pdf} & \hspace*{-4mm}
    \includegraphics[page=10,width=42.6mm,trim=62 0 0 2,clip] 
    {Figs/moment0.pdf} 
  \end{tabular}
  \caption{Continuum-subtracted moment-0 maps from the equilibrated two-zone model after beam convolution. The integrated line flux values are reported in Table~\ref{tab:result_linefluxes}. Apparent asymmetries are because of the molecular emissions coming from some height over the midplane, the line emissions from the far disk side being partially absorbed by the dust in the midplane, disk inclination, object position angle, and the beam convolution. Beam parameters are visualized by an ellipse in the bottom left corners.}
  \label{fig:model_all_mol}
  \vspace*{-2mm}
\end{figure}

\subsection{Line flux results}
Table~\ref{tab:result_linefluxes} summarizes the line flux results from the one/two zone models in comparison to the left/right observational data.  Fitting a large number of ALMA lines from different molecules, which truly probes the disk chemistry, by a single thermo-chemical disk model is a challenging endeavor that has rarely been attempted, as discussed in the Introduction.  Moreover, trying to fit radial intensity profiles instead of just the integrated line fluxes is also a new approach.  Therefore, we consider our results as a major step forward to understand disk chemistry and line formation, despite the remaining mismatches between model and observations.  All line fluxes fit the observations better than a factor of two while the median percentage difference in line fluxes is less than $60\%$. Outliers are the C$^{18}$O line which is about $4\times$ too strong, and the pair of \ce{C2H} and \ce{c-C3H2}, which are $4\times$ too weak and $4\times$ too strong respectively. This latter set of lines is discussed in more detail in Section~\ref{sec:CO}. Isotope-selective chemistry (beyond deuterium) is not yet implemented in ProDiMo and could reduce the $^{13}$CO and C$^{18}$O line fluxes.

The original models, where we assumed the gas to be in heating/cooling balance, clearly show too strong illuminated/shadowed line flux contrasts.
The equilibrated models obtain significantly lower contrast ratios and hence better overall fits. These contrast ratios are actually a bit too low, which could mean that our cooling relaxation timescales are too long. Therefore, we have computed another pair of models where we have arbitrarily decreased the cooling timescales by a factor of five, called the ``one/two fast-cooling models'' in Table~\ref{tab:result_linefluxes}.  These are showing the best agreement with the observed line flux contrasts. However, we decided to stay with the equilibrated models for the remainder of this paper, as these models are physically most consistent with our treatment of heating and cooling in ProDiMo. Disregarding any observational errors, the mean value of $2(m-o)/(m+o)$, where $m$ stands for a model line flux and $o$ for its observed value, is $0.36\pm0.67$, i.e.\ the model lines are slightly too bright (by one third) on average, but the general agreement with the observations is excellent across all 14 molecules.

\subsection{Molecular line maps and radial line intensity profiles}
Figure~\ref{fig:model_all_mol} shows the simulated continuum-subtracted moment-0 maps of all fitted ALMA lines on a logarithmic intensity scale from the two-zone equilibrated model, which can be directly compared to Fig.\,\ref{fig:all_mol}. The model manages to explain the general appearances of all lines, which show centrally peaked images for CO and \ce{HCO+}, whereas the CN, HCN and CS lines show ring-like structures.  The rings of CN, HCN and CS seem a little too narrow though, located at about 30-40\,au, which coincides with the location on the inner wall of the outer disk.  In fact, the inner disk plays a quite significant role in the optical appearance as seen by the human eye, which naturally focuses on image centers.  Physically, it means that the inner disk produces substantial CO and \ce{HCO+} emissions above the continuum level whereas it does not for CN, HCN, or CS.

\begin{figure*}
  \hspace*{3mm}\vspace*{-3mm}
  \begin{tabular}{cc}
  one-zone equilibrated model & two-zone equilibrated model\\
    \hspace*{-4mm}
    \includegraphics[page=1,width=86mm,height=59mm,trim=35 60 55 350,clip] 
    {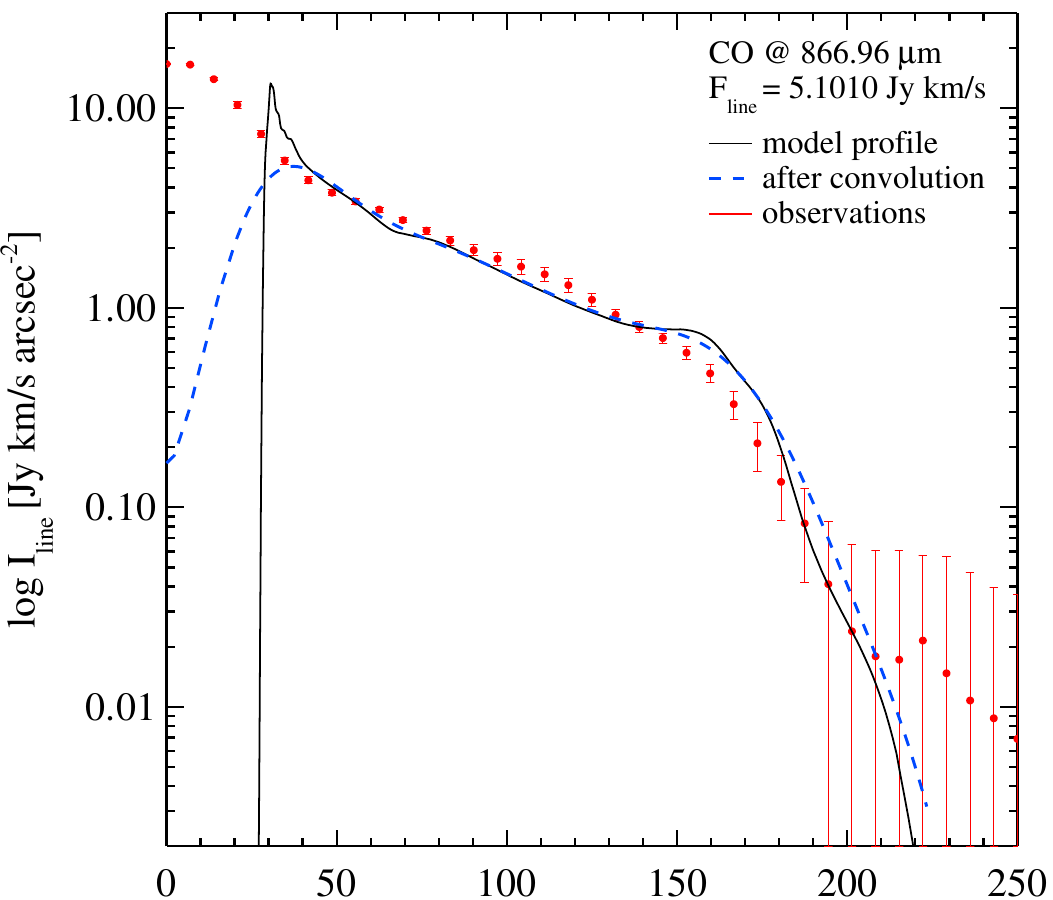} &
    \hspace*{-6mm}
    \includegraphics[page=1,width=86mm,height=59mm,trim=35 60 55 350,clip] 
    {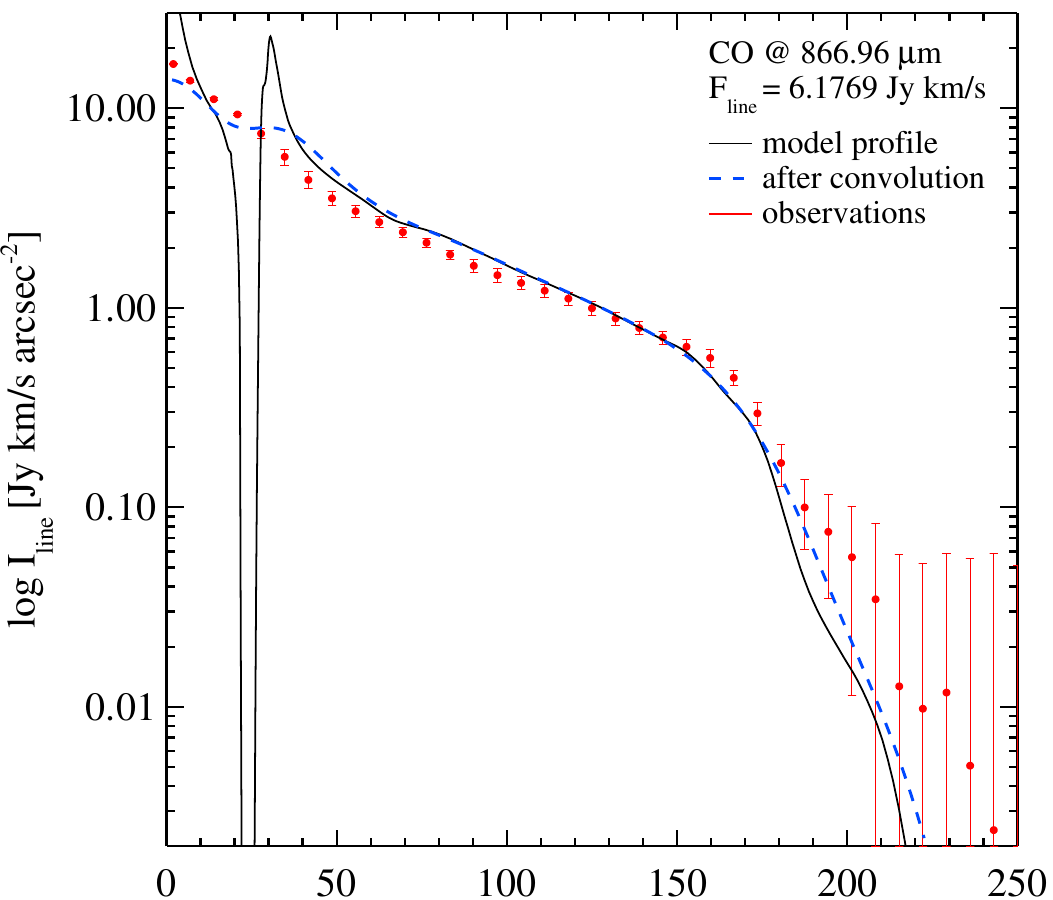} \\[-1mm]     
    \hspace*{-4mm}
    \includegraphics[page=2,width=86mm,height=59mm,trim=35 60 55 350,clip] 
    {Figs/out_oneWeight_Fig13.pdf} &
    \hspace*{-6mm}
    \includegraphics[page=2,width=86mm,height=59mm,trim=35 60 55 350,clip] 
    {Figs/out_twoWeight_Fig13.pdf} \\[-1mm]
    \hspace*{-4mm}
    \includegraphics[page=3,width=86mm,height=62mm,trim=35 30 55 350,clip] 
    {Figs/out_oneWeight_Fig13.pdf} &
    \hspace*{-6mm}
    \includegraphics[page=3,width=86mm,height=62mm,trim=35 30 55 350,clip] 
    {Figs/out_twoWeight_Fig13.pdf} \\[0mm]
  \end{tabular}
  \caption{Fit of radial line intensity profiles by the one-zone (left) and two-zone (right) $T$-equilibrated models for $^{12}$CO, HCO$^+$ and HCN. The black lines are the azimuthally averaged radial line intensity profiles from the continuum-subtracted model images, and the blue dashed lines show the same after convolving the images with the ALMA beams. The red errorbars show the radial line intensity data as retrieved from the left and the right side of the ALMA moment-0 maps of HD\,143006, respectively. This figure is continued in the Appendix for CN, $^{13}$CO and C$^{18}$O in Fig.\,\ref{fig:LineFit2} and for CS, \ce{N2H+} and \ce{C3H2} in Fig.\,\ref{fig:LineFit3}.}  
  \label{fig:LineFit1}
  \vspace*{-2mm}
\end{figure*}

\begin{figure*}
  \hspace*{-3mm}\vspace*{-3mm}
  \begin{tabular}{cc}
  \small CO $3\!\to\!2$\ \ \ one-equilibrated &  \small two-equilibrated\\
    \hspace*{-4mm}
    \includegraphics[page=1,width=90mm,trim=30 48 95 572,clip] 
    {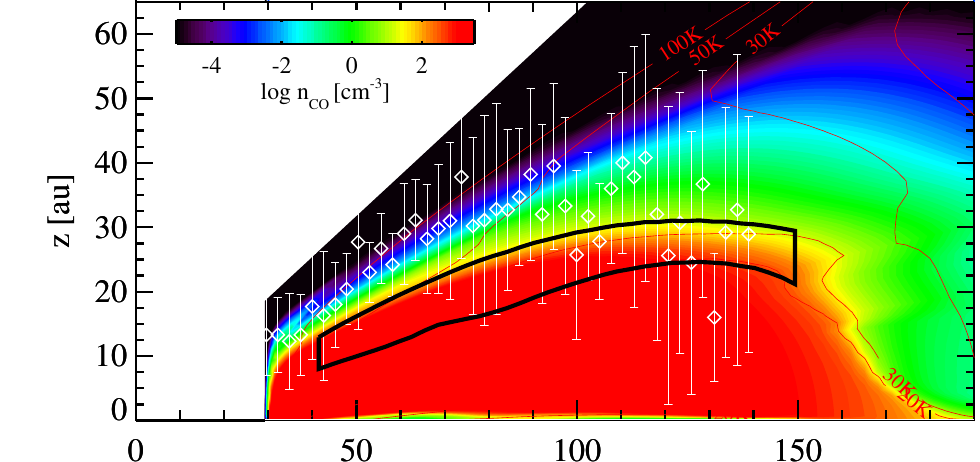} &
    \hspace*{-6mm}
    \includegraphics[page=1,width=90mm,trim=30 48 95 572,clip] 
    {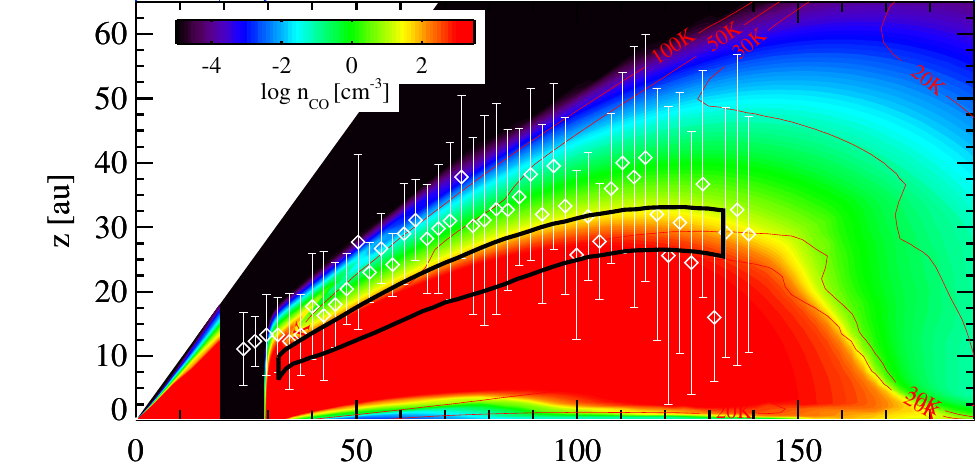} \\[0mm]
  \small $^{13}$CO $3\!\to\!2$\ \ \ one-equilibrated & \small two-equilibrated\\
    \hspace*{-4mm}
    \includegraphics[page=2,width=90mm,trim=30 48 95 572,clip] 
    {Figs/out_oneWeight_Fig14.pdf} &
    \hspace*{-6mm}
    \includegraphics[page=2,width=90mm,trim=30 48 95 572,clip] 
    {Figs/out_twoWeight_Fig14.pdf} \\[0mm]
  \small C$^{18}$O $3\!\to\!2$\ \ \ one-equilibrated & \small two-equilibrated\\
    \hspace*{-4mm}
    \includegraphics[page=3,width=90mm,trim=30 48 95 572,clip] 
    {Figs/out_oneWeight_Fig14.pdf} &
    \hspace*{-6mm}
    \includegraphics[page=3,width=90mm,trim=30 48 95 572,clip] 
    {Figs/out_twoWeight_Fig14.pdf} \\[0mm]
  \small CN $(3,5/2)\!\to\!(2,3/2)$\ \ \ one-equilibrated & \small two-equilibrated\\  
    \hspace*{-4mm}
    \includegraphics[page=4,width=90mm,trim=30 30 95 572,clip] 
    {Figs/out_oneWeight_Fig14.pdf} &
    \hspace*{-6mm}
    \includegraphics[page=4,width=90mm,trim=30 30 95 572,clip] 
    {Figs/out_twoWeight_Fig14.pdf} \\[1mm]
  \end{tabular}
  \caption{Molecular particle densities and line formation in the illuminated one-zone (left) and the two-zone (right) $T$-equilibrated models for $^{12}$CO, $^{13}$CO, C$^{18}$O and CN.  The black boxes mark the disk regions from which the model produces about 50\% of the respective line flux. The red contour lines mark where the gas temperature equals 20\,K, 30\,K, 50\,K and 100\,K, respectively -- these lines are identical in all plots in one column.  The white errorbars in the top two rows show the molecular emission surface data for $^{12}$CO and $^{13}$CO from \cite{Galloway2025}.  The figure is continued in the Appendix for HCN, \ce{HCO+}, \ce{N2H+} and CS in Fig.\,\ref{fig:LineFormation2} and for \ce{C3H2}, \ce{C2H} and \ce{H2CO} in Fig.\,\ref{fig:LineFormation3}.}
  \label{fig:LineFormation1}
  \vspace*{-2mm}
\end{figure*}

A clearer picture is provided by the continuum-subtracted radial line intensity profiles shown in Fig.\,\ref{fig:LineFit1}, which is continued as Figs.\,\ref{fig:LineFit2}, and \ref{fig:LineFit3} in the Appendix.  The $\chi^2$-deviations between the model and the line data is computed solely based on these results, not on the integrated line fluxes.  One can clearly see that the CO is radially much more extended than \ce{HCO+}.  The overall $\chi_{\rm lin}$ of the equilibrated models is 3.3 for the illuminated left side and 3.5 for the right shadowed side.  This result and the various contributions of the other observational data components (continuum images, SED, etc.) are outlined in Table~\ref{tab:Mdisk_var}. The radial intensity profiles provide a very hard test for the models. The observational data between about 40\,au and 130\,au, where the lines of $^{12}$CO, $^{13}$CO, CN, HCN, CS and \ce{HCO+} are optically thick, directly probe the gas temperatures at different radii and heights in the disk.  As shown in Figs.\,\ref{fig:LineFit1}, \ref{fig:LineFit2} and \ref{fig:LineFit3}, the model often struggles to find a disk setup where the gas is sufficiently cool in the line-forming regions, resulting in too high line intensities at these radii for these molecules. The best fit model is hence featured by parameter values that minimize the heating mechanisms: a low heating efficiency of exothermal reactions $\gamma^{\rm chem}$ and a low heat per photodissociation $\Delta E_{\rm pd}$. The model also assumes very low disk flaring $\beta$, which means that every new radial shell in the illuminated case intersects only little new star light. The role of the polycyclic amorphous hydrocarbon molecules (PAHs) is ambivalent. On the one hand side, photoeffect on PAHs is one of the most important heating processes for the gas in the line emitting regions, but on the other hand side, electron attachment to PAHs can reduce the electron concentration in the disk, which helps to make the \ce{N2H+} and \ce{HCO+} lines stronger. Lastly, the rather low turbulent line width $v_{\rm turb}$ helps to reduce the integrated line intensities to values that are more consistent with the observations.    

The gas temperatures seem too high especially around 30-40\,au, where the outer disk starts, and around 170\,au in consideration of CN and HCN. The beginning of the outer disk at 30\,au naturally intersects a lot of fresh star light, so warmer conditions are indeed expected there. At 170\,au, there is a weak PDR-structure in the model due to the interaction with the interstellar UV background radiation field, which causes a slight warming and intensified photo-chemical production of CN and HCN.  The data shows none of these features, in fact the line data is entirely smooth even between 20\,au and 40\,au, where the ALMA continuum images show a deep minimum which we interpret as the gap between the inner and the outer disk zones, laying the foundation of our modeling approach.  In summary, our model can satisfactorily fit the fluxes of 14 different ALMA lines on both the illuminated and the shadowed disk sides, the radial extension of the disk as seen in these lines, and the radial intensity slopes observed in the different molecules. There are, however, some radial shape mismatches which remain.

\subsection{Line formation}
Figures~\ref{fig:LineFormation1}, which is continued as Figs.~\ref{fig:LineFormation2} and \ref{fig:LineFormation3} in the Appendix, provides additional insight into the molecular concentrations and the line formation in the disk.  We visualize the molecular particle densities in the illuminated one-zone and the shadowed two-zone equilibrated disk models side by side. The red contour lines show the gas temperature structure, which is only slightly different in the one/two zone equilibrated models.  The black boxes in these figures mark the line forming regions.  The left and right black boundaries are where the cumulative line flux reaches 15\% and 85\%, respectively, and in every vertical column, the lower and upper black boundaries mark where the integrated line intensity reaches 15\% and 85\% at that radius, respectively. Thus, the black boxes encircle the regions in the disk from which altogether about 50\% of the line fluxes originate.

We note that the gas temperature structure is different from the dust temperature structure. In order to explain the line observations, we need gas temperatures as low as 20\,K to 60\,K in the line forming regions between about 40 and 130\,au, where the dust temperatures are significantly higher, about 40\,K to 80\,K.  Above $z\!\approx\!10$\,au, the gas in the model first cools by line emission as it thermally decouples from the dust. Eventually, it warms up again as the height increases and more starlight reaches these layers directly. Most line-forming regions are situated in this exact temperature minimum.  

HCN and \ce{N2H+}, shown in Fig.\,\ref{fig:LineFormation2}, are examples of molecules that show strong dependencies on chemical details. In particular, HCN is found to react strongly to the UV irradiation.  We only discovered this behavior after including the observed IUE stellar UV data (see Sect.\,\ref{sec:UV}). Analysis shows that the HCN photo-cross sections overlap well with Ly$\alpha$, whereas those of \ce{H2}, CO, CN and \ce{HCO+} do not.  Since the UV data shows that approximately half of the stellar UV radiation is emitted in form of Ly$\alpha$ photons, this behavior of HCN is understandable.  \ce{N2H+} is one of the molecules that is notoriously difficult to predict.  A recent model comparison organised by the DECO program (priv. comm.) showed variations of two orders of magnitude in the predictions of \ce{N2H+} line fluxes between different thermochemical models for the same physical disk setup. In our model, we observe the classical cold formation path via \ce{H3+} and \ce{N2} close to the deep (almost horizontal) CO ice line, since gaseous CO tends to "steal" the proton from \ce{N2H+}, i.e. $\rm\ce{N2H+} + CO \to \ce{HCO+} + \ce{N2}$. This region is larger when the disk is in the shadow.  However, there is another region, where the \ce{H3+} is produced by X-rays in the disk surface, above and around the inner rim of the outer disk. The \ce{H3+}-production does not work in that region when the X-rays are blocked by the inner disk.  These two effects almost perfectly cancel each other in our model concerning the \ce{N2H+} line flux.  Another remarkable result shown in Fig.\,\ref{fig:LineFormation3} is that while the \ce{C2H} and \ce{C3H2} line fluxes do not change much between the illuminated and the shadowed cases, their line emission regions shift outward and upward in the disk shadow.

\begin{figure*}
  \hspace*{0mm}\vspace*{-3mm}
  \begin{tabular}{cc}

    \hspace*{0mm}
    \includegraphics[page=1,width=85mm,trim=40 54 60 390,clip] 
    {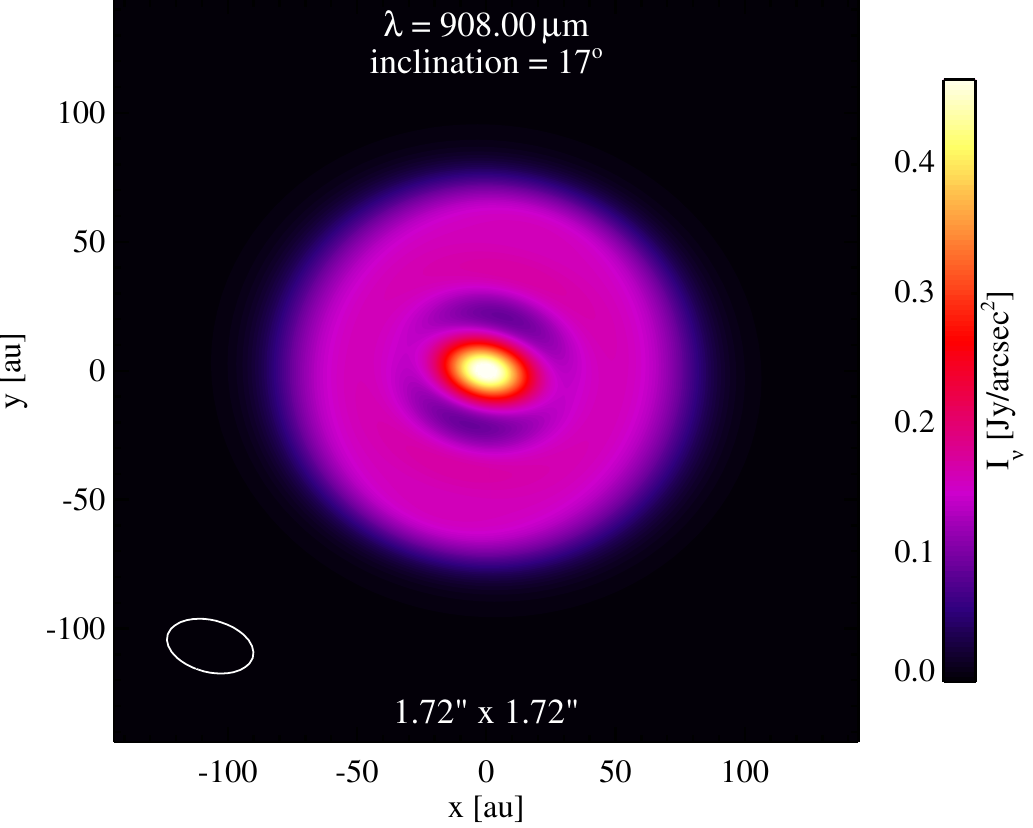} &
    \hspace*{-2mm}
    \includegraphics[page=2,width=85mm,trim=40 54 60 390,clip] 
    {Figs/out_twoZone_Fig15.pdf} \\[0mm] 
    \hspace*{-4mm}
    \includegraphics[page=3,width=85mm,height=62mm,trim=40 78 93 380,clip] 
    {Figs/out_twoZone_Fig15.pdf} &
    \hspace*{-6mm}
    \includegraphics[page=4,width=85mm,height=62mm,trim=40 78 93 380,clip] 
    {Figs/out_twoZone_Fig15.pdf} \\[0mm] 
    \hspace*{-4mm}
    \includegraphics[page=1,width=85mm,height=67mm,trim=40 40 93 380,clip] 
    {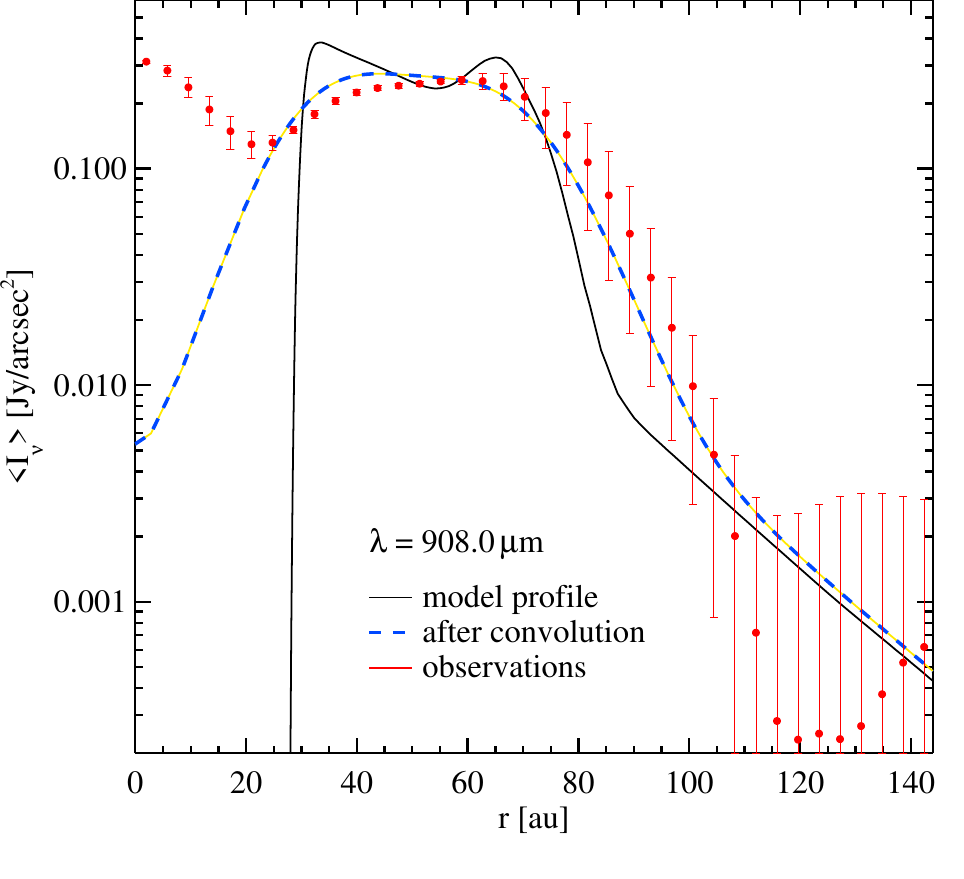} &
    \hspace*{-6mm}    
    \includegraphics[page=2,width=85mm,height=67mm,trim=40 40 93 380,clip] 
    {Figs/out_oneZone_Fig15.pdf} \\[-2mm] 
  \end{tabular}
  \caption{Fit of two ALMA continuum images of HD\,143006 at 908\,$\mu$m (left) and 1255\,$\mu$m (right). The top row shows the synthetic images from the two-zone model after beam convolution. The middle row shows the azimuthally averaged radial intensity profile (dashed) from these model images in comparison to the continuum data from the shaded right disk side. The bottom row shows the radial intensity profiles from the one-zone model in comparison to the data from the illuminated left disk side. The full black lines in these plots are the original model radial intensities before beam convolution. The synthetic images from the one-zone model are not shown here.}
  \label{fig:ContFit}
  \vspace*{0mm}
\end{figure*}

\subsection{Fitted continuum images}
Figure \ref{fig:ContFit} shows our fits of the two ALMA images at $908\,\mu$m and $1255\,\mu$m, including the radial intensity profiles. The inclusion of radial drift in our approach (see Sect.\,\ref{sec:RadialDrift}) produces a secondary intensity maximum at about 65-70\,au due to the accumulation of large dust grains $\ga\!80\,\mu$m.  Beyond that radius, the continuum intensities rapidly diminish, until the disk disappears in the noise between about 90\,au to 110\,au, depending on beam resolution. The model fits this sudden transition fairly well in both images. The inner disk in the two-zone model produces millimeter intensities at small radii up to 20\,au that fit the observations quite well, although the gap is slightly too deep in the high-resolution DSHARP image at $1255\,\mu$m.  Additional experiments show that our modeling approach would allow us to fit the double-humped shape of the continuum intensities more convincingly if we considered only the continuum data shown here.  However, the shape of the outer disk strongly impacts the line results, and the position of the inner wall of the outer disk is important for the SED fit as well. Therefore, the model strikes a balance to fit both the continuum and the line data as good as possible. The models predict a slightly too bright shadowed disk between 30 and 70\,au, especially in the 1255\,$\mu$m image, which could mean that too much starlight reaches the shadowed disk side despite the inclined inner disk in the model.

\section{Discussion}
\label{sec:discussion}

\subsection{Dust thermal relaxation}
\label{sec:TdustRelax}

\begin{figure}
  \vspace*{-1mm}
  \includegraphics[page=1,width=87mm,height=75mm,trim=40 30 65 335,clip] 
  {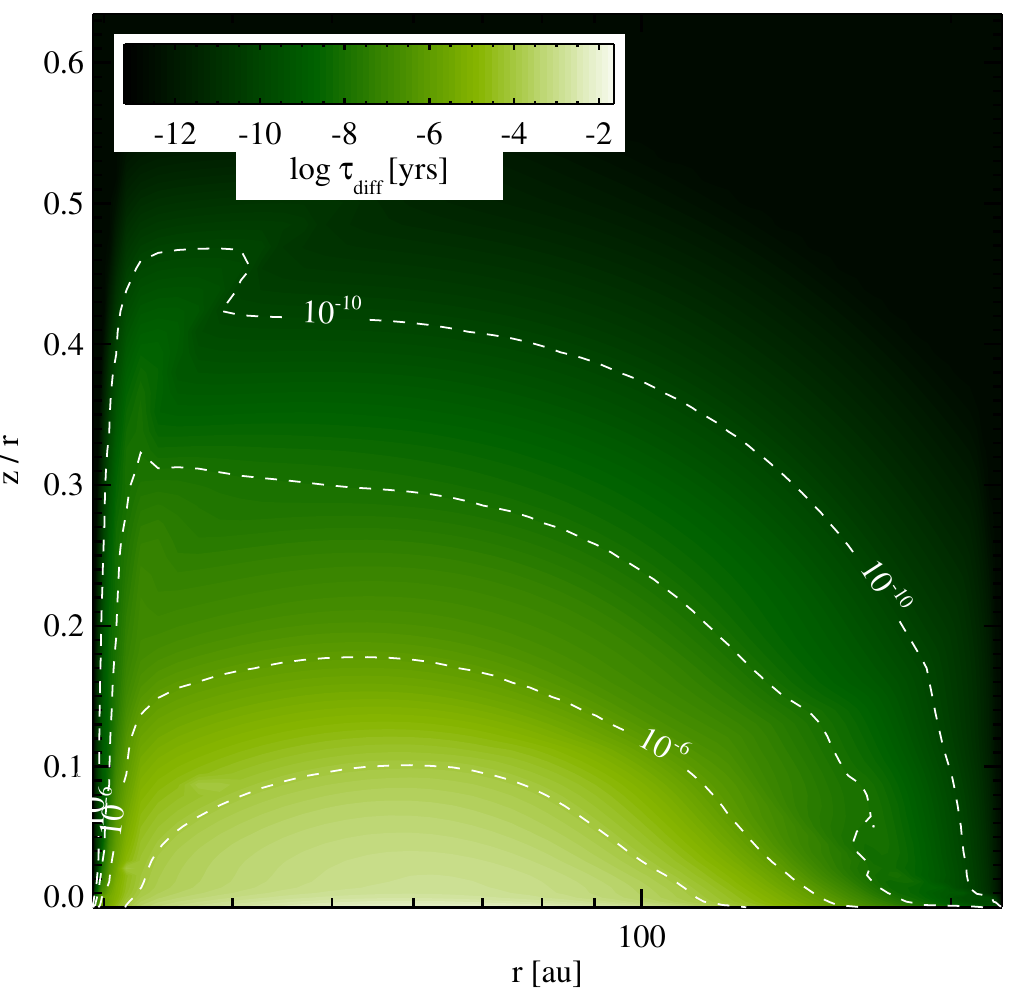}
  \vspace*{-2mm}
  \caption{Radiative diffusion timescales in the one-zone model.}
  \label{fig:tauDust}
  \vspace*{-2mm}
\end{figure}

The scenario of gas parcels passing periodically through the shadow casted by an inclined inner disk raises a number of fundamental questions for the disk modeling.  In our ``equilibrated'' models we have taken into account a potentially slow relaxation of the gas temperature, but there are two more processes to be considered: the thermal relaxation of the dust and the time-dependence of the chemistry.

The following simple estimate shows that the thermal relaxation of the dust in the entire disk can be expected to be quite fast.  According to the photometric observations, the outer disk of HD\,143006 radiates away about $L_{\rm\,farIR}\!\approx\!0.6\,L_\odot$ in the far IR.  The mass of the outer disk is about $10^{-3}\,M_\odot$, so the disk has about $N\!=\!5\times10^{53}$ \ce{H2} molecules with a total thermal energy of $E\!=\!N\,(5/2)\,k\bar{T}$ (we assume the \ce{H2} rotational states to be populated here). Assuming $\bar{T}\!\approx\!50$\,K, this results in $E\!\approx\!10^{40}$\,erg.  If the star would instantly stop shining, the disk would cool down on a timescale given by $E/L_{\rm\,farIR}\approx 0.1$\,yr, which is much shorter than the Keplerian orbital timescale.

One could argue, however, that this is only true when the energy within the disk is quickly transported to the disk surface, so that the cooling might be slower in the optically thick midplane.  However, as Fig.\,\ref{fig:tauDust} shows, the radiative diffusion timescale in this disk is even shorter, $\tau_{\rm diff}\!<\!0.01$\,yr, everywhere in the disk.  Here we have calculated the diffusion timescale as
\begin{equation}
\tau_{\rm diff} = \frac{u_{\rm gas}+u_{\rm rad}}{u_{\rm rad}}
                  \frac{3\,\tau_{\rm Ross}^2}{c\,\kappa_{\rm Ross}}
\end{equation}                  
where $u_{\rm gas}\!=\!(5/2)\,n\,kT$ is the thermal energy density, $u_{\rm rad}\!=\!\frac{4\pi}{c}J$ the radiative energy density, and $J$ is the bolometric mean intensity. $\kappa_{\rm Ross}$ is the Rosseland-mean dust opacity, which we calculate from our frequency-dependent dust opacities after the model has determined the local dust temperature, and $\tau_{\rm Ross}$ is the vertical integral over $\kappa_{\rm Ross}$ down to the point of interest.  

The diffusion timescales can be substantially longer in more massive disks that have larger column densities than our model for HD\,143006 (see Fig.\,\ref{fig:coldens}), however, even if our disk mass is only correct by order of magnitude, the dust temperature in the entire outer disk of HD\,143006 can still be expected to react almost instantly to changes of the irradiation, within less than one year. This justifies our choice of changing only the gas temperatures in the model approach.

\subsection{Chemical relaxation}
\label{sec:ChemRelax}
The chemical relaxation is more difficult to assess. Astrochemical networks are featured by an enormous dynamical range of chemical relaxation timescales that can be different by many tens of orders of magnitudes, which makes the problem {\em stiff} and numerically hard to solve.  Some chemical processes occur extremely fast, for example the production/destruction of trace species by photo-reactions on timescales of seconds, whereas the relaxation of the more abundant species can take longer than the lifetime of the universe in cold and perfectly shielded conditions \citep{Kanwar2025}. The formation and desorption of ices is a medium-fast process (operating on the timescale required for a molecule to collide with a dust grain), but converting one ice species into another can take tens of millions of years, driven by cosmic ray chemistry \citep{Helling2014}. 

\begin{figure}
  \vspace*{0mm}
  \centering
  \includegraphics[width=83mm,trim=20 20 10 10,clip] 
  {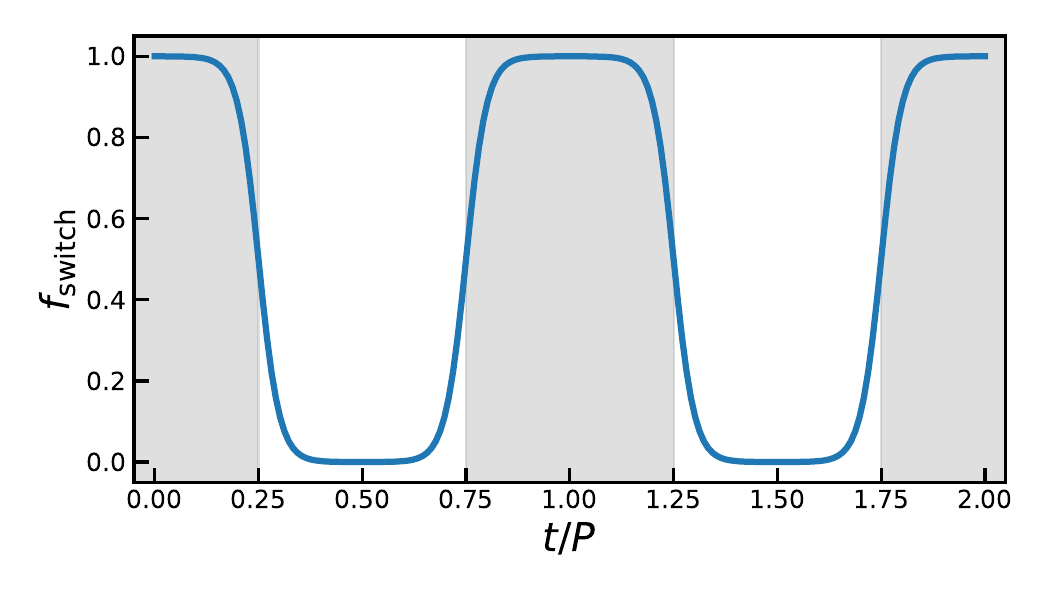}
  \vspace*{-2mm}
  \caption{The function $f_{\rm switch}(t)$ used to smoothly switch between the rate coefficients in the shadowed (grey) and the illuminated (white) disk sides.}
  \label{fig:fswitch}
\end{figure}

\begin{figure*}
\sidecaption
\begin{minipage}{12cm}
  \includegraphics[page=1,width=\linewidth,height=45mm,trim=20 20 10 15,clip] 
  {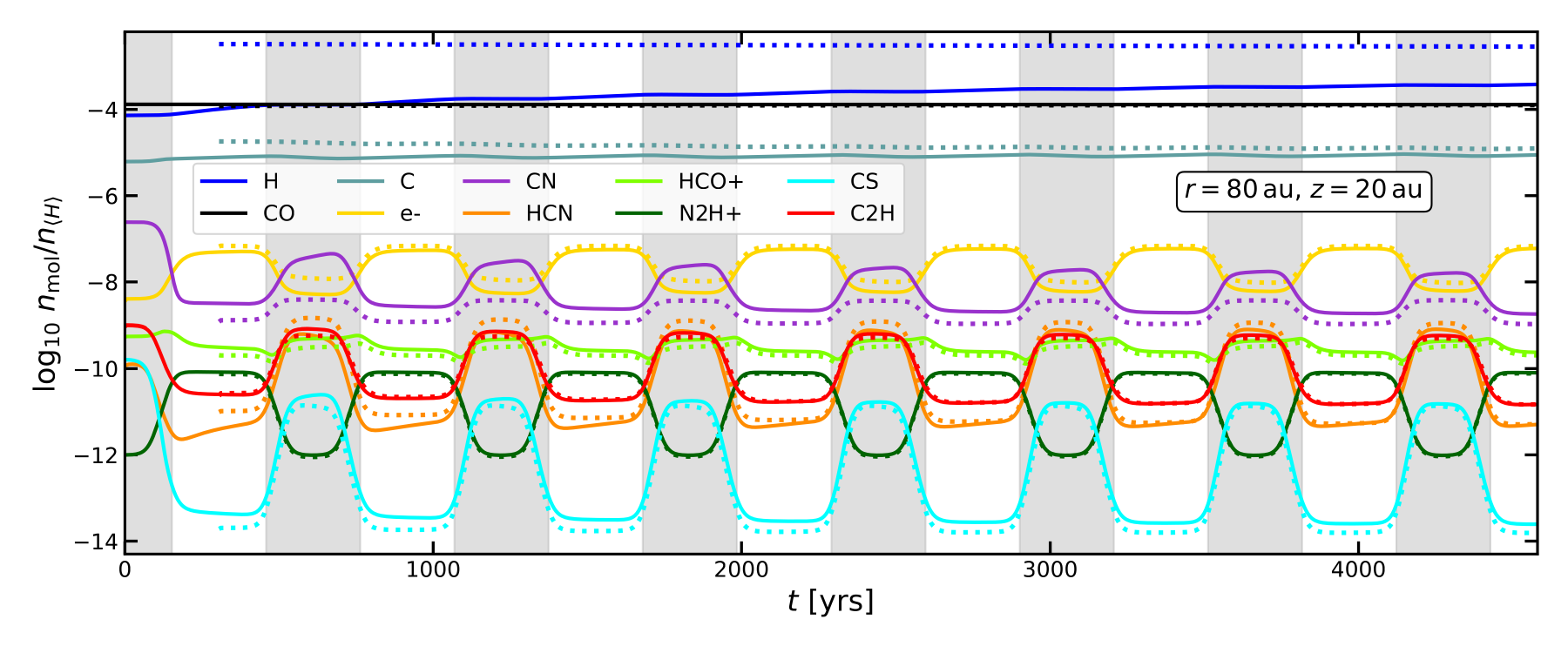}\\
  \includegraphics[page=1,width=\linewidth,height=45mm,trim=20 20 10 15,clip] 
  {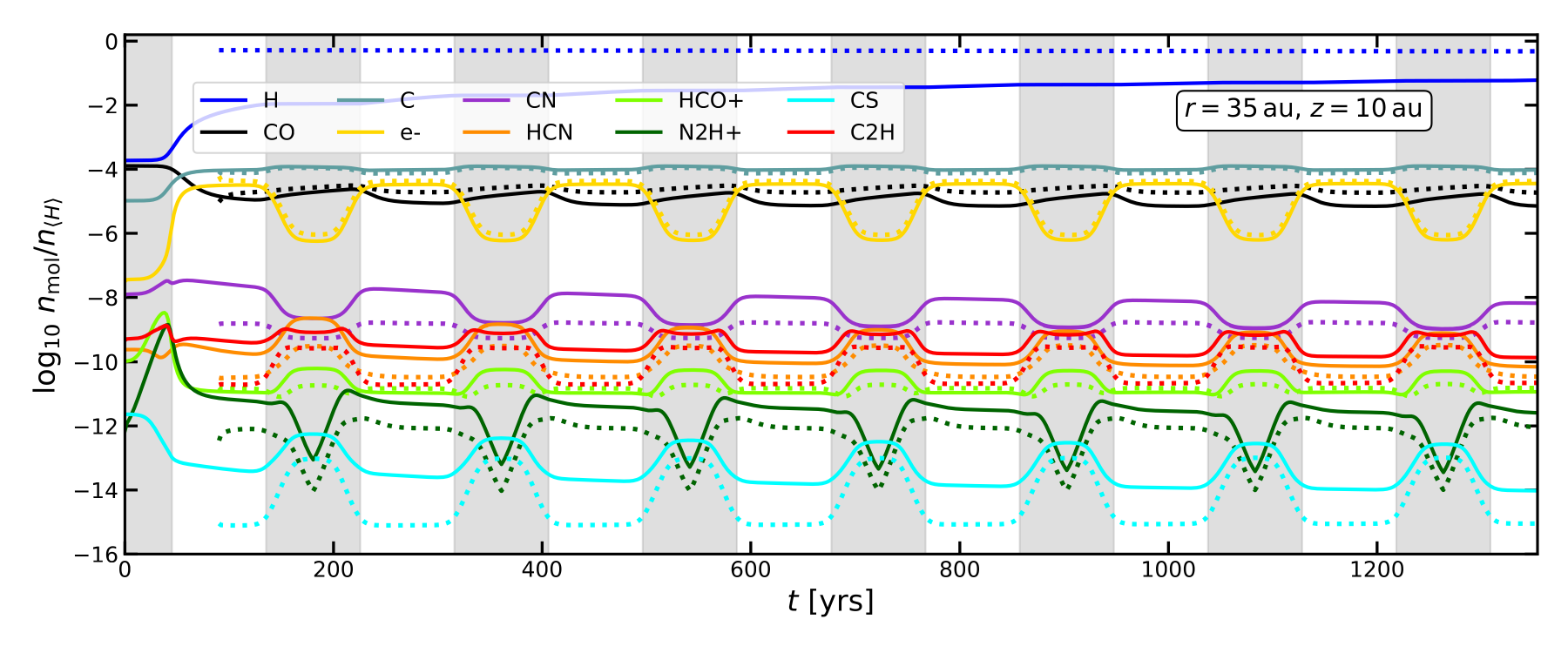}\\
  \includegraphics[page=1,width=\linewidth,height=45mm,trim=20 20 10 15,clip] 
  {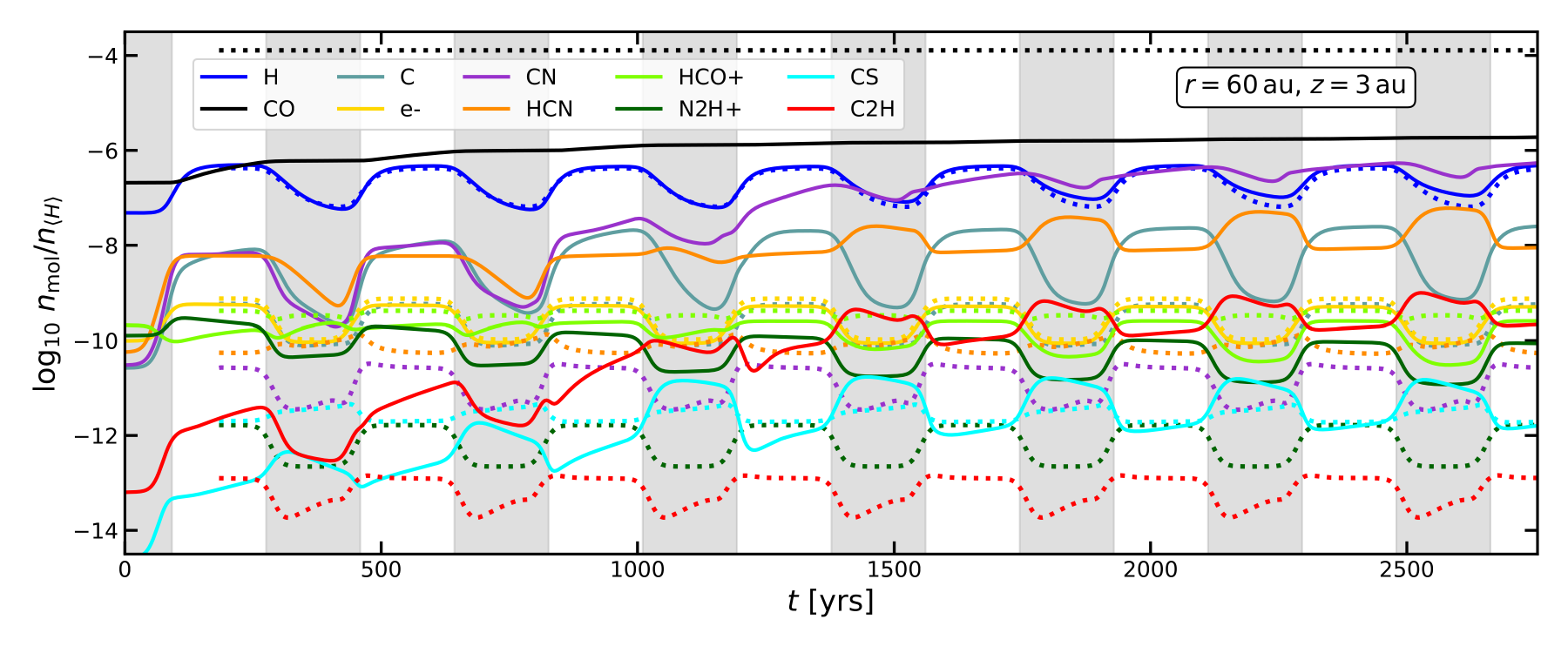}\\[-1mm]
\end{minipage}
  \caption{Time-dependent chemical concentrations at three different spatial points in the disk, simulating gas parcels that pass periodically through the shadow casted by an inclined inner disk. 
  Full lines: the model starts in the middle of the shadow with initial chemical concentrations from the time-independent two-zone equilibrated model. 
  Dotted lines: the model starts half a period later in the illuminated disk part, with initial chemical concentrations from the one-zone equilibrated model. The Keplerian orbital periods are $P\!=\!610\,$yr at ($r\!=\!80\,$au, $z\!=\!20\,$au), $P\!=\!179\,$yr at ($r\!=\!35\,$au, $z\!=\!10\,$au), and $P\!=\!367\,$yr at ($r\!=\!60\,$au, $z\!=\!3\,$au). The phases where the gas parcels are in the shadow are underlaid with gray areas.}
  \label{fig:PeriodicChemistry}
  \vspace*{-2mm}
\end{figure*}

\begin{figure*}
    \centering
    \includegraphics[width=1\linewidth]{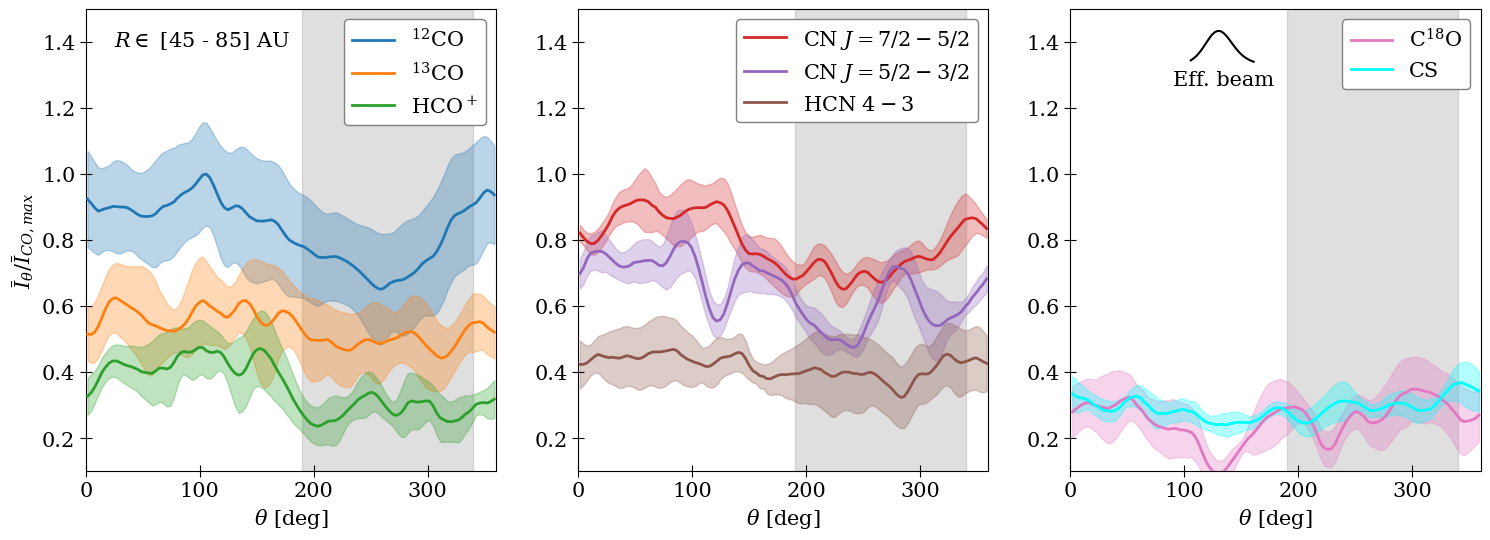}
    \vspace*{-6mm}
    \caption{Observed azimuthal intensity profiles for the Band 7 lines (except N$_2$H$^+$), averaged between radii $45 - 85$\,au and normalized to the peak intensity of $^{12}$CO $J=3-2$. 
    \label{fig:azimuthalprofile}}
\end{figure*}

In order to obtain some insight into the time-dependent chemical behavior, when the irradiation and the temperatures are varied periodically, we have advanced our chemical network at a few selected spatial points in the disk while periodically switching between the rate coefficients taken from the equilibrated one-zone model $R_k^{(1)}$ and the equilibrated two-zone model $R_k^{(2)}$. These rate coefficients have the full local information about the radiation field, gas and dust temperatures imprinted in them.  We use a smoothly switching, periodic function $f_{\rm switch}(t)$ as shown in Fig.\,\ref{fig:fswitch} for this purpose and re-compute the actual rate coefficients at every time-step as
\begin{equation}
R_k(t) = f_{\rm switch}(t)\;R_k^{(2)}
       \,+\, \big(1-f_{\rm switch}(t)\big)\;R_k^{(1)} \ .
\end{equation}
We start the model at $t\!=\!0$ when the gas parcel is in the middle of the disk side in the shadow. As Fig.\,\ref{fig:fswitch} shows, we advance the chemistry in the gas parcel for another quarter period in the conditions prevailing in the shadowed part, then for half a period in the conditions prevailing in the illuminated part, then back to the shadowed part for another quarter period, and so on. The initial concentrations are taken from the results of the two-zone equilibrated model.

Figure~\ref{fig:PeriodicChemistry} shows the resulting time-dependent chemical concentrations of such periodic models at three selected points in the disk.  The upper panel shows the results for ($r\!=\!80\,$au, $z\!=\!20\,$au), which is a central point in the line emitting regions of a number of molecules, such as CO, CN, HCN, \ce{HCO+} and \ce{C2H}, see Figs.\,\ref{fig:LineFormation1}, \ref{fig:LineFormation2} and \ref{fig:LineFormation3}. The hydrogen nucleus density is about $4\times10^{7}\rm\,cm^{-3}$ here, and the dust is optically thin in the UV in both directions, vertically and radially (the latter changes to radially optically thick once the inner disk blocks the direct star light).  The gas temperature is well equilibrated and stays constant at about 28\,K, whereas the dust temperature periodically changes between 46\,K and 95\,K. At the same time the UV flux between 91.2\,nm and 205\,nm changes between $2\times10^{9}\rm\,cm^{-2}s^{-1}$ ($\chi_{\rm UV}\!=\!19$) and $1.7\times10^{11}\rm\,cm^{-2}s^{-1}$ ($\chi_{\rm UV}\!=\!2400$).  The dust grains are small and rare, the dust/gas ratio is about $5\times10^{-6}$, and there is no ice.  The shapes of the resulting chemical concentrations as function of time resemble our switching function $f_{\rm switch}(t)$, with different positive or negative amplitudes depending on the molecule. 

When we start the model half a period later, with initial conditions taken from the illuminated part, we obtain approximately the same graph just shifted by half a period, see dotted lines. These results suggest that the chemical response to the periodical irradiation is nearly instantaneous at this point. While the electron density increases during the phases of illumination, the concentrations of the neutral molecules CN, HCN, CS and \ce{C2H} are reduced when illuminated. This is because the main destruction reactions for these molecules are UV-photo reactions, which become instantly larger when illuminated, whereas the main formation reactions require free atoms, which are too abundant to follow the periodic chemistry.  The strong responses of HCN, CS and \ce{C2H} explain why we see enhancements of their concentrations in the upper disk regions on the right side of Figs.\,\ref{fig:LineFormation1} and \ref{fig:LineFormation2}.  This is why the line fluxes of HCN, CS, and \ce{C2H} are stronger in the two-zone equilibrated model, see Table~\ref{tab:result_linefluxes}.  

Other molecules such as CO and \ce{HCO+} are less affected by the UV-photons and their responses are weaker. In contrast, \ce{N2H+} becomes more abundant when irradiated, unlike all other depicted molecules, and despite the higher electron concentrations present during these phases.  This is a consequence of the stellar X-ray irradiation being blocked by the inner disk.  When illuminated, the X-rays create more \ce{H3+} and other protonated molecules.  However, \ce{HCO+} is less affected by the irradiation as there is another photo-chemical formation pathway via \ce{CH5+} that does not require X-rays, see further details in Sect.\,\ref{sec:pathways}.  The top panel of Fig.\ref{fig:PeriodicChemistry} furthermore shows that the concentrations of CO and the free atoms only change on much longer timescales, if at all.  In fact, we would expect that once the full and dotted lines of CO and the free atoms unite, the model should become perfectly periodic.

Fig.~\ref{fig:azimuthalprofile} shows the observational data as function of azimuth $\theta$.  Here, we have averaged the moment-0 line intensity data $I(r,\theta)$ over radii 45-85\,au and have plotted them as function of $\theta$, normalized to the peak average intensity of $^{12}$CO $J$=3-2.  The figure shows that the disk's CN emission is brightest around $\theta\!=\!0^\circ-180^\circ$, whereas HCN has a major and a secondary peak at similar $\theta$ values.  The \ce{HCO+} line shows the strongest relative azimuthal variations (about 40\%) with a maximum on the illuminated side that is shifted to slightly larger $\theta$-values.  The $^{12}$CO line is optically thick in this radial range and therefore a good tracer of the (slow) azimuthal temperature changes.  In fact, Fig.~\ref{fig:PeriodicChemistry} supports this interpretation: CO is too abundant to be affected much by the periodical chemistry on orbital timescales, hence the $^{12}$CO line is a temperature indicator, not a concentration indicator.  The other molecules show less azimuthal variations in the data, in fact those variations are surprisingly small.

\begin{figure*}[htp]
  \centering
  \vspace*{-2mm}
  \begin{tabular}{|c|c|}
    \hline
    one-zone equilibrated model & two-zone equilibrated model\\
    \hline
    &\\[-2.2ex]
    \resizebox{86mm}{!}{
    $\nH\!=\!4\times10^{7}\rm cm^{-3}$, 
    $T_{\rm gas}\!=\!29\,$K, 
    $T_{\rm dust}\!=\!95\,$K, 
    $\chi_{\rm UV}\!=\!2400$,
    $\zeta_{\rm X}\!=\!1.5\times10^{-15}\rm\,s^{-1}$} &
    \resizebox{86mm}{!}{
    $\nH\!=\!4\times10^{7}\rm cm^{-3}$, 
    $T_{\rm gas}\!=\!28\,$K, 
    $T_{\rm dust}\!=\!46\,$K, 
    $\chi_{\rm UV}\!=\!19$,
    $\zeta_{\rm CR}\!=\!1.5\times10^{-17}\rm\,s^{-1}$} \\
    \hline
    &\\[-1.5ex]
    \hspace*{-3mm}\includegraphics[height=82mm]{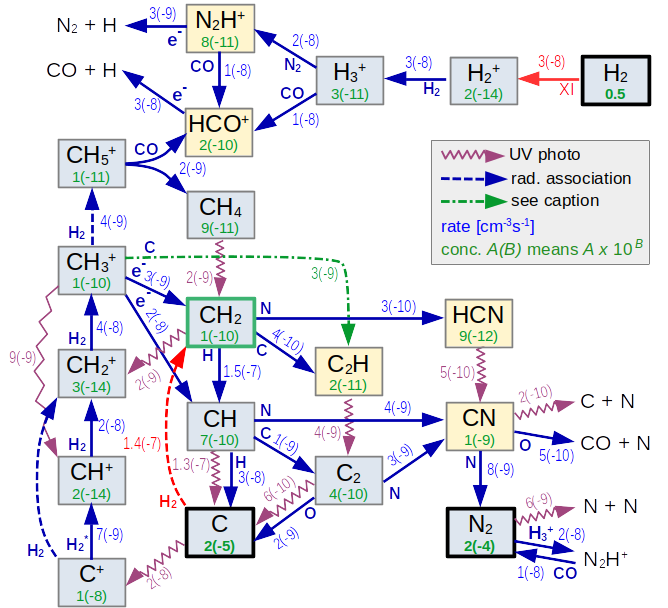} 
    \hspace*{-4mm} & 
    \hspace*{-2mm}\includegraphics[height=82mm]{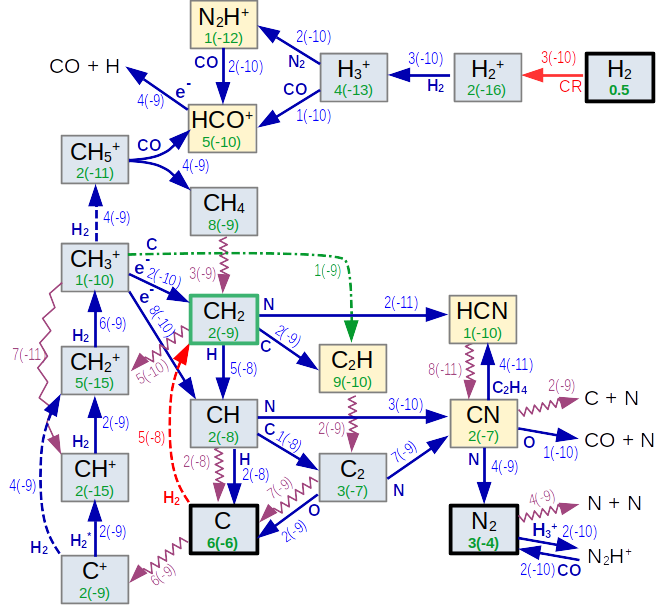}
    \hspace*{-3mm}\\[1mm]
    \hspace*{-3mm}\includegraphics[height=27mm]{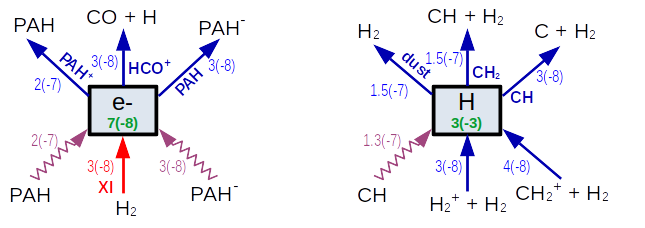}
    \hspace*{-3mm}&
    \hspace*{-2mm}\includegraphics[height=27mm]{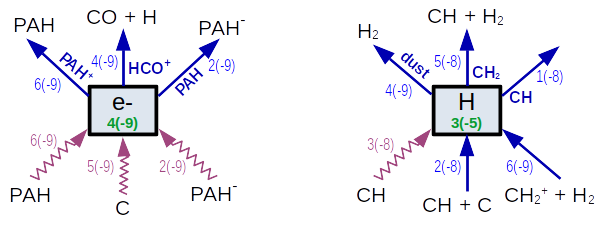}
    \\
    \hline
  \end{tabular}
  \caption{Reaction pathways creating some of the observable molecules, marked in yellow, in the illuminated disk side (left) and in the disk side in the shadow (right). The point selected for this chemical analysis is at $r\!= \!80\,$au and $z\!=\!20\,$au. The main element carriers are \ce{H2}, CO, and \ce{N2}. The density, temperature and irradiation conditions are listed, where $\nH$ is the hydrogen nuclei density, $T_{\rm gas}$ and $T_{\rm dust}$ are the gas and dust temperatures, $\chi_{\rm UV}$ the UV field strength in Draine units, and $\zeta_{\rm X}$ and $\zeta_{\rm CR}$ are the relevant \ce{H2} ionization rates due to X-rays and cosmic rays. $\zeta_{\rm X}$ is tiny on the right side. The green dash-dotted arrow represents an almost linear (non-branching) chain of fast ion-neutral reactions forming \ce{C2H} from \ce{CH3+} via 
  $\rm\ce{CH3+} + C \to \ce{C2H+} + \ce{H2}$ followed by
  $\rm\ce{C2H+} + \ce{H2} \to \ce{C2H2+} + H$ followed by
  $\rm\ce{C2H2+} + \ce{H2} \to \ce{C2H3+} + H$ followed by
  $\rm\ce{C2H3+} + \ce{e-} \to \ce{C2H} + H + H$.}
  \label{fig:pathway}
  \vspace*{-1mm}
\end{figure*}

The second panel of Fig.\,\ref{fig:PeriodicChemistry} shows the modeling results at the point ($r\!=\!35\,$au, $z\!=\!10\,$au) situated above the dust over the inner wall of the outer disk.  This region is important for the formation of the apparent rings in the CN, HCN and CS simulated ALMA moment-0 maps, see Fig.\,\ref{fig:model_all_mol}, and contributes to the \ce{N2H+} line flux in the illuminated case, see left side of Fig.\,\ref{fig:LineFormation2}. Here, $\nH\!\sim\!7\times10^{7}\rm\,cm^{-3}$ and the dust is optically thin. The cooling timescales are shorter here, so the gas temperature varies between 104\,K and 108\,K, while the dust temperature changes between 63\,K and 132\,K. The UV-flux varies between $\chi_{\rm UV}\!=\!120$ and $\chi_{\rm UV}\!=\!14000$.  The latter is by far the most important factor driving the periodic chemistry.  At first glance, the responses of the neutral molecules look similar to the case shown in the top panel, however, the CN-variations are actually inversed here.  The CN concentration is largest in the illuminated phases, similar to \ce{N2H+}.  
After 8 periods, there are still some deviations between the two models started with different initial conditions. Indeed, we see a slow downward drift of the periodical variations started in the shadow, which seems connected to the slow but steady increase of the atomic H density in this model. Chemical reactions with \ce{H2} are mostly constructive (building complexity) whereas reactions with atomic H are often destructive (reducing complexity), see Fig.~\ref{fig:pathway}.  Therefore, the dotted model started with high concentration of atomic H shows overall lower concentrations of the neutral molecules.  Another observation is that the concentrations in the illuminated phases prevail a little longer than in the shadowed phases, this asymmetry is particularly visible for \ce{N2H+}. Possibly, the chemical relaxation timescales are somewhat shorter under illumination.  However, all these deviations from periodicity are relatively modest, and we can conclude that -- in both points discussed so far -- our modeling strategy to calculate the chemistry in the time-independent approximation seems justified. 

The third panel of Fig.\,\ref{fig:PeriodicChemistry} shows the results at the point ($r\!=\!60\,$au, $z\!=\!3\,$au), which is situated at the horizontal transition zone into the icy region, where the following ices start to occur: \ce{CO2}\#, \ce{C2H4}\#, \ce{HCN}\#, \ce{HNC}\#, and \ce{NH3}\#.  Since the model is carbon-rich, there is no \ce{H2O}\#. This zone is featured by a maximum of the \ce{N2H+} concentration, see right side of Fig.\,\ref{fig:LineFormation2}, where the \ce{CO} gas is already gone, and the \ce{N2} gas is about to go, caused by the combined effect of the formation of these ices. This zone is mainly responsible for the \ce{N2H+} line flux. Here, $\nH\!\sim\!6\times10^{8}\rm\,cm^{-3}$, and the dust is optically thick in the radial direction, and borderline optically thin in the vertical direction ($A_{\rm V,ver}\!=\!1.2$). The UV-flux varies between $\chi_{\rm UV}\!=\!0.4$ and $\chi_{\rm UV}\!=\!5$. The gas is thermally coupled to the dust, hence both temperatures vary in similar ways, the gas temperature between 39\,K and 43\,K and the dust temperature between 37\,K and 45\,K. The ionization of \ce{H2} is mostly driven by X-rays and is hence constant, because the X-rays reach this point along the midplane that is not blocked by the inclined inner disk. However, the UV-illumination that reaches this point via scattering on small grains in the disk surface is just about strong enough in the illuminated case to mostly prevent the ice formation by photo-desorption. Therefore, the dust particles here have only 66 ice layers in the illuminated case, but 8000 layers in the shadowed case according to the time-independent equilibrated one-zone and two-zone models, which provide the initial conditions for this simulation.  Consequently, the model started on the shadowed side (full lines) evolves very differently from the model started on the illuminated side (dotted lines). The two models do not relax toward a common solution during the first 8 periods as we have seen in the other two cases shown in Fig.\,\ref{fig:PeriodicChemistry}. In particular, the CO concentration is very different by initially three orders of magnitude, and that concentration does not change much during the first eight periods. The reason for the initial differences is that, on very long timescales, CO can be converted into \ce{CO2} ice, or can be unblocked by X-rays or cosmic rays to subsequently form hydro-carbon molecules such as \ce{C2H4} that can freeze out \citep{Helling2014}, but these processes require millions of years or more. Such times are not available here, and therefore, the chemistry develops a periodic behavior only in the fast modes, under the constraint of fixed concentrations of the element carriers such as CO, \ce{N2} and the ices. 

Interestingly, the CN, HCN and \ce{C2H} concentrations in the bottom panel of Fig.\,\ref{fig:PeriodicChemistry} increase rapidly in the case where the model is started in the shadow.  The CN concentration reaches values up to $10^{-7}-10^{-6}$ which is larger as in any of the two time-independent models.  \ce{C2H}, CS and HCN show similar trends.  We attribute this effect to the UV photo-desorption of \ce{NH3}-ice and hydro-carbon ices, which injects fresh nitrogen, carbon and sulfur into an oxygen-poor environment.
For HCN, the effects are similar, but since HCN has photo-dissociation cross sections also at softer UV wavelengths, the available UV is sufficient to limit this trend.  The response of CS is similar to HCN, driven by the UV-desorption of \ce{H2S}-ice.  In the shadowed phases, the ices re-build again. \ce{N2H+} is found to be way more abundant in the model started from the dark icy disk side which lacks CO. The low CO abundance increases the \ce{N2H+}-lifetime, since \ce{N2H+} is mainly destroyed by the reaction $\ce{N2H+} + \ce{CO} \to \ce{HCO+} + \ce{N2}$, see Sect.\,\ref{sec:pathways}.  This explains why \ce{N2H+} is more prominent towards the midplane in the two-zone model shown on the right side of Fig.\,\ref{fig:LineFormation2}.   We expect that even these two time-dependent models would eventually become periodical, too, but that could take millions of years (tens of thousands of periods), which is impossible to calculate.  This is a complicated case, where our modeling approach (see Sect.\,\ref{sec:model_approach}) is not fully capable of predicting the periodic changes of the gas and ice composition.  

\subsection{Chemical pathways}
\label{sec:pathways}

Figure~\ref{fig:pathway} shows the dominant chemical pathways to form the observable molecules CN, HCN, \ce{C2H}, \ce{HCO+} and \ce{N2H+}, again at $r\!=\!80\,$au and $z\!=\!20\,$au as already discussed in the top panel of Fig.\,\ref{fig:PeriodicChemistry}, but here we analyze the time-independent results from the equilibrated one-zone and two-zone models, where every chemical species is created and destroyed at the same rate.  Since we are discussing here a slightly carbon-rich model ($\rm C/O\!\sim\!1.05$) with plenty of nitrogen ($\rm N/C\!\sim\!4$) see Table~\ref{tab:parameter}, there are a few surprises.  The chemical activity starts  with the two photo-reactions
\begin{equation}
  \ce{CO} ~+~ h\nu ~\longrightarrow~ \rm C ~+~ O 
  \quad\mbox{and}\quad
  \ce{N2} ~+~ h\nu ~\longrightarrow~ \rm N ~+~ N \ , 
\end{equation}
to create some free atoms, followed by the radiative association reaction
\begin{equation}
  \ce{C} ~+~ \ce{H2} ~\longrightarrow~ \ce{CH2} ~+~ h\nu \ .
\end{equation}
These reactions, together with the ionization of \ce{H2} by either X-rays or cosmic rays (both marked with red arrows in Fig.\,\ref{fig:pathway}), are the reactions that activate the chemistry. The neutral molecule \ce{CH2} plays a key role for the production of all observable molecules but \ce{N2H+}.  Either it (i) reacts with a N-atom to form HCN, which photodissociates to CN, or it (ii) reacts with a C-atom to form \ce{C2H}.  Most frequently, however, it (iii) reacts with a H-atom to form \ce{CH}, which can form CN by reactions with N-atoms either directly or indirectly via \ce{C2}. Alternatively, \ce{CH2} can (iv) be photoionized to form \ce{CH2+} which undergoes fast reactions with \ce{H2} to form first \ce{CH3+} and then \ce{CH5+}.  \ce{CH2+} is also formed by $\rm C + h\nu \to C^+ + e^-$ followed by another radiative dissociation reaction $\rm C^+ + H_2 \to CH_2^+ + h\nu$.   \ce{CH5+} then reacts with CO to form \ce{HCO+} via the following reaction
\begin{equation}
  \ce{CH5+} ~+~ \ce{CO} ~\longrightarrow~ \ce{HCO+} + \ce{CH4} \ .
  \label{eq:HCO+formation}
\end{equation}
Reaction (\ref{eq:HCO+formation}) is one of the main production channels to form \ce{HCO+}, besides $\ce{H3+} + \ce{CO} \to \ce{HCO+} + \ce{H2}$.  In contrast, the molecule \ce{N2H+} forms only via the \ce{H3+} pathway that starts with an ionization of \ce{H2}.  In the illuminated case shown on the left side of Fig.\,\ref{fig:pathway}, the \ce{H2}-ionization is driven by X-rays (XI means a reaction with a fast secondary electron created by the primary X-rays processes).  On the shadowed side shown on the right, those X-rays are blocked by the inner disk, and hence the \ce{H2}-ionization is driven by cosmic rays alone, which is slower by a factor of about 100 in this case.  Consequently, the \ce{N2H+}-density is also lower by a factor of about 100 is the shadowed disk side. Concerning the observed neutral molecules, the photo-reactions are slower in the shadow, which allows these molecules to reach higher concentrations.

One can directly read off some chemical relaxation timescales from Fig.\,\ref{fig:pathway}. For CN, for example, we find on the left side a molecular density of $n_{\rm CN}\approx 10^{-9}\nH \approx 0.04\rm\,cm^{-3}$ and a total formation ($=$ total destruction) rate of $R\approx 8\times10^{-9}\rm\,cm^{-3}s^{-1}$ which results in $\tau_{\rm CN}=n_{\rm CN}/R\approx 0.2\,$yr. The relation $\tau\ll P$ holds for all molecules shown, where $P$ is the orbital period.  However, when we apply the same recipe to atomic hydrogen we find $\tau_{\rm H}\approx 10000\,{\rm yrs}\approx 17\,P$, which is an example for an intermediate timescale that in fact leads to long-term trends in Fig.\,\ref{fig:PeriodicChemistry}.

Figure~\ref{fig:pathway} also shows the impact of the PAH chemistry on the electron concentration. The total PAH concentration in this model is $f_{\rm PAH}\cdot 3\times10^{-7}\approx 2\times 10^{-9}$ which is similar to the sulfur abundance assumed ($5\times 10^{-9}$). The electron attachment to PAHs \citep{Thi2019} can hence reduce the concentration of free electrons significantly, even at a height $h/r\!=\!0.4$ as in this case. This allows our model to generate the high \ce{HCO+} and \ce{N2H+} concentrations needed to explain the line observations of these protonated molecules. 

\begin{figure}[h!]
\centering
    \vspace*{-1mm}
    \includegraphics[width=85mm,trim=15 20 20 15,clip]{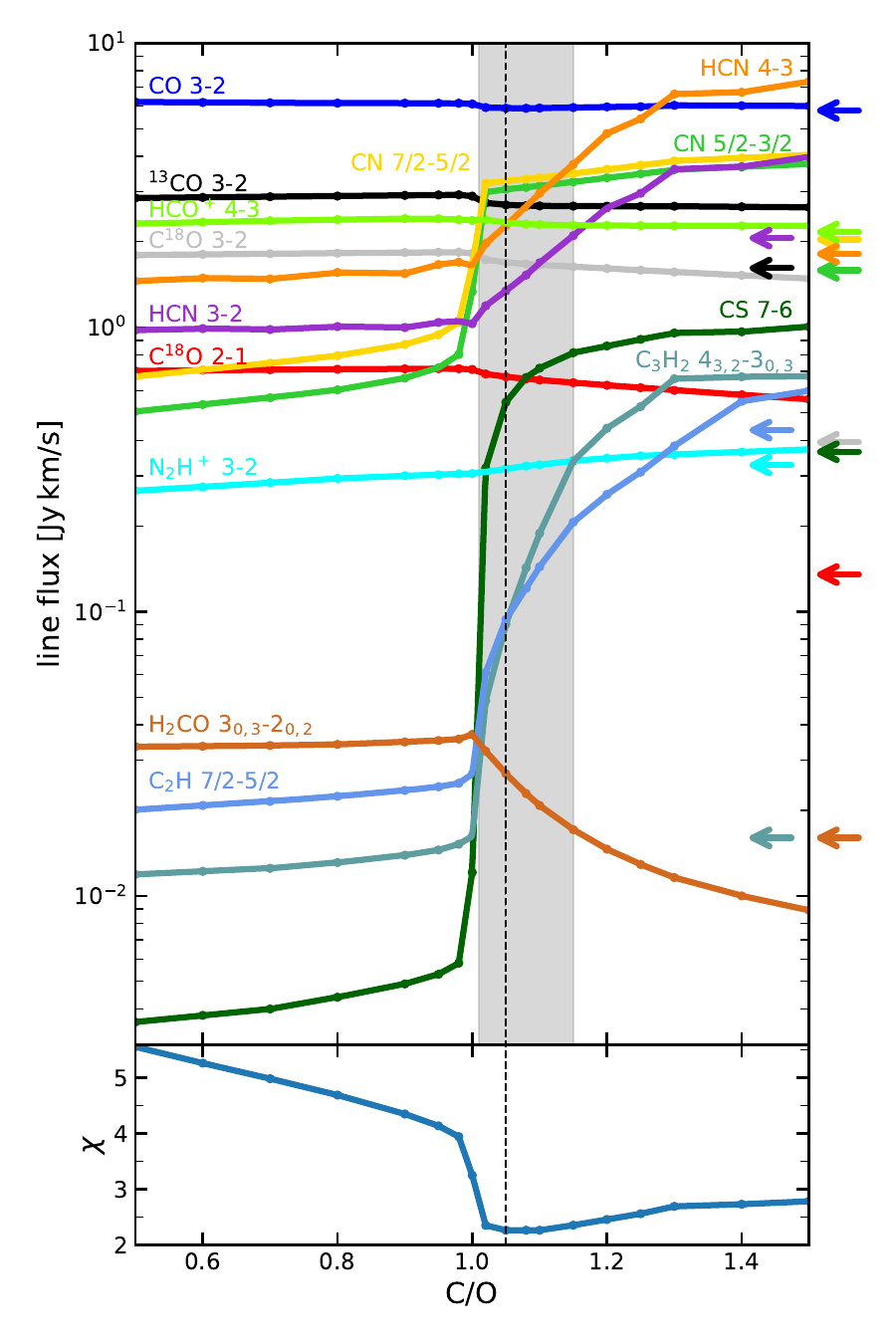}
    \vspace*{-1mm}
    \caption{Variation of the carbon-to-oxygen ratio in the best-fitting model, and its effect on predicted line fluxes and $\chi^2$. The arrows on the right represent the observational data. The vertical dashed line represents our best-fitting model with $\rm C/O=1.05$, and the gray shaded area visualizes the range of C/O-ratios compatible with most observations but the $^{13}$CO, C$^{18}$O and C$_2$H data. The lower plot shows the total model fit quality $\chi$ with respect to all observational data.}
    \label{fig:LineFluxes_CtoO}
\end{figure}

\subsection{The C/O ratio}
\label{sec:CO}
The C/O element abundance ratio\footnote{Element abundances include gas and ice in ProDiMo. Thus, when for example water freezes out, it does not change C/O according to our definition. Unfortunately, the definition of C/O is not unique in the literature. Often, the C/O ratio is simply used for the gas, without giving a clear definition, as e.g.\ in \cite{Keyte2023}. In that case, C/O obviously becomes a spatially dependent quantity, unsuitable for a model parameter.  In thermo-chemical equilibrium, where CO is converted into the thermodynamically more favorable \ce{CH4} and \ce{H2O} at low temperatures, the gas phase C/O in a solar composition gas can be as high as several 1000 when water freezes out \citep{Woitke2018}. According to our definition, only element-selective transport processes, such as the drift of icy pebbles, can change element abundances. Diffusion (as long as gas and icy grains have the same diffusion coefficient, see \citealt{Woitke2022}) and any local chemical processes, including freeze-out, cannot.\label{foot:C/O}} has a strong influence on our modeling results once it gets close to $\rm C/O\!=\!1$.  Figure~\ref{fig:LineFluxes_CtoO} shows a series of disk simulations, based on our best fitting model, with C/O ranging from 0.5 to 1.5. We see that the \ce{C2H} and \ce{C3H2} lines, and in particular the CS line, increase their line fluxes by order of magnitudes once $\rm C/O\!>\!1$, which is a natural consequence of CO-blocking. The CN and HCN lines react in similar ways but less pronounced.  In contrast, the \ce{H2CO} line is the only line in our sample that shows an inverse trend.  The other lines (CO, \ce{HCO+} and \ce{N2H+}) show little effect.  Similar results were found in the disk models of \cite{Williams2025}, where the CS column density was found to be higher by three orders of magnitude when the C/O ratio was increased to 1.4.

Our best-fitting model has $\rm C/O\!=\!1.05$ which strikes a balance between all lines concerning how they vary with C/O.  In particular, the best-fit model has to balance bright \ce{C2H} with very faint \ce{C3H2}, and ends up compromising between both of these. Models with $\rm C/O\!<\!1$ can clearly be ruled out -- those would have too weak CS and \ce{C2H} lines. Models with $\rm C/O\!\gtrsim\!1.15$ produce too strong HCN, CN, CS and \ce{C3H2} lines, and a too weak \ce{H2CO} line.  We conclude that our 14-line fit models suggest $\rm CO\!=\!1.05_{-0.03}^{+0.1}$. We note that $\rm C/O\!\to\!1$ is the expected result of chemical disk evolution inside of the CO-iceline after millions of years. This occurs once all excess oxygen not bound in CO is converted into molecules that can freeze out, and that ice is subsequently transported away by gravitational settling and radial drift \citep{Oberg2011, Helling2014}.

\subsection{The disk mass}
\begin{table}
\caption{Additional fitting run with fixed disk mass.}
\label{tab:Mdisk_var}
\vspace*{-3mm}
\begin{tabular}{c|cc|c}
\hline
&&\\[-2.2ex]
& main run & $M_{\rm gas}\!=\!0.021\,M_\odot$\\ 
&&\\[-2.2ex]
\hline   
&&\\[-2.2ex]
$\chi$            & 2.23\,, 2.27   & 3.03\,, 2.85  & \\
$\chi_{\rm ph}$   & 1.39\,, --     & 1.84\,, --    & \\
$\chi_{\rm Sp}$   & 3.98\,, --     & 4.42\,, --    & \\
$\chi_{\rm im}$   & 1.24\,, 1.17   & 1.93\,, 1.12  & \\
$\chi_{\rm line}$ & 3.26\,, 3.50   & 4.59\,, 4.54  & \\
$\chi_{\rm hei}$  & 1.16\,, 1.09   & 1.16\,, 1.14  & \\
&&\\[-2.2ex]
\hline
& \multicolumn{2}{c}{left\ , right line fluxes [Jy km/s]}
  & observed\\
\hline
&&&\\[-2.2ex]
$^{12}$CO 3-2     & 5.62\,, 6.18   & 5.69\,, 5.83  & 6.15\,, 5.43\\
$^{13}$CO 3-2     & 2.74\,, 2.64   & 3.09\,, 2.86  & 1.68\,, 1.55\\  
C$^{18}$O 3-2     & 1.77\,, 1.60   & 2.58\,, 2.25  & 0.40\,, 0.39\\
C$^{18}$O 2-1     & 0.71\,, 0.63   & 1.16\,, 0.99  & 0.14\\
HCO$^+$ 4-3       & 2.69\,, 1.97   & 2.67\,, 2.39  & 2.42\,, 1.90\\
\ce{N2H+} 3-2     & 0.36\,, 0.27   & 0.45\,, 0.51  & 0.32\,, 0.33\\
CN 5/2-3/2        & 3.50\,, 2.63   & 3.50\,, 1.88  & 1.72\,, 1.46\\
CN 7/2-5/2        & 3.69\,, 2.86   & 3.65\,, 2.04  & 2.17\,, 1.91\\
HCN 3-2           & 1.23\,, 1.48   & 2.31\,, 1.42  & 2.06\\
HCN 4-3           & 2.13\,, 2.45   & 3.92\,, 2.27  & 1.94\,, 1.68\\
CS 7-6            & 0.57\,, 0.55   & 1.42\,, 0.25  & 0.35\,, 0.38\\
\ce{C3H2}         & 0.09\,, 0.10   & 0.44\,, 0.07  & 0.02\,, 0.01\\
\ce{C2H}          & 0.09\,, 0.10   & 0.12\,, 0.04  & 0.44\\
\ce{H2CO}         & 0.02\,, 0.03   & 0.08\,, 0.09  & 0.02\\
\hline
\end{tabular}\\[1mm]
\tablefoot{Double entries are for the one-zone\,, two-zone equilibrated models. The meaning of the different components of $\chi$ are explained in Sect.\,\ref{sec:model_approach}}.
\end{table}

The disk gas mass of HD\,143006 is of particular interest. \cite{Trapman2025a} have recently determined the disk gas mass of HD\,143006 to be $M_{\rm disk}\!=\!0.021^{+0.024}_{-0.009}\,M_\odot$, which is about $17\times$ larger than our preferred value of $0.0012\,M_\odot$. \cite{Trapman2025b} have generated a grid of generic DALI models for T\,Tauri stars to calculate the densities, the dust and gas temperature structures and the internal UV-radiation field.  The chemical results are then partly replaced by an external code that uses a small chemical network to simulate only the \ce{CO}, \ce{HCO+} and \ce{N2H+} chemistry for different initial CO-abundances $x_{\rm CO}$ and different cosmic ray ionization rates \citep{vantHoff2017}.  These resulting chemical concentrations are then re-injected into the DALI models to calculate the line fluxes.  Since this is a general-purpose model grid, these models do not use a two-zone disk structure, or take into account the shadow casted by the inclined inner disk. The spectral type and the UV- and X-ray properties of HD\,143006 were not considered. Only the three total line fluxes of $\rm^{13}CO$, $\rm C^{18}O$ and \ce{N2H+}, and one mm-continuum flux are used to determine the disk gas mass. In contrast to the modeling presented here, no spatial information, such as continuum images or radial line intensity profiles were used. 

Using their $M_{\rm disk}$-value in our best-fitting model increases the total $\chi$ from 2.3 to 3.6.  An increased disk mass changes the dust settling and therefore the fits of the continuum fluxes and images become slightly worse. The $\chi_{\rm lin}$ increases from 3.4 to 6.0 (mainly due to larger line emitting radii). It is actually quite surprising that the fit quality worsens so little, which means that the disk gas mass belongs to those parameters that are hard to determine. Indeed, we find that changing many of the other 30 free parameters in our model (see Table~\ref{tab:parameter}) by a factor of 17 (either up or down) leads to much worse fitting results.  

In order to assess how robust our $M_{\rm disk}$-determination really is, we have repeated the fitting experiment with a fixed value for $M_{\rm disk}\!=\!0.021\,M_\odot$, to see in how far the model can adjust to such a given constraint.  The automated fitting algorithm was observed to find an alternative $\chi^2$-minimum and some of the results of this experiment are reported in Table~\ref{tab:Mdisk_var}.  This time, the total $\chi$ only increased from 2.3 to 2.9. The re-fitted model could indeed almost entirely compensate the effects of a $17\times$ higher disk mass by rather subtle changes in other parameters, remarkably the reference gas scale height $H(30\rm\,au)$ (from 3.8\,au to 2.5\,au), and the tapering off power exponent $\gamma$ (from 1.02 to 0.89).

At the same time, the parameters regulating some of the heating processes ($f_{\rm PAH}$ and $\Delta E_{\rm pd}$) were found to be re-adjusted to 30-40\% lower values, making the gas in the outer disk cooler. By making the disk more compact this way, both radially and vertically, and slightly cooler in the gas, the line fluxes and radial intensity profiles of the adjusted high-mass model became hardly distinguishable from the main low-mass model, although the mathematical $\chi$-determinations still clearly favor the low-mass model in all aspects.  Considering, on top of that, the principal uncertainties in thermo-chemical disk modeling, we conclude that it is virtually impossible to derive the disk gas mass to better than an order of magnitude, even when fitting 14 ALMA lines. 

\begin{table*}
\caption{Attenuation factors $r_{\rm line}$ and $r_{\rm cont}$ for contributions from the far disk surface to the line and continuum fluxes.} \label{tab:line_attenuation}
\vspace*{-3mm}
\centering
\begin{tabular}{ccccccccccccc}
\hline
&&&&&&&&&&&\\[-2ex]
&
$^{12}$CO & $^{13}$CO & C$^{18}$O
  & CN $J$=5/2-3/2 & CN $J$=7/2-5/2 & HCN
  & \ce{HCO+} & CS & \ce{N2H+} 
  & \ce{C3H2} & \ce{C2H} & \ce{H2CO} \\
\hline  
&&&&&&&&&&&\\[-2ex]
$r_{\rm line}$ & 
0.08 & 0.16 & 0.36
  & 0.13 & 0.12 & 0.28
  & 0.26 & 0.55 & 0.50
  & 0.78 & 0.85 & 0.91 \\
$r_{\rm cont}$ &
0.67 & 0.67 & 0.67 
  & 0.67 & 0.67 & 0.68 
  & 0.67 & 0.67 & 0.68 
  & 0.67 & 0.68 & 0.69 \\ 
\hline
\end{tabular}
\tablefoot{Attenuation factors are measured from additional radiative transfer experiments on top of the two-zone equilibrated model.}
\end{table*}

Another caveat in this respect is that we have chosen to fix the carbon abundance to our standard value $\rm C/H\!=\!1.38\times10^{-4}$, while freely varying C/O, C/N, and C/S. Our motivation for this choice was that carbon is the element among $\rm\{C,N,O,S\}$ that requires the lowest temperatures to freeze out completely (\ce{NH3} freezes out at similar temperatures as \ce{H2O}).  The fitting procedure, however, resulted in $\rm C/N\!=\!0.25$, see Table~\ref{tab:parameter}, which means that N/H is about $7\times$ larger than solar in our best-fitting model.  In retrospect, anticipating the result $\rm C/N\!<\!solar$, it would have been better to fix the nitrogen abundance to solar and vary C/N, O/N and S/N instead.  However, a simple scaling does not work.  We have studied  a model with $7\times$ lower C/H, $7\times$ higher disk mass, $7\times$ lower $f_{\rm PAH}$, and $7\times$ lower $\alpha_{\rm set}$ to compensate the effect of the higher gas densities on dust settling. That model indeed has the same absolute amounts of carbon, nitrogen, oxygen and dust distributed in the same way, it just has more hydrogen. However, the chemical structure and line results are quite different, for example because of modified heating/cooling, modified \ce{H2}-shielding, and X-ray absorption. The total $\chi$ increases from 2.3 to 2.8.

In fact, all three element abundances, C/H, O/H and N/H, are disputed for disks in the literature. For example, \citep{Bosman2021a} suggested that both C and O are depleted in AS\,209, HD\,163296, and MWC\,480, which matches our results for HD\,143006 in consideration of the previous paragraph.

It seems principally very difficult with thermochemical disk models and current data to pinpoint the absolute element abundances, unless we can observe a pure H-tracer such as HD in the far-infrared, whereas some conclusions about relative element abundances, such as C/O and C/N, are becoming more and more possible with improved ALMA data. In this respect, higher-resolution C$_2$H observations, deeper C$_3$H$_2$ data, and the observations of another oxygen-tracer such as SO \citep{Rivieremarichalar2026} would all be steps into the right direction.

\subsection{The contribution of emission from the rear of the disk}
\label{sec:backillum}
On the right side in Fig.\,\ref{fig:all_mol}, where the disk surface facing the observer is in the shadow, we expect the rear disk surface to be illuminated.  If the line emissions from these two opposite disk surfaces simply add up, it would not matter which surface is closer and which is further, and the left/right results would be the same even if the two disk surfaces are intrinsically very different in brightness. This could explain the rather modest left/right line contrasts observed. In reality, however, line photons emitted from the rear disk surface will partly be re-absorbed by the molecules in the front side, and the dust is also in the way. 

In order to estimate the magnitude of this effect, we have performed additional line radiative transfer calculations which only trace the front side of the disk by letting all rays start at $z\!=\!0$ in the midplane, instead of well behind the disk, which is the default. This way, we can find out about the relative contribution from the rear disk surface as:
\begin{equation}
   F_L = F_{L,1} + r_{\rm line} F_{L,2} \ ,
\end{equation}
where $F_{\rm L}$ is the observed line flux, and $F_{L,1}$ and $F_{L,2}$ are the intrinsic line flux contributions from the front and rear disk surfaces, and $r_{\rm line}$ is the line attenuation factor.  In the additional line radiative transfer experiments, we measure $F_{\rm L}$ from the full rays and $F_{L,1}$ from the half rays. Assuming $F_{L,1} = F_{L,2}$, we can determine $r_{\rm line}$, see Table~\ref{tab:line_attenuation}.  We see $r_{\rm line}$-values as small as $0.08$ for the optically thick lines like $^{12}$CO and CN, but other lines which are more optically thin and/or probe the more tenuous outer disk regions can achieve values as large as 0.9.

In reality, when one disk surface is illuminated ($\to F_{L,1}$) and the other one is not ($\to F_{L,2}$), the left/right line flux contrast is expected to be
\begin{equation}
\frac{F_L^{\rm left}}{F_L^{\rm right}}
= \frac{F_{L,1} + r_{\rm line} F_{L,2}}
       {F_{L,2} + r_{\rm line} F_{L,1}} \ .
\end{equation}
For example, if we assume $F_{L,1}\!=\!2\times F_{L,2}$, then an attenuation factor of $r_{\rm line}\!=\!0.1$ would lead to a left/right line flux contrast of $F_L^{\rm left}/F_L^{\rm right}\!=\!1.75$, which is still much higher than observed, but an attenuation factor of $r_{\rm line}\!=\!0.6$ already results in $F_L^{\rm left}/F_L^{\rm right}\!=\!1.18$, which is close to the observed left/right line flux contrasts, see Table~\ref{tab:result_linefluxes}.
The continuum attenuation factors reported in Table~\ref{tab:line_attenuation} are not as relevant, since the midplane dust rings are physically narrow and are generally located interior to where most of the line emission originates (cf. Fig.~\ref{fig:LineFit1} and Fig.~\ref{fig:ContFit}).
We conclude that while this effect is not very important for the optically thick lines, it may have a significant influence on the prediction of line flux contrasts for the optically thin lines that probe the outer disk parts. We have not included this effect in our analysis because of the technical challenges involved in doing so. 
Mostly affected are the lines for which only unresolved line data were used (\ce{C2H} and \ce{H2CO}) and the very faint \ce{C3H2} line). The high S/N of the optically thick lines means that the slow heating and cooling of the molecular disk layer is still necessary for a good fit to the data.

\subsection{Dependencies on the chemical network}

\begin{table}
\caption{Comparison of models using different chemical networks.}
\label{tab:ChemNetwork}
\vspace*{-6mm}
\begin{center}
\resizebox{80mm}{!}{\begin{tabular}{c|ccc}
\hline
&&&\\[-2.2ex]
&\!main model\!&\!KIDA\,2024\!&\!\sc ChaiTea\,2025\!\\
\hline
&&&\\[-2.2ex]
\ce{H2}   & 0.50    & 0.50    & 0.50    \\
\ce{CO}   & 1.2(-4) & 1.2(-4) & 1.2(-4) \\
\ce{N2}   & 2.4(-4) & 2.4(-4) & 2.4(-4) \\
\hline
&&&\\[-2ex]
H         & 3.3(-3) &     1.5(-3) &     2.9(-3) \\
C         & 1.8(-5) &     1.5(-5) &     1.4(-5) \\
N         & 6.2(-5) &     6.8(-5) &     7.1(-5) \\
\ce{e-}   & 6.6(-8) &     5.5(-8) &     9.5(-8) \\
\hline
&&&\\[-2ex]
\ce{CH2}  & 1.2(-10)&     2.3(-10)&     9.9(-11) \\ 
CH$_3^{\,+}$ & 1.3(-10)&     1.0(-10)&     7.3(-11) \\
&&&\\[-2ex]
\hline
&&&\\[-2ex]
\ce{HCO+} & 2.0(-10)&     1.5(-10)&     1.1(-10) \\
\ce{N2H+} & 7.5(-11)&     4.5(-11)&     8.4(-11) \\
\ce{CN}   & 1.2(-9) &     2.2(-9) &     1.2(-9)  \\
\ce{HCN}  & 9.9(-12)&     2.4(-11)&     1.1(-11) \\
\ce{CS}   & 1.9(-14)&     3.1(-14)&     1.3(-14) \\
\ce{C2H}  & 2.4(-11)&     3.1(-11)&     1.0(-11) \\
\ce{C3H2} & 4.7(-11)&     1.3(-11)&     1.6(-11) \\
\ce{H2CO} & 3.2(-13)&     1.6(-13)&     1.2(-13) \\
&&&\\[-2.2ex]
\hline
&&&\\[-2ex]
$^{12}$CO 3-2    & 5.9  &     5.6  &     6.1  \\
$^{13}$CO 3-2    & 2.7  &     2.6  &     2.8  \\        
C$^{18}$O 3-2    & 1.7  &     1.7  &     1.8  \\
HCO$^+$ 4-3      & 2.3  &     2.1  &     2.3  \\
\ce{N2H+} 3-2    & 0.32 &     0.89 &     0.18 \\
CN 5/2-3/2       & 3.1  &     3.2  &     3.0  \\
HCN 4-3          & 2.3  &     2.7  &     1.9  \\
CS 7-6           & 0.57 &     0.89 &     0.58 \\
\ce{C2H}         & 0.10 &     0.11 &     0.10 \\
\ce{C3H2}        & 0.09 &     0.06 &     0.06 \\
\ce{H2CO}        & 0.03 &     0.02 &     0.02 \\[0.2ex]
\hline
\end{tabular}}
\end{center}
\vspace*{-3mm}
\tablefoot{The first 17 rows show molecular concentrations in the equilibrated one-zone model at the point $r\!=\!80\,$au and $z\!=\!20\,$au, where $\nH\!=\!4\times10^7\rm\,cm^{-3}$, $\Tg\!=\!28\,K$, $\Td\!=\!95\,K$ and where the dust is optically thin ($A_{\rm V,rad}\!=\!0.02$ and $A_{\rm V,ver}\!=\!6\times10^{-4}$). The subsequent 11 rows show the predicted line fluxes [Jy\,km/s]. The notation $a(-b)$ means $a\times 10^{-b}$.}

\end{table}

Table~\ref{tab:ChemNetwork} shows some results from our best-fitting model when we use different chemical rate networks.  The main model uses the UMIST\,2022 database as base chemical network, see further explanations in Sect.\,\ref{sec:ProDiMo}.  In the KIDA\,2024 model, we have exchanged that base network by reactions from the official KIDA release file {\tt kida.uva.2024.zip} provided at the \href{https://kida.astrochem-tools.org/networks}{KIDA online database}. In the {\sc ChaiTea} model \citep{Kanwar2025}, the base network consists of a mixture of the UMIST~2022 and the STAND~2020 \citep{Rimmer2016} network, which automatically generates all gas-kinetic backward reactions and calculates their rates from the $T$-dependent Gibbs free energies of the molecules, and includes termolecular and other three-body reaction rates, also in the high-pressure limit.  In all cases, the base network is then completed with our selection of ProDiMo-specific reactions, see Sect.\,\ref{sec:ProDiMo}, which includes e.g. the UV-photo and X-ray reactions, and the ice chemistry. The latter reactions are hence identical in all three networks.  Table~\ref{tab:ChemNetwork} illustrates the differences to be expected when using different base chemical networks.  While the molecular concentrations can differ by factors up to three at the examined point, the uncertainties in line fluxes are typically smaller, better than a factor of two, often much less.  We note two interesting details here: (i) The KIDA network results in $n_{\ce{C2H}}/n_{\ce{C3H2}}\!\!\sim\!2.4$ whereas the main UMIST model has $\sim\!0.5$, which can be traced back to slightly less efficient gas-kinetic production rates of \ce{C3H2} in the KIDA network. This might help to understand why our main model seems to overpredict the observed \ce{C3H2} line while underpredicting the observed \ce{C2H} line. (ii) The KIDA network leads to a \ce{N2H+} line flux that is about $2.8\times$ larger than the UMIST network, and about $5\times$ larger than the {\sc ChaiTea} network, which shows that this line depends crucially on the chemical network used.

\subsection{The significance of electron donors and PAH molecules}
\begin{table}
\caption{Line fluxes affected by metal abundances and PAHs}
\label{tab:donors}
\vspace*{-3mm}
\resizebox{90mm}{!}{\begin{tabular}{c|cccc}
\hline
& DIANA\tablefootmark{(2)}\!\! & 
  \!low metal\! & \!no PAH\#\! & \!main model\!\\
\hline
&&&&\\[-2ex]
\!\!\ce{HCO+} $J$=4-3 & 0.95 & 1.86 & 1.41 & 1.97\\
\!\!\ce{N2H+} $J$=3-2 & 0.14 & 0.15 & 0.27 & 0.27\\
\hline
\end{tabular}}\\[1mm]
\tablefoot{Fluxes in Jy km/s, from the two-zone equilibrated model.\\
\tablefoottext{2}{The original DIANA standard \citep{Kamp2017} uses $\epsilon_{\rm Mg}\!=\!1.1\times10^{-8}$, $\epsilon_{\rm Na}\!=\!2.3\times10^{-9}$, $\epsilon_{\rm Si}\!=\!1.7\times10^{-8}$, $\epsilon_{\rm Fe}\!=\!1.7\times10^{-9}$, with PAH ice included.}
}
\end{table}

The \ce{HCO+} lines are usually quite strong in disk observation, often comparable to $^{13}$CO, see for example \cite{Pegues2023}.  In our ProDiMo models, we  achieved such \ce{HCO+} line fluxes only after reducing the abundances of potential electron donors, such as Na, Mg, Si and Fe, to a minimum, see Sect.\,\ref{sec:ProDiMo}, assuming that these elements are almost entirely locked up in solid phases. Another important step was to remove PAH ice from the list of chemical species in ProDiMo, so that the PAH molecules (with concentrations $\sim\!10^{-9}$) do not freeze out and can continue to charge up negatively.  Both measures led to considerably lower electron densities in the midplane regions, an increase of the \ce{HCO+} and \ce{N2H+} lifetimes against dissociative recombination, and hence higher concentrations and stronger line fluxes.  In the particular model for HD\,143006 presented in this paper, both measures led to an increase of the \ce{HCO+} and \ce{N2H+} line fluxes by a factor of about two, where the lower metal abundances seem more important for \ce{HCO+} but the PAHs more important for \ce{N2H+}, see Table~\ref{tab:donors}. More details about the impact of X-ray and stellar energetic particles (SEPs) on the \ce{HCO+} and \ce{N2H+} lines can be found in \cite{Rab2017} and \cite{Rab2020a}.  Electron attachment to PAHs can be considered as a proxy for dust charging \citep{Balduin2023}, which is not included in this work. However, for typical dust particle concentrations of order $10^{-14} - 10^{-13}$ the grains would need to pick up $10^4 - 10^5$ charges per grain on average to have the same effect on the electron density.

\subsection{Comparison to previous modeling work}
\label{sec:prevModel}
Although HD 143006 is a particularly interesting disk from the perspective of its physical structure, it is also noteworthy that the overall sample of protoplanetary disks for which consistent models spanning many lines (including less-abundant species such as CS, HCN, and c-C$_3$H$_2$) have been made is relatively small. It is therefore instructive to also see how the results derived here compare, albeit generally, to other such efforts. The modeling of HD 100546 by~\citet{Leemker2024} is similar in the number of lines covered, although their best model is approximated differently from the models here, and the authors used the DALI code~\citep{Bruderer2009,Bruderer2012}. Comparing their fiducial model to the data, it is interesting to note that the radius at which the CN emission reaches its highest value tends to be closer-in than the observations show, a feature our models share, and similarly struggle to jointly fit $^{12}$CO and C$^{18}$O despite using isotope-selective processes.

Another interesting point of comparison is the analysis of AB Aurigae presented by \citet{Rivieremarichalar2020,  Rivieremarichalar2026}. AB Aurigae is a Herbig Ae star, and has a larger disk than HD 143006. It is likely actively accreting~\citep{Speedie2025}, has a high mass, and seems to be both very warm and vertically extended from modeling~\citep{Woitke2019} and thus is be quite different from HD 143006. The literature observations focus on S-bearing species, of which we only have CS, but include other molecules such as HCO$^+$, HCN and H$_2$CO which are also discussed in this work. The ~\citet{Rivieremarichalar2020} models (using the NAUTILUS code,~\citealp{Wakelam2016}) generally favor a C/O ratio around one, and tend to overpredict the radii at which HCO$^+$ is found, which suggests that the electron donor abundances may be lower in other disks as well, even in such drastically different systems.

\section{Summary and Conclusions}
\label{sec:conclusion}

The inclined inner disk of HD\,143006 casts a broad shadow over its outer disk, creating a unique UV laboratory to study the effects of irradiation on chemistry in protoplanetary disks, where high-resolution observations of molecular lines are crucial. In this paper, we have presented new ALMA observations of this disk for CO, $^{13}$CO, C$^{18}$O, CN, HCN, HCO$^+$, N$_2$H$^+$, CS and C$_3$H$_2$.  Combined with previously published C$_2$H and H$_2$CO observations, two continuum ALMA images, and auxiliary data including photometric fluxes, Spitzer/IRS spectrum and observed CO emission heights, this extensive dataset of 14 ALMA lines from nine different molecular species provides a holistic view of the disk's structure, dust evolution, and chemistry.

We subdivided the ALMA data into two disk halves: the illuminated part and the shadowed part.  We developed ProDiMo models for both disk halves by optionally including an inclined inner disk that blocks the stellar illumination of the upper layers of the outer disk while it does not hinder the stellar illumination of the outer disk along the midplane.  The observed line flux contrasts between the illuminated and the shadowed disk sides are not very pronounced.  We explain this phenomenon by introducing ``equilibrated models'' which consider a slow heating/cooling relaxation of gas parcels orbiting in and out of the shadow. This way, the gas retains its heat while passing through the shadow, smoothing out the expected thermal gradients.  We also introduced a new recipe to account for the effects of radial drift in ProDiMo, where dust grains beyond an adjustable size are moved inward to a specified radius, creating a bump in millimeter-opacity as observed in HD\,143006 around 70\,au.

We then fitted a 31-parameter ProDiMo model to the full suite of new and archival observational data. This approach is unique in the literature, aiming to determine the entire structure of the protoplanetary disk around HD\,143006, its element abundances, chemical composition, gas and dust temperature, and the spatial dust structure. The fitting strategy included comparisons between the predicted and the observed radial intensity profiles for both lines and continuum, separately for the illuminated and shadowed disk parts. This way, every model results in an overall reduced $\chi^2$ with adjustable weights for SED, continuum images, line profiles and CO emission heights. We used a genetic algorithm to minimize $\chi^2$ to find the best-fitting model parameters.

The model can fit the double-ring structure observed in two ALMA continuum images reasonably well, where the first ring around 30\,au is identified as the inner rim of the outer disk, and the second ring around 70\,au is interpreted as an accumulation of large ($>\!80\,\mu$m) dust grains from the outer disk via radial drift. In contrast, the gas distribution is smooth and radially more extended. The gas column density first increases around 30\,au $(\epsilon\!\approx\!-4)$ before it decreases exponentially beyond 70\,au $(\gamma\!\approx\!1)$.  The inner rim of the outer disk at 30\,au has a relative height of $h/r\!\approx\!0.15$. This height creates a collecting area that absorbs about the right amount of starlight to produce the mid-far IR bump by dust thermal re-emission as observed in the SED around 30\,$\mu$m. The stellar illumination of that rim along the disk midplane remains about constant over time, i.e.\ it is not blocked by the inclined inner disk. This constant illumination is clearly necessary in the models to heat the disk and fit the ALMA continuum images on both disk sides.

Our model can satisfactorily fit the 14 ALMA line fluxes from both the illuminated and the shadowed disk sides with the exception of C$^{18}$O, which is about $4\times$ too strong, and C2H, which is about $4\times$ too weak. The predicted general shape of the line intensity profiles and the different radial extensions of the disk in these lines match the observations well.  In the central part of the outer disk between about 40\,au and 140\,au, most of the ALMA lines are optically thick and probe the gas temperature structure at different heights.  Here, our fits of the observed radial line intensity profiles require a combination of low gas temperatures (20-70\,K) and low turbulent line width ($\sim\rm\!0.04\,km/s$). These gas temperatures are substantially lower than the local dust temperatures in the model, which requires low densities and a settled dust structure to allow the gas to decouple thermally from the dust. Outside of this radial domain, around 150\,au, the model is featured by a weak outer PDR-region caused by the external UV irradiation, with significant photo-chemical production of CN, HCN and \ce{C2H}, leading to enhanced CN and HCN line intensities at $\sim\!180\,$au. This feature is not seen in the observations.

Most lines observed with ALMA are emitted from relative heights $h/r\!\approx\!0.15-0.3$, i.e.\ from layers that are substantially higher than the inner rim of the outer disk. In the model, such a scale height relation between gas and dust is only possible when we assume a strong dust settling ($\alpha_{\rm settle}\!\approx\!2\times10^{-4}$) that affects even the small grains at these heights according to our settling method adopted from \cite{Riols2018}. These layers hence receive the full starlight when the inclined inner disk in not in the way. They host only a few very small dust grains, are virtually ice-free, and controlled by photo-chemistry, making molecular UV-shielding a key process. 

Some ALMA lines do probe the deeper disk layers as well, in particular \ce{N2H+} and CS, although this is not so clear-cut, because the height of the line emission can change with illumination -- we observe this effect also for HCN, \ce{C2H}, \ce{C3H2} in the models. These lines can hence be affected by ice UV photo-desorption and behave differently in time-dependent periodical chemical models -- they can become brighter in the shadows.

The dust in the outer disk is expected to react on short timescales ($\tau_{\rm dust}\!<\!0.1\,$yr) to the periodic illumination. In contrast, the chemistry has a wide spectrum of relaxation timescales. The photochemical processes active in the observable disk surface are fast ($\tau^{\rm photo}_{\rm chem}$), whereas the chemical processes that build up the ice and alter the primary element carriers ($\tau^{\rm ice}_{\rm chem}$) are slow. Gas cooling timescales fall somewhere in the middle, establishing a clear hierarchy of relaxation timescales relative to the orbital period $P$
\begin{equation}
  \tau_{\rm dust}<\tau^{\rm photo}_{\rm chem}<\tau_{\rm cool}\approx P \ll \tau^{\rm ice}_{\rm chem} \ .
\end{equation}
Taking the slow gas cooling into account was a crucial step toward a successful fitting of the line emissions from both disk sides with our model. The relation $\tau^{\rm photo}_{\rm chem} < P \ll \tau^{\rm ice}_{\rm chem}$ was confirmed by time-dependent chemical models for selected points in the disk where we periodically change between the temperature and radiation field conditions in the illuminated and the shadowed disk sides, simulating periodic orbits through a fixed shadow. These models show a number of surprising effects in deeper layers where the thermal and photo-desorption of ices injects fresh carbon, nitrogen and sulfur into an oxygen-poor gas.

The paper also presents a few additional studies concerning the chemical pathways to form the observable molecules and how the main results change when different chemical networks and element abundances are used:
(1) The element abundances of Fe, Si, Mg and Na, which act as electron donors, needed be to reduced to a minimum to fit the \ce{N2H+} and \ce{HCO+} observations.
(2) \ce{N2H+} is primarily created via X-ray ionization of \ce{H2}, whereas for \ce{HCO+} there is an additional photo-chemical formation path via \ce{CH5+} that does not require \ce{H2}-ionizations.\linebreak
(3) The ion-neutral and neutral-neutral chemistry of small hydro-carbon molecules is crucial for the production of all observable molecules but \ce{N2H+}. \ce{CH2} and \ce{CH3+} constitute important branching points in the H-C chemistry. 
(4) The inclusion of X-ray chemistry seems mandatory, whereas using different chemical rate networks (UMIST/KIDA/ChaiTea) appears less crucial, although the KIDA-based models result in about $2-3\times$ stronger \ce{CS} and \ce{N2H+} lines.

Our simultaneous modeling of 14 ALMA lines allowed us to constrain the C/O-ratio in the outer disk of HD\,143006 to be $\rm CO\!=\!1.05_{-0.03}^{+0.1}$. The model clearly favors C/O-values larger than one, with the CS and \ce{C2H} lines relative to CO and CN providing the strongest constraints for a carbon-rich environment. Conversely, the total disk mass of HD\,143006 remains poorly constrained. Our models favor a value of $0.0012\,M_\odot$, but models with $0.021\,M_\odot$ (as suggested by \citealt{Trapman2025b}) result in only slightly worse line fits after re-adjusting a few disk shape and heating/cooling parameters.

\begin{acknowledgements}
We thank Manuel G{\"u}del for a discussion about the ROSAT observations of HD\,143006, and his assistance to estimate its X-ray luminosity. We are grateful to the referee, whose insightful comments helped to considerably improve this manuscript.
This paper makes use of the following ALMA data: ADS/JAO.ALMA\#2016.1.00484.L, ADS/JAO.ALMA\#2019.1.01683.S, ADS/JAO.ALMA\#2021.1.01123.L, and ADS/JAO.ALMA\#2023.1.00334.S. ALMA is a partnership of ESO (representing its member states), NSF (USA) and NINS (Japan), together with NRC (Canada), NSTC and ASIAA (Taiwan), and KASI (Republic of Korea), in cooperation with the Republic of Chile. The Joint ALMA Observatory is operated by ESO, AUI/NRAO and NAOJ.
Based on archival observations collected at the European Organisation for Astronomical Research in the Southern Hemisphere under ESO program 0103.C-0470 (PI: M. Benisty).

\end{acknowledgements}

\bibliographystyle{aa}
\bibliography{references}

\begin{appendix}
\section{Observational details}

\noindent\begin{minipage}{\textwidth}
    \centering
    \begin{tabular}{cccccccc}
    \hline
    Spectral setup & Date & Antennas & Baselines & Time on source & Bandpass calibrator & Phase calibrator & Flux calibrator \\
                   &      &          &    [m]    &     [m:s]      &                     &                  &      \\ 
    \hline 
    &&&&&&&\\*[-2ex]
    A & 2019-12-17 & 45 & 15 - 313 & 07:05  & J1337-1257 & J1626-2951 & J1337-1257 \\
      & 2020-01-09 & 45 & 15 - 500 & 14:10 & J1517-2422 & J1626-2951 & J1517-2422 \\
      & 2021-05-03 & 43 & 15 - 2000 & 27:13 & J1517-2422 & J1553-2422 & J1517-2422 \\
      & 2021-05-06 & 45 & 15 - 2500 & 27:13 & J1517-2422 & J1553-2422 & J1517-2422 \\
    \hline
    &&&&&&&\\*[-2ex]
    B & 2019-12-16 & 42 & 15 - 324 & 13:10 & J1517-2422 & J1626-2951 & J1517-2422 \\
      & 2019-12-17 & 45 & 15 - 313 & 13:10 & J1337-1257 & J1517-2422 & J1337-1257 \\
      & 2021-04-21 & 44 & 15 - 1400 & 49:34 & J1337-1257 & J1553-2422 & J1337-1257\\
      & 2021-06-08 & 46 & 15 - 2400 & 49:32 & J1517-2422 & J1553-2422 & J1517-2422\\
    \hline
    \end{tabular}
    \captionof{table}{Calibrators, baselines, and observation dates for project 2019.1.01683.S}
    \label{tab:calibration}
\end{minipage}

\section{Examples of masks used for the observational data}

\noindent\begin{minipage}{\textwidth}
    \centering
    \includegraphics[width=\linewidth]{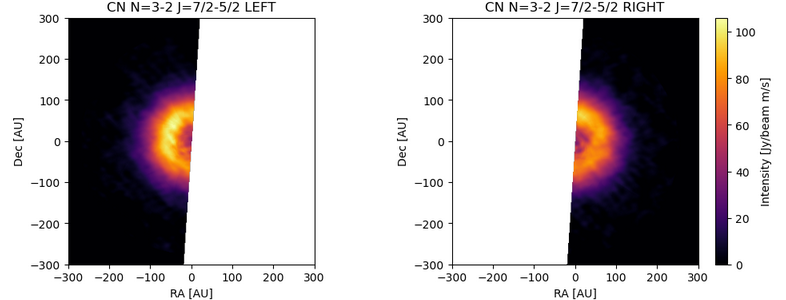}
    \captionof{figure}{An example for the left and right masking for the integrated flux calculation. The molecular transition shown here is CN (N=3-2 J=7/2-5/2).}
    \label{fig:example_flux_mask}

\end{minipage}

\begin{figure}[b]
    \centering
    \includegraphics[width=1\linewidth]{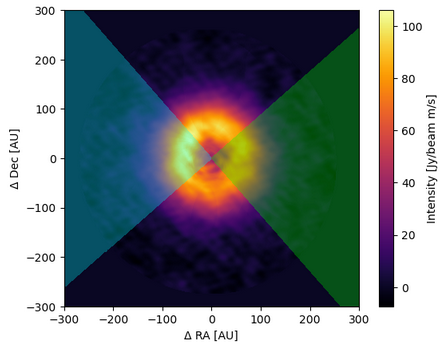}
    \caption{An example for the masking of the disk for the radial profile extraction. The blue quarter visualizes the left mask and the green quarter the right mask. The molecular transition shown here is CN ($N=3-2 J=7/2-5/2$).}
    \label{fig:example_radialprofile_mask}
\end{figure}

\FloatBarrier

\clearpage
\section{Continued Figures}

\noindent\begin{minipage}{\textwidth}
  \hspace*{3mm}\vspace*{-3mm}
  \begin{tabular}{cc}
  one-zone equilibrated model & two-zone equilibrated model\\
    \hspace*{-4mm}
    \includegraphics[page=1,width=88mm,height=59mm,trim=35 60 55 350,clip] 
    {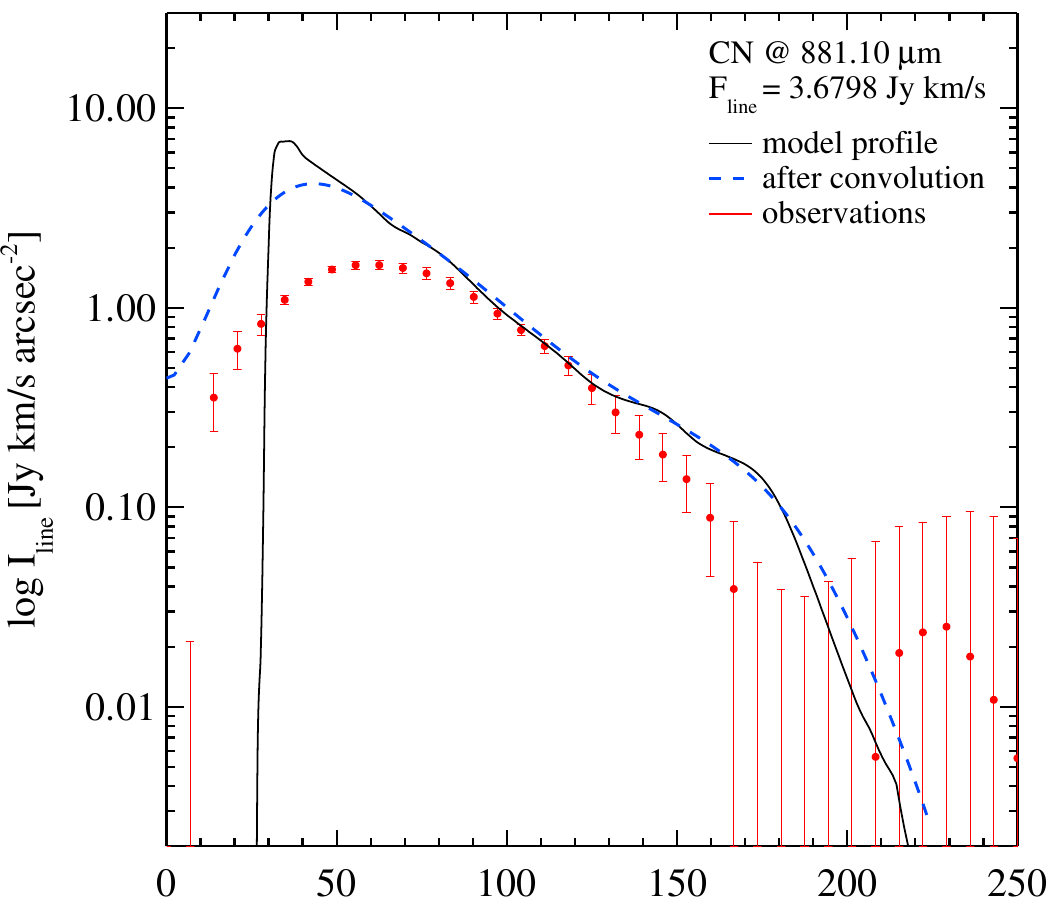} &          
    \hspace*{-6mm}
    \includegraphics[page=1,width=88mm,height=59mm,trim=35 60 55 350,clip] 
    {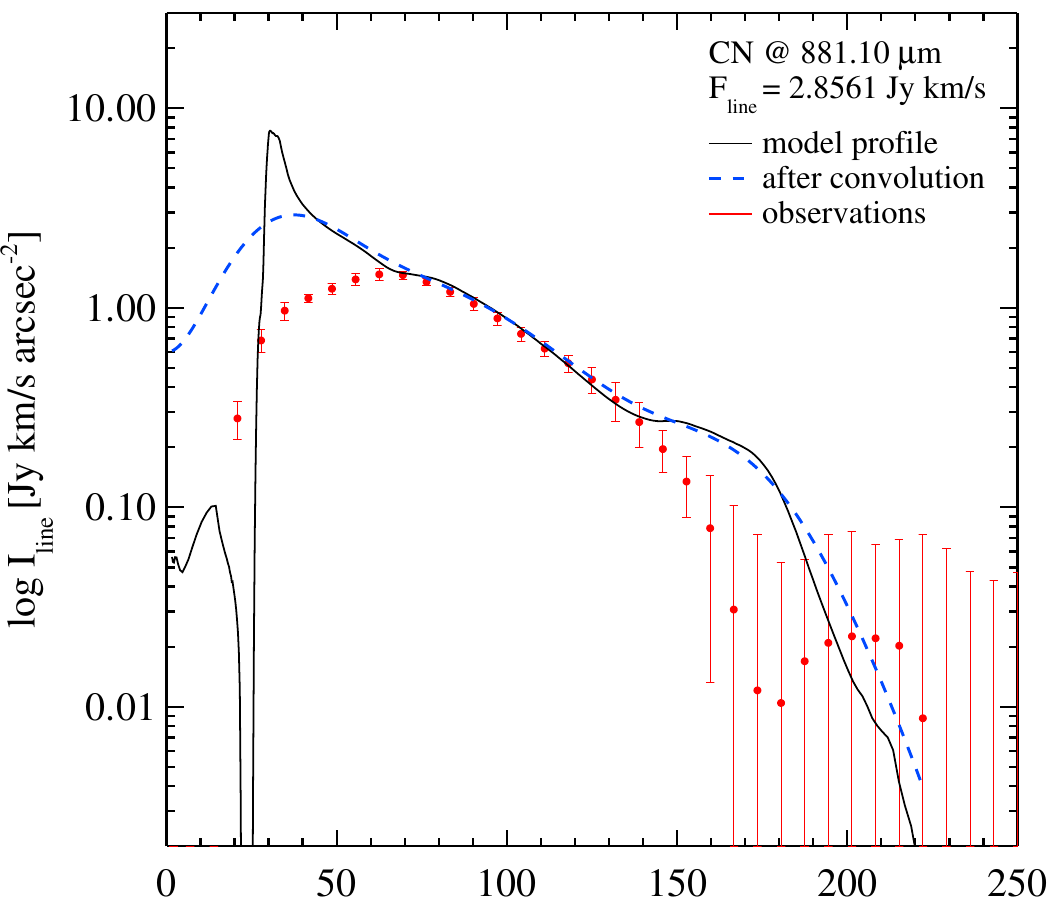} \\[-1mm]
    \hspace*{-4mm}
    \includegraphics[page=2,width=88mm,height=59mm,trim=35 60 55 350,clip] 
    {Figs/out_oneWeight_C1.pdf} &         
    \hspace*{-6mm}
    \includegraphics[page=2,width=88mm,height=59mm,trim=35 60 55 350,clip] 
    {Figs/out_twoWeight_C1.pdf} \\[-1mm]
    \hspace*{-4mm}
    \includegraphics[page=3,width=88mm,height=62mm,trim=35 30 55 350,clip] 
    {Figs/out_oneWeight_C1.pdf} &         
    \hspace*{-6mm}
    \includegraphics[page=3,width=88mm,height=62mm,trim=35 30 55 350,clip] 
    {Figs/out_twoWeight_C1.pdf} \\[0mm]
  \end{tabular}
  \captionof{figure}{Continued from Fig.\,\ref{fig:LineFit1} for CN, $^{13}$CO and C$^{18}$O lines.}
  \label{fig:LineFit2}
  \vspace*{-3mm}
\end{minipage}

\clearpage

\noindent\begin{minipage}{\textwidth}
  \hspace*{3mm}\vspace*{-3mm}
  \begin{tabular}{cc}
  one-zone equilibrated model & two-zone equilibrated model\\
    \hspace*{-4mm}
    \includegraphics[page=1,width=88mm,height=59mm,trim=35 60 55 350,clip] 
    {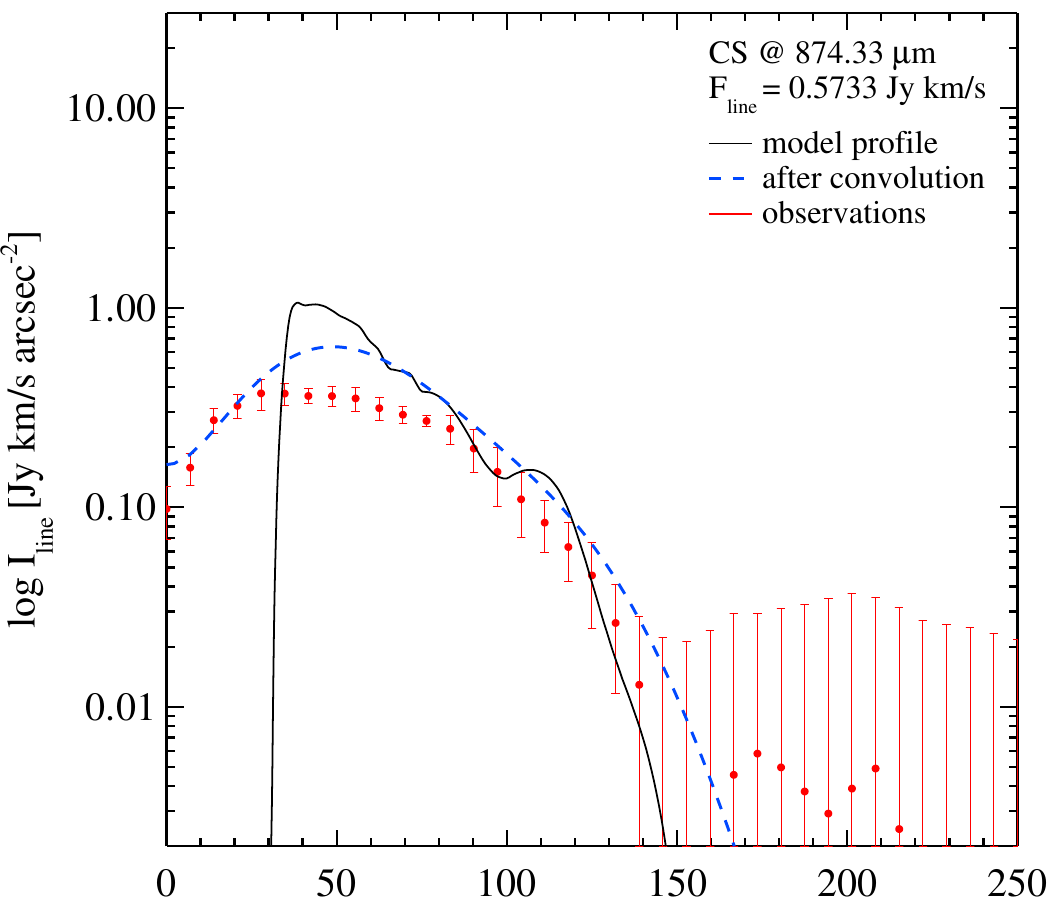} &         
    \hspace*{-6mm}
    \includegraphics[page=1,width=88mm,height=59mm,trim=35 60 55 350,clip] 
    {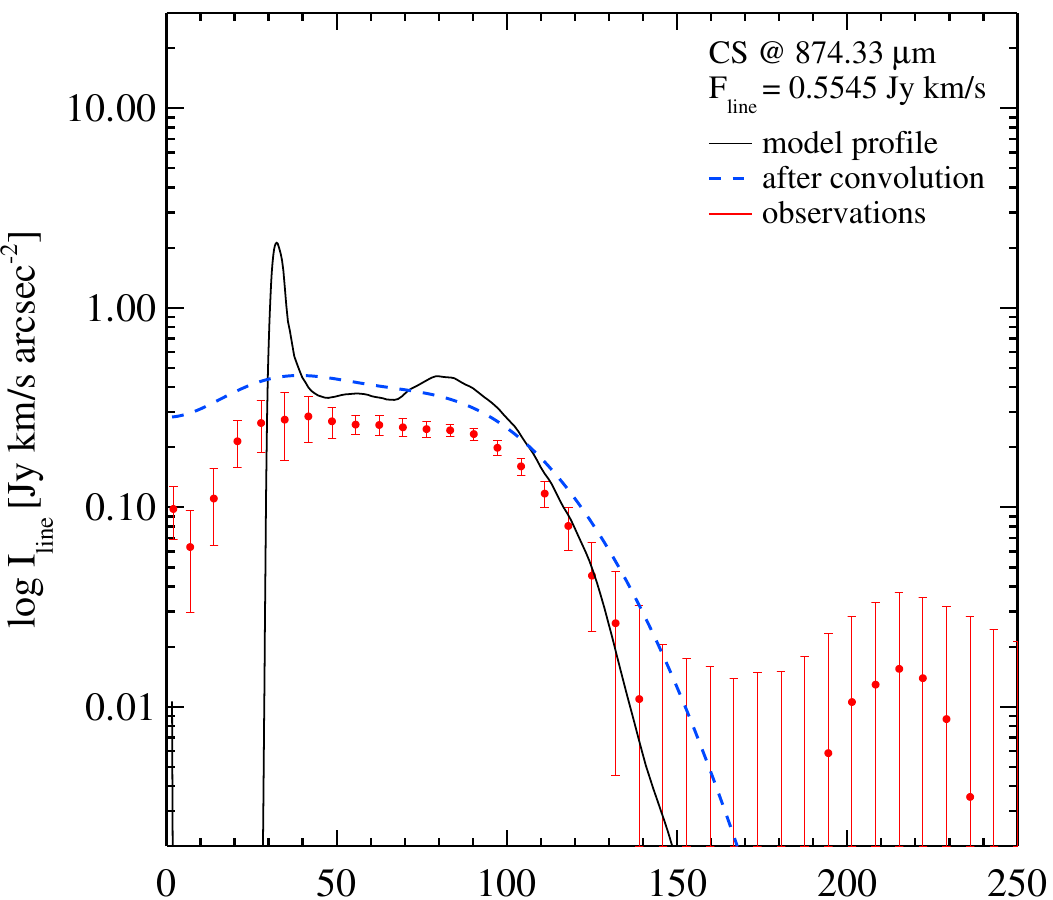} \\[-1mm]
    \hspace*{-4mm}
    \includegraphics[page=2,width=88mm,height=59mm,trim=35 60 55 350,clip] 
    {Figs/out_oneWeight_C2.pdf} &         
    \hspace*{-6mm}
    \includegraphics[page=2,width=88mm,height=59mm,trim=35 60 55 350,clip] 
    {Figs/out_twoWeight_C2.pdf} \\[-1mm]
    \hspace*{-4mm}
    \includegraphics[page=3,width=88mm,height=62mm,trim=35 30 55 350,clip] 
    {Figs/out_oneWeight_C2.pdf} &         
    \hspace*{-6mm}
    \includegraphics[page=3,width=88mm,height=62mm,trim=35 30 55 350,clip] 
    {Figs/out_twoWeight_C2.pdf} \\[1mm]
  \end{tabular}
  \captionof{figure}{Continued from Fig.\,\ref{fig:LineFit1} for CS, N$_2$N$^+$ and C$_2$H$_3$ lines.}
  \label{fig:LineFit3}
  \vspace*{-2mm}
\end{minipage}

\begin{figure*}
  \hspace*{-3mm}\vspace*{-3mm}
  \begin{tabular}{cc}
  \small HCN $4\!\to\!3$\ \ \ one-equilibrated & \small two-equilibrated\\
    \hspace*{-4mm}
    \includegraphics[page=1,width=90mm,trim=30 48 95 572,clip] 
    {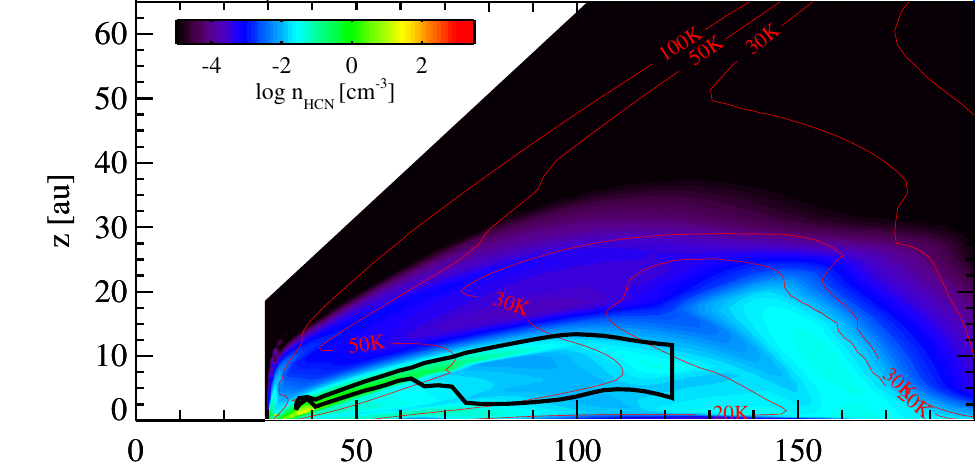} &
    \hspace*{-6mm}
    \includegraphics[page=1,width=90mm,trim=30 48 95 572,clip] 
    {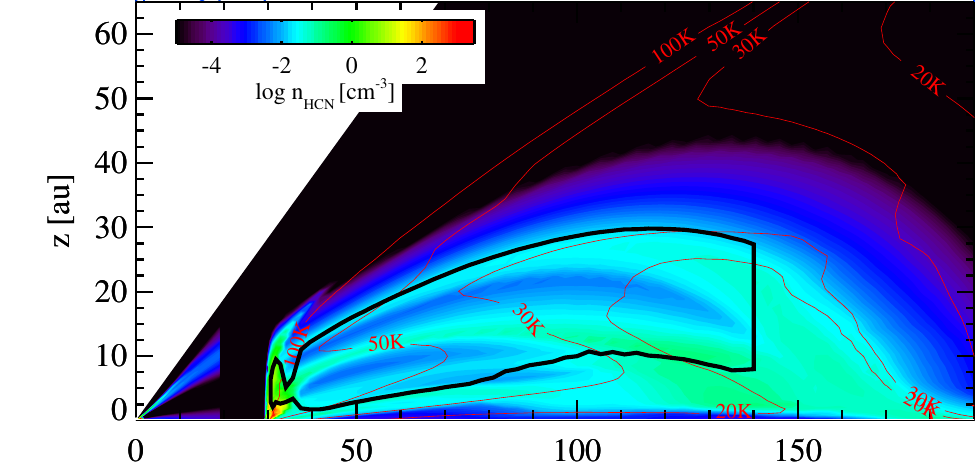} \\[0mm]
  \small HCO$^+$ $4\!\to\!3$\ \ \ one-equilibrated & \small two-equilibrated\\
    \hspace*{-4mm}
    \includegraphics[page=2,width=90mm,trim=30 48 95 572,clip] 
    {Figs/out_oneWeight_C3.pdf} &
    \hspace*{-6mm}
    \includegraphics[page=2,width=90mm,trim=30 48 95 572,clip] 
    {Figs/out_twoWeight_C3.pdf} \\[0mm]
  \small N2H$^+$ $3\!\to\!2$\ \ \ one-equilibrated & \small two-equilibrated\\
    \hspace*{-4mm}
    \includegraphics[page=3,width=90mm,trim=30 48 95 572,clip] 
    {Figs/out_oneWeight_C3.pdf} &
    \hspace*{-6mm}
    \includegraphics[page=3,width=90mm,trim=30 48 95 572,clip] 
    {Figs/out_twoWeight_C3.pdf} \\[0mm]
  \small CS $7\!\to\!6$\ \ \ one-equilibrated & \small two-equilibrated\\
    \hspace*{-4mm}
    \includegraphics[page=4,width=90mm,trim=30 30 95 572,clip] 
    {Figs/out_oneWeight_C3.pdf} &
    \hspace*{-6mm}
    \includegraphics[page=4,width=90mm,trim=30 30 95 572,clip] 
    {Figs/out_twoWeight_C3.pdf} \\[1mm]
  \end{tabular}
  \caption{Continuation of Fig.~\ref{fig:LineFormation1} for HCN, \ce{HCO+}, \ce{N2H+} and CS.}
  \label{fig:LineFormation2}
  \vspace*{-3mm}
\end{figure*}
\begin{figure*}
  \hspace*{-3mm}\vspace*{-3mm}
  \begin{tabular}{cc}
  \small c-C$_3$H$_2$ $4_{3,2}\!\to\!3_{0,3}$\ \ \ one-equilibrated & two-equilibrated\\ 
    \hspace*{-4mm}
    \includegraphics[page=1,width=90mm,trim=30 48 95 572,clip] 
    {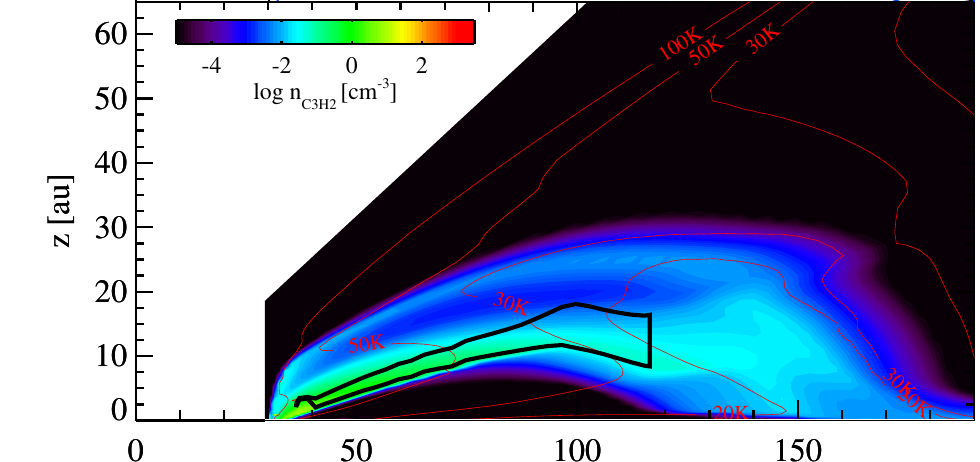} &
    \hspace*{-6mm}
    \includegraphics[page=1,width=90mm,trim=30 48 95 572,clip] 
    {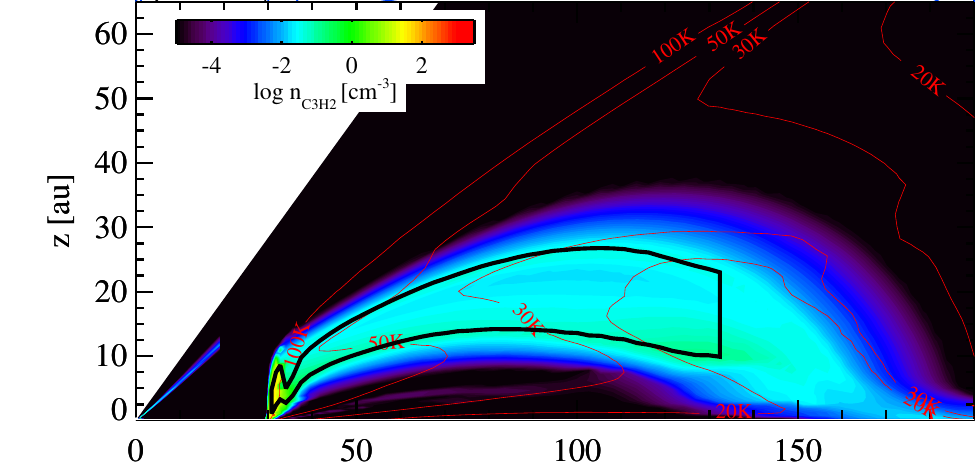} \\[0mm]
  \small C$_2$H\ \ $N$=3-2, $J$=7/2-5/2, $F$=4-3\ \ \ one-equilibrated & two-equilibrated\\ 
    \hspace*{-4mm}
    \includegraphics[page=2,width=90mm,trim=30 48 95 572,clip] 
    {Figs/out_oneWeight_C4.pdf} &
    \hspace*{-6mm}
    \includegraphics[page=2,width=90mm,trim=30 48 95 572,clip] 
    {Figs/out_twoWeight_C4.pdf} \\[0mm]
  \small H$_2$CO\ \ $3_{0,3}\to2_{0,2}$\ \ \ one-equilibrated & two-equilibrated\\ 
    \hspace*{-4mm}
    \includegraphics[page=3,width=90mm,trim=30 30 95 572,clip] 
    {Figs/out_oneWeight_C4.pdf} &
    \hspace*{-6mm}
    \includegraphics[page=3,width=90mm,trim=30 30 95 572,clip] 
    {Figs/out_twoWeight_C4.pdf} \\[1mm]
  \end{tabular}
  \caption{Continuation of Fig.~\ref{fig:LineFormation1} for \ce{C3H2}, \ce{C2H} and \ce{H2CO}.}
  \label{fig:LineFormation3}
  \vspace*{-3mm}
\end{figure*}

\FloatBarrier

\end{appendix}

\end{document}